\documentclass[%
 aip,
 amsmath,amssymb,
 reprint,%
]{revtex4-1}

\usepackage{graphicx}
\usepackage{dcolumn}
\usepackage{bm}

\usepackage[utf8]{inputenc}
\usepackage[T1]{fontenc}
\usepackage{mathptmx}
\usepackage{textcomp}
\usepackage{gensymb}
\usepackage[none]{hyphenat}

\begin{document}

\title{Evaluation of the Thermal Stability of TiW/Cu Heterojunctions Using a Combined SXPS and HAXPES Approach}

\author{C.~Kalha}
\affiliation{Department of Chemistry, University College London, 20 Gordon Street, London, WC1H~0AJ, United Kingdom.}

\author{M.~Reisinger}
\affiliation{Kompetenzzentrum Automobil- und Industrie-Elektronik GmbH, Europastraße 8, 9524 Villach, Austria.}

\author{P.~K.~Thakur}%
\author{T.~-L.~Lee}%
\affiliation{Diamond Light Source Ltd., Harwell Science and Innovation Campus, Didcot, Oxfordshire, OX1 3QR, United Kingdom.}

\author{S. Venkatesan}
\affiliation{Infineon Technologies AG, Am Campeon 1-15, 85579 Neubiberg, Germany.}

\author{M.~Isaacs}%
\affiliation{Harwell XPS, Research Complex at Harwell (RCaH), Didcot, OX11 0FA}
\affiliation{Department of Chemistry, University College London, 20 Gordon Street, London, WC1H~0AJ, United Kingdom.}

\author{R.~G.~Palgrave}
\affiliation{Department of Chemistry, University College London, 20 Gordon Street, London, WC1H~0AJ, United Kingdom.}

\author{J.~Zechner}
\author{M.~Nelhiebel}
\affiliation{Kompetenzzentrum Automobil- und Industrie-Elektronik GmbH, Europastraße 8, 9524 Villach, Austria.}

\author{A.~Regoutz}
 \email{a.regoutz@ucl.ac.uk}
\affiliation{Department of Chemistry, University College London, 20 Gordon Street, London, WC1H~0AJ, United Kingdom.}

\date{\today}

\begin{abstract}
Power semiconductor device architectures require the inclusion of a diffusion barrier to suppress, or at best prevent the interdiffusion between the copper metallisation interconnects and the surrounding silicon substructure. The binary pseudo-alloy of titanium-tungsten (TiW), with $>$70~at.\% W, is a well established copper diffusion barrier but is prone to degradation via the out-diffusion of titanium when exposed to high temperatures ($\geq$400$\degree$C). Here, the thermal stability of physical vapour deposited (PVD) TiW/Cu bilayer thin films in Si/SiO\textsubscript{2}(50~nm)/TiW(300~nm)/Cu(25~nm) stacks were characterised in response to annealing at 400$\degree$C for 0.5~h and 5~h, using a combination of soft and hard X-ray photoelectron spectroscopy (SXPS and HAXPES) and transmission electron microscopy (TEM). Results show that annealing promoted the segregation of titanium out of the TiW and interdiffusion into the copper metallisation. Titanium was shown be driven toward the free copper surface, accumulating there and forming a titanium oxide overlayer upon exposure to air. Annealing for longer timescales promoted a greater out-diffusion of titanium and a thicker oxide layer to grow on the copper surface. However, interface measurements suggest that the diffusion is not significant enough to compromise the barrier integrity and the TiW/Cu interface remains stable even after 5~h of annealing.

\end{abstract}

\maketitle

\section{Introduction}

Power semiconductor device technologies are at the core of the power-train and battery-management systems in hybrid and electric vehicles.~\cite{Shen_2007, 2020_Do, Iannaccone_2021} With the continuing global efforts to popularise the electrification of the automotive industry, there is incentive towards developing devices with smaller feature sizes, improved reliability, and higher power capacity. These targets are being realised through the replacement of aluminium topside metallisation connections in these devices with copper. Copper (Cu), compared to aluminium (Al) is shown to exhibit improved thermal and electrical conductivity, superior electromigration resistance, low resistivity, and enhanced lifetime.~\cite{IBM, Behrens_2013} \par

However, the application of copper metallisations has been under intense investigation due to the material's tendency to drift into silicon substructures at relatively low temperatures,~\cite{Souli_2017} forming deep level traps and intermetallic silicide compounds, both of which severely impact the reliability of the device.~\cite{STOLT1991147, Shacham_Diamand_1993, Sachdeva_2001} To overcome this limitation, a diffusion barrier (DB) is implemented between the metallisation and silicon, designed to suppress the diffusion of copper, reducing the likelihood of the Cu-Si interaction.\par

The binary pseudo-alloy of titanium-tungsten (TiW) has long been considered a viable copper diffusion barrier candidate and has a well-established history. TiW was first proposed by Cunningham~\textit{et al.} as a means of improving the corrosion resistance and metallurgical properties of metal-Au systems in integrated circuits.~\cite{Cunningham_1970} They highlighted that titanium (Ti) has a 10~at.\% solubility limit in tungsten (W) at temperatures below 600{\textdegree}C, and therefore should excess of this concentration be present, a pseudo-alloy of TiW can be created, where the excess titanium is not dissolved or alloyed with the tungsten, and is therefore able to impart desirable characteristics (i.e. adhesion and corrosion resistance). A TiW film consisting of 10-30~at.\% Ti has been shown to offer an increased barrier failure temperature compared to a single tungsten or titanium barrier.~\cite{Shen_1986, OLOWOLAFE199337} Since the introduction of TiW and its associated benefits, TiW has been applied between various metal contacts and metallisation layers, all with moderate to good success.~\cite{Christou_1977, NICOLET_1978, GHATE1978117, CANALI19829, Babcoock_1982, Olowolafe_1985, Babcock_1986, Krautz_1988, ASHKENAZI1993746, Wang_1993, Chiou_1995, Chang_2000, BHAGAT20061998, PETROVIC20102099, WANG2011979, Plappert_2012, FUGGER20142487} However, these studies predominately focus on determining the conditions that are required for the barrier to fail (i.e. temperature and duration), with less emphasis on defining the mechanism that promotes the failure or studying the chemical states across the critical diffusion barrier/metallisation interface.\par

Thermal exposure during production or service has shown to initiate and propagate the segregation of titanium out of the barrier, and the subsequent diffusion into the adjacent metallisation, given sufficient time.~\cite{FUGGER20142487} This migration of titanium is assumed to occur via defect sites, such as grain boundaries.~\cite{Olowolafe_1985} Depletion of titanium is thought to encourage inter-layer delamination, but also allow for copper to bypass the TiW system and interact with the underlying silicon substructure. Moreover, at high temperatures intermetallic compounds can form, leading to the consumption of the TiW barrier and subsequently allowing for the Cu-Si interaction.~\cite{Wang_1993} Therefore, the performance of these devices is heavily constrained by the reliability of these metallisation layers, and therefore it is imperative to improve the understanding of these systems when subjected to high temperatures for prolonged timescales.\par

In order to obtain a greater insight into the thermal stability of the TiW/Cu system, we employ a combination of soft and hard X-ray photoelectron spectroscopy (SXPS and HAXPES) to study the chemical states and electronic structure across post-deposition annealed Si/SiO\textsubscript{2}/TiW/Cu thin film stacks. HAXPES extends the probing depth of conventional SXPS by utilising a higher photon energy, thus turning the technique from surface- to bulk-sensitive.~\cite{Woicik2016, HAXPES_Big_Boy} Additionally, the films were annealed for varying timescales to simulate the different thermal budgets experienced by a real device during the many heat treatments within the manufacturing stages. The combination of SXPS and HAXPES provides the opportunity to gain information from two depth perspectives and so offers a non-destructive approach for characterising the buried TiW/Cu interface in its true state. Complimentary transmission electron microscopy (TEM) with energy-dispersive X-ray spectroscopy (EDS) is employed in an effort to better map out the titanium diffusion process and assess any changes to the microstructure in response to annealing. Additionally, laboratory-based SXPS was utilised to conduct depth profile measurements and quantify the elemental distribution across the bilayer, complimenting the synchrotron XPS measurements.

\section{\label{sec:Method}Methodology}
Un-patterned Si(100)/SiO\textsubscript{2}(50~nm)/TiW(300~nm)/Cu(25~nm) thin film stacks were manufactured using a standard industrial process. Both the TiW and copper films were deposited via DC magnetron physical vapour deposition (PVD) using an argon (Ar) discharge. Both TiW and Cu deposition stages were conducted in the same deposition chamber and without a vacuum break. The deposition chamber operated under a base pressure of 10\textsuperscript{-8}-10\textsuperscript{-7}~mbar. The TiW film was deposited using a composite TiW sputter target containing 30~at.\% Ti concentration. A detailed description of the deposition process used can be found in Refs.~\cite{Plappert_2012, SAGHAEIAN2019137576}. Samples were directly annealed to 400{\textdegree}C using a 10{\textdegree}C/s heating rate and held there for two timescales - 0.5 and 5~h. Annealing was conducted in an inert forming gas atmosphere (95\% N\textsubscript{2}, 5\% H\textsubscript{2}). To provide a reference, the annealed samples are compared to an as-deposited (AD) sample (i.e.~no anneal). 400{\textdegree}C was selected as the target annealing temperature as it is a common temperature experienced by a device during the manufacturing stages.~\cite{Plappert_2012}\par

Synchrotron-based SXPS and HAXPES measurements were conducted at beamline I09 at the Diamond Light Source, UK, employing 1587.4~eV (1.6~keV) and 5927.4~eV (5.9~keV) photon excitation energies, respectively.~\cite{Duncan_18} The maximum inelastic mean free path ($\lambda$) for copper at 1.6~keV and 5.9~keV is 2.2~nm and 6.4~nm, respectively (calculated using QUASES~\cite{TPP-2M_QUASES}). Following the assumption that the signal decay of a core level is exponential, the probing depth can be estimated as 95\% of the signal (i.e.~3$\lambda$). Therefore, the probing depth is estimated to be 6.6~nm for SXPS and 19.2~nm for HAXPES. The 1.6~keV photon energy was selected using a plane-grating monochromator with a grating of 400~lines/mm. The HAXPES 5.9~keV photon energy was selected using a liquid nitrogen cooled Si(111) double crystal monochromator and an additional Si(004) channel cut post monochromator. The end station is equipped with a high-voltage VG Scienta EW4000 electron analyzer, which has an angular acceptance of $\pm$28{\textdegree}. Measurements were conducted at a base pressure of $\approx$3.5x10$\textsuperscript{-10}$~mbar, and carried out in grazing incidence and near normal emission geometry. Even with HAXPES, the 25~nm metallisation overlayer is too thick for photoelectrons from the TiW layer to escape. Therefore, careful in-situ thinning (but not removal) of the copper was carried out using a de-focused Ar\textsuperscript{+} ion sputter beam. Ar\textsuperscript{+} sputtering was conducted step-wise, with a total of four iterations carried out before the copper was suitably thinned while ensuring the TiW/Cu interface was not compromised. The argon gun operated at a 1~keV energy and 5$\times$10\textsuperscript{-5}~mbar pressure, providing a current of 10.5~$\mu$A. Survey, key core level, and valence band spectra were collected at the centre of the sputter crater with both SXPS and HAXPES. A pass energy of 100~eV was used to acquire the valence band spectra, and 70~eV for survey and core levels for both SXPS and HAXPES measurements. The experimental resolution when using a pass energy of 70~eV was found to be 390~meV and 240~meV for the 1.6~keV and 5.9~keV photon energies, respectively, as determined from measuring one standard deviation either side of the Fermi edge of a gold polycrystalline foil (16/84\% method). The binding energy scale of core level spectra was calibrated to the HAXPES recorded Cu~2\textit{p}\textsubscript{3/2} BE position measured at the TiW/Cu interface.\par
SXPS elemental depth profiles across the copper metallisation were conducted using a Kratos Axis Supra laboratory-based instrument (EPSRC National Facility for X-ray Photoelectron Spectroscopy, Harwell, UK). The instrument uses a monochromatic Al~K$\alpha$ (1486.7~eV) X-ray source, and operated with a 15~mA emission current, 12~kV anode voltage (180~W), 700$\times$300~$\mu$m\textsuperscript{2} elliptical spot size, and a base pressure of 1.2$\times$10\textsuperscript{-8}~mbar. Depth profiles were carried out using a focused 500~eV Ar\textsuperscript{+} ion gun beam, with an 11~mA emission current, rastering over a 3$\times$3~mm\textsuperscript{2} area. A total of 15 etch (or sputter) steps were taken for all samples, with an etching interval of 700~s. Survey, key core level and valence band spectra were acquired at the centre of the sputter crater after each etch step, using a pass energy of 160, 20 and 10~eV, respectively. Charge neutralisation was achieved using an electron flood gun with a filament current of 0.38~A, charge balance of 2~V, and a filament bias of 4.2~eV. The instrument was calibrated to the binding energy position of Au~4\textit{f}\textsubscript{7/2} (83.95~eV). The instrumental resolution was determined to be 290~meV at 10~eV pass energy. The collected depth profile data was quantified through peak-fit analysis using the CasaXPS software package,~\cite{CASA} which implements a F~1\textit{s} Kratos-modified Scofield photoionisation cross section database.\par


In order to investigate the microstructural changes and the Ti diffusion process, an as-deposited and a 5~h annealed sample were analysed using a state-of-the-art FEI Titan (G3) Cubed Themis 60-300~kV TEM (with probe and C\textsubscript{s} corrector). The instrument is equipped with a X-FEG source and CMOS-based FEI CETA 16-megapixel camera. Images were acquired using a 300~kV accelerating voltage. Energy dispersive spectroscopy (EDS) maps were obtained with Super-X EDX System with four windowless Silicon Drift Detector (SSD) of 120~mm\textsuperscript{2}. The microstructural images were acquired via conventional TEM mode and the elemental distribution maps of Cu, Ti, W and O signals were recorded using Scanning TEM (STEM) mode. TEM lamellas are prepared using Helios NanoLab 450 – FEI focused ion beam in-situ lift out technique.

\section{\label{sec:Results}Results}

\subsection{Copper Surface} \label{Copper_Surface}

The microstructure of the TiW/Cu bilayer was first explored with bright-field TEM to ascertain whether the interface is stable during annealing. Cross sectional images of the 5~h annealed sample are displayed in Fig.~\ref{fig:HRTEM_5h}, and images of the as-deposited sample can be viewed in Supplementary Information I. The as-deposited sample shows evidence that the TiW consists of a columnar grain boundary structure. Typically an as-deposited grain structure of a thin film alloy either resembles a ``frozen'' structure of fine precipitates or can evolve during deposition and form columnar grain boundaries with much larger grain sizes, both of which are dependent on the temperature conditions during deposition.~\cite{Harper_1997} This structure becomes more distinct in the 5~h annealed sample, as shown in the bright-field TEM image in Fig.~\ref{fig:HRTEM_5h}(a) and the high-angle annular dark field (HAADF) image in Fig.~\ref{fig:HRTEM_5h}(b), with the latter emphasising this observation. This agrees well with past studies and is ubiquitous with sputter-deposited TiW films.~\cite{DIRKS1990201, SAGHAEIAN2019137576} The fact that minimal changes occur to the microstructure with annealing suggests that the TiW barrier and the TiW/Cu interface are stable after annealing for prolonged timescales.\par

\begin{figure}[ht!]
\centering
    \includegraphics[keepaspectratio, width= \linewidth]{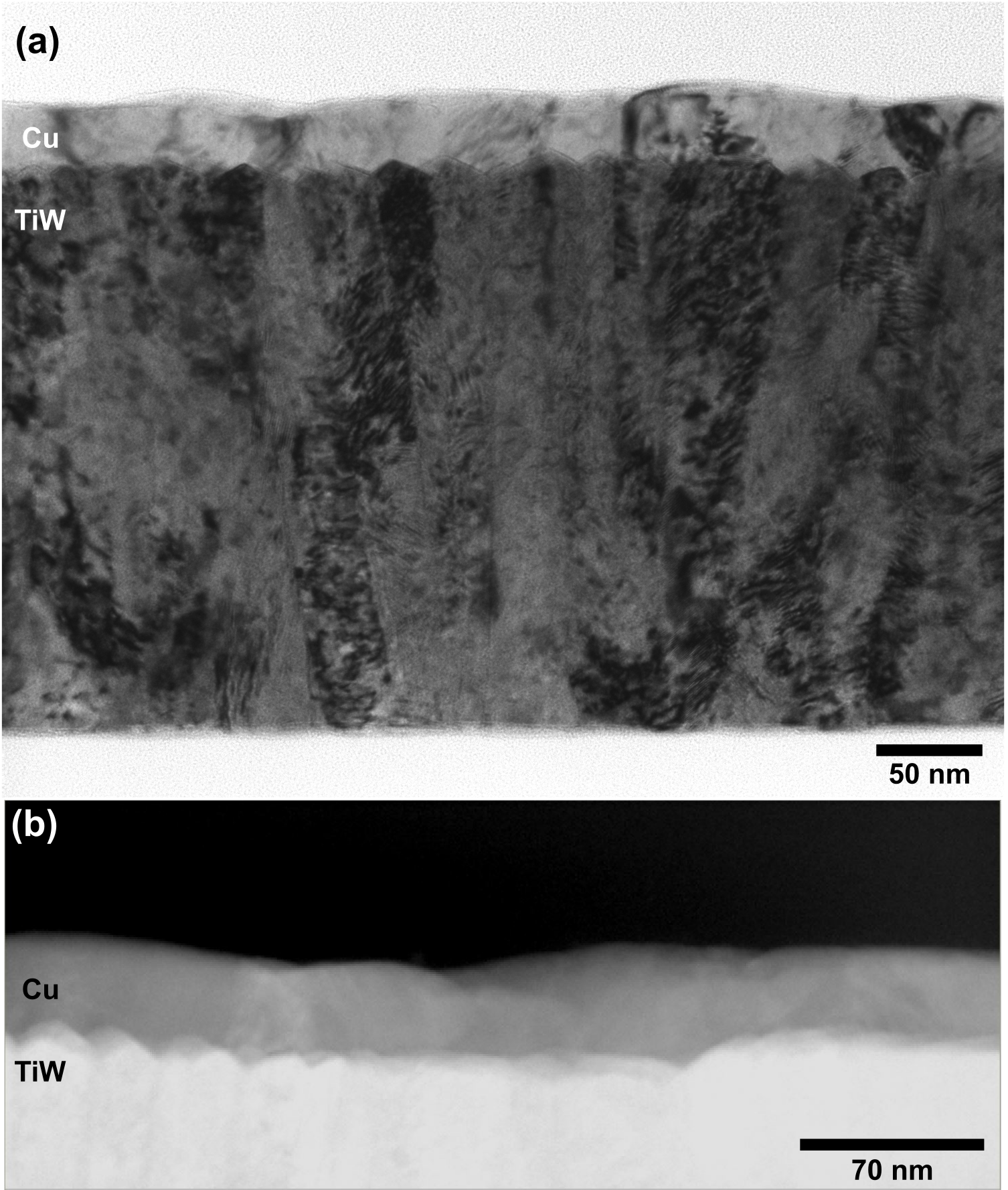}
    \caption{Cross-sectional microstructural images taken with TEM for the 5~h annealed sample, including (a) bright-field TEM image and (b) an enlarged HAADF image of the interface.}
    \label{fig:HRTEM_5h}
\end{figure}

\begin{figure}[ht!]
\centering
    \includegraphics[keepaspectratio, width= \linewidth]{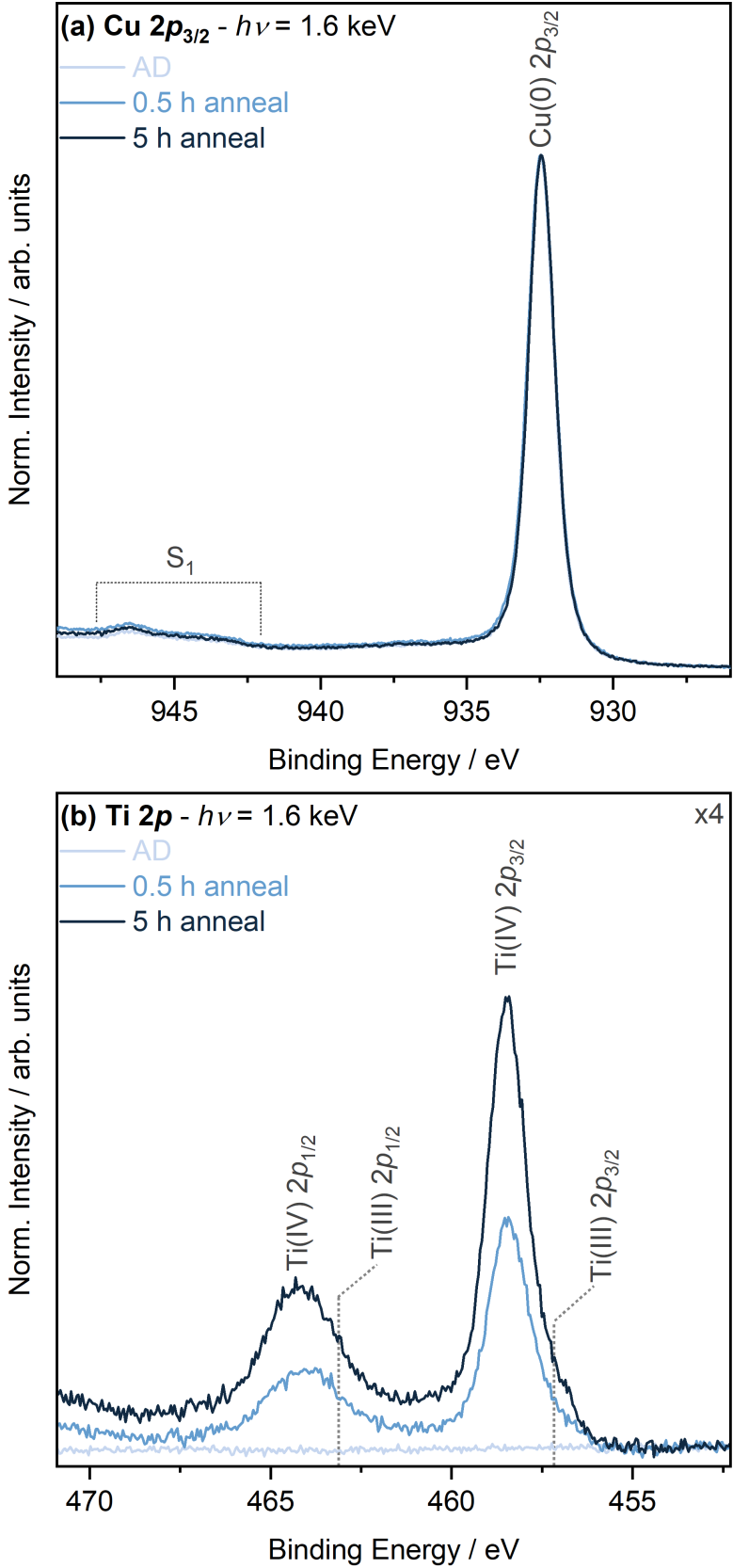}
    \caption{Key core levels collected with SXPS at the copper surface for all as-received samples, with (a) displaying the Cu~2\textit{p}\textsubscript{3/2} and (b) the Ti~2\textit{p} core level spectra. Data was normalised to the area of the Cu~2\textit{p}\textsubscript{3/2} signal intensity, with the Ti~2\textit{p} spectra four times magnified relative to the Cu~2\textit{p}\textsubscript{3/2} spectra to aid with viewing the data.}
    \label{fig:Ti2p_surface}
\end{figure}

However, the primary mode of degradation of this system is the out-diffusion of Ti and this is difficult to observe in the TEM images. To address this as-received samples were characterised using synchrotron-based SXPS ($h\nu$ = 1.6 keV). The collected survey spectra are displayed in Supplementary Information II, and show, as expected, a dominant copper signal, with minor signals from oxygen, carbon and nitrogen due to exposure to the ambient environment. Additionally, in the annealed samples only, a subtle titanium signal appears in the survey spectra, confirming that at 400{\textdegree}C, titanium diffuses out of the barrier layer, through the metallisation and accumulates at the copper surface. To address this further, high-resolution Cu~2\textit{p}\textsubscript{3/2} and Ti~2\textit{p} core level spectra collected with SXPS for each sample are displayed in Fig.~\ref{fig:Ti2p_surface}.

The Cu~2\textit{p}\textsubscript{3/2} core level spectra displayed in Fig.~\ref{fig:Ti2p_surface}(a) appear near-identical for all samples, with a binding energy (BE) position of 932.5~eV, attributed to copper metal. A low intensity satellite structure between 943 and 948~eV is detected, labelled with the notation S\textsubscript{1}. This satellite is attributed to the presence of a thin native Cu(I) oxide (Cu\textsubscript{2}O) layer on the copper metal surface and is generated due to the charge transfer between the ligand and metal antibonding orbitals of Cu\textsubscript{2}O.~\cite{Karlsson_1992, SCHON197396} As the spectral lines of Cu\textsubscript{2}O and copper metal overlap in the main photoionisation peak, it is difficult to isolate the contributions from peak fit analysis, however, this is not of interest to this study and from the core level alone the chemical state can be safely assumed to be primarily metallic copper with some Cu(I) oxide.~\cite{KAUSHIK1989581, Cu_Miller, Vasquez_1998, Tahir_2012}\par

The Ti~2\textit{p} core level spectra are displayed in Fig.~\ref{fig:Ti2p_surface}(b). The spectra show that after annealing for as little as 0.5~h at 400{\textdegree}C titanium is observed at the surface having diffused across the 25~nm-thick copper metallisation. In contrast, the as-deposited sample shows no indication of titanium diffusion, confirming that the diffusion mechanism is thermally initiated. This result falls inline with previous studies that also observe the upward diffusion of titanium from TiW into a metallisation overlayer in response to thermal treatments.~\cite{GHATE1978117, CANALI19829, Olowolafe_1985, Wang_1993, Plappert_2012, FUGGER20142487} Souli~\textit{et al.} report that Ti diffusion in TiW is initiated and facilitated not only by the thermal stress during annealing but also the resultant residual stress in the TiW layer during annealing. The authors observed that during annealing Si/TiW~(100~nm, 15.8~at.\% Ti)/Cu~(1.5~$\mu$n) stacks to 673~K, the residual stress within the TiW layer decreases, lowering the activation energy needed for diffusion.~\cite{Souli_2017}

Between the 0.5 and 5~h annealed samples, a 65\% increase in the total Ti~2\textit{p} signal intensity was observed, meaning that a thicker titanium overlayer is present. The signal increase is not linear in relation to annealing duration as a ten-fold increase in intensity would be expected between the 0.5~h and 5~h annealed samples. A similar finding was put forward by Olowolafe~\textit{et al.} for a PVD sputtered TiW/Al bilayer.~\cite{Olowolafe_1985} They conducted RBS depth profile measurements on Si/SiO\textsubscript{2}/Ti\textsubscript{22}W\textsubscript{78}/Al and Si/SiO\textsubscript{2}/Ti\textsubscript{22}W\textsubscript{78}/Al(2~at.\% Cu) stacks, annealing them between 450{\textdegree}C and 530{\textdegree}C for up to 4.5~h and reported the out-diffusion of Ti from the TiW layer and the subsequent Ti accumulation on the Al and Al(Cu) surface. Olowolafe~\textit{et al.} showed that the Ti accumulation was proportional to the square root of time representing a diffusion-limited growth. Moreover, the authors were able to determine that the activation energy of the diffusion process in their system was 2.4~eV, which they state is in keeping with the assumption that Ti proceeds to migrate through aluminium via grain boundary diffusion. As only two annealing timescales were tested in our study, it is difficult to confirm (a) the exact relationship between annealing duration and titanium accumulation and (b) the kinetic regime of diffusion.\par

The BE position of the main Ti~2\textit{p} peaks (458.5 and 464.2~eV) and the spin orbit splitting (SOS) distance of the doublet peaks (5.7~eV) suggest that the primary contribution in both SXPS and HAXPES spectra is from the Ti(IV) oxidation state.~\cite{SEN1976560, SAYERS1978301, Diebold_1996, Kurtz_1998} Additionally, the lower BE side of the main doublet peaks display a small shoulder attributed to the presence of the Ti(III) oxide state.~\cite{SAYERS1978301, Kurtz_1998} The BE position of the Ti~2\textit{p} core lines of titanium metal are at approximately 454 and 460~eV,~\cite{BERTOTI1995357} and are absent from the core level spectra. This confirms that the surface accumulated titanium has completely converted to its oxide state. It is assumed that after annealing and while the sample is confined in the deposition chamber under vacuum, diffused titanium is present at the copper surface in its metallic form and, when the sample is exposed to air, it oxidises.\par

\begin{figure*}[ht!]
\centering
    \includegraphics[keepaspectratio, width= 0.8\textwidth]{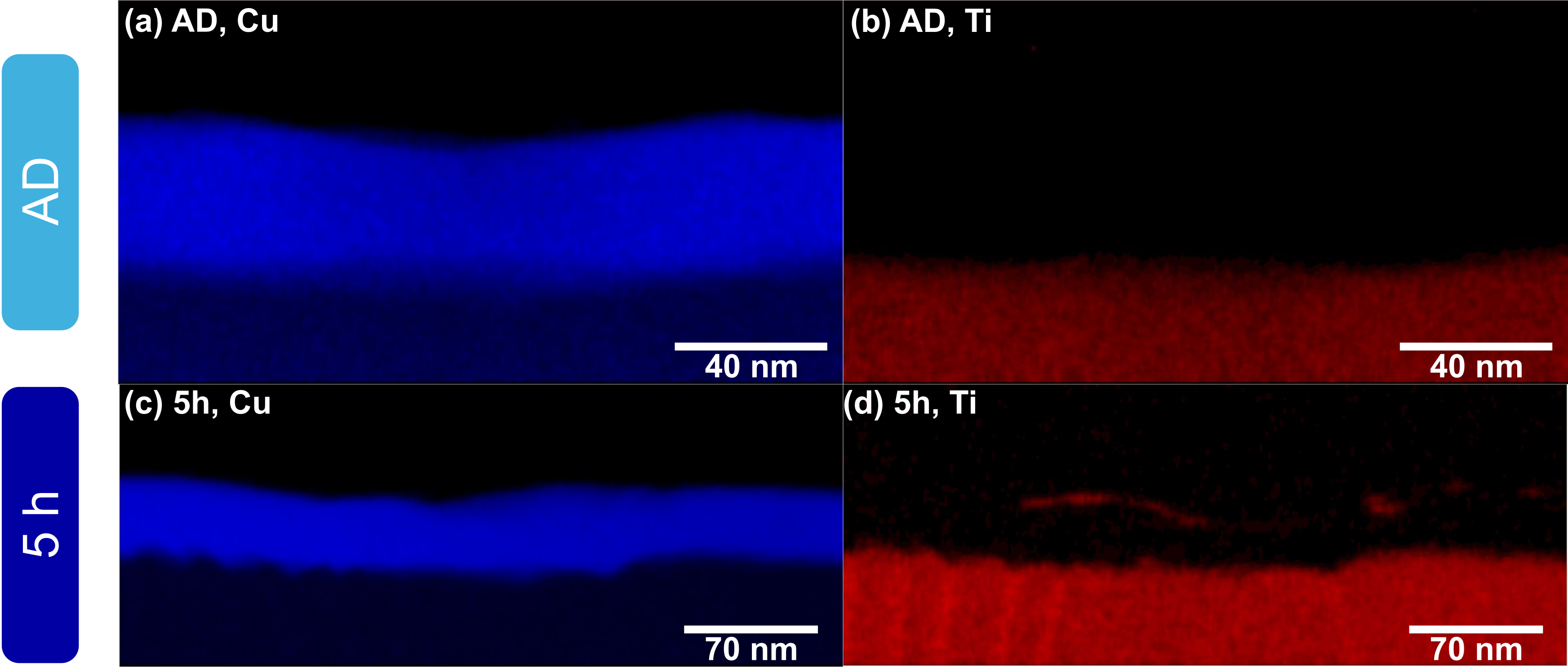}
    \caption{TEM images and overlaid EDS maps for the AD (top row) and 5~h annealed (bottom row) samples. The left column shows the Cu maps, whereas the right column shows the Ti maps.}
    \label{fig:EDS}
\end{figure*}

The out-diffusion of Ti and accumulation at the Cu surface is further corroborated with EDS. Fig.~\ref{fig:EDS} displays the results from EDS elemental mapping of Cu and Ti on the TEM cross section images (the complete set of EDS maps can be viewed in Supplementary Information III). Fig.~\ref{fig:EDS}(b) shows that for the as-deposited sample, the titanium is well-confined to the TiW diffusion barrier, however, after annealing for 5~h, titanium is observed on the copper surface (Fig.~\ref{fig:EDS}(d)), aligning with the results from SXPS. This confirms the out-diffusion of Ti and that this degradation mode is prominent in this TiW(300~nm)/Cu(25~nm) system. Additionally, from the EDS maps copper is not observed to diffuse downward after annealing for 5~h, suggesting that the interface and the TiW barrier is stable under the conditions tested. These images show a well-defined interface, but cannot provide information on the chemical state at or across the interface. This prompted measurements using a combination of SXPS and HAXPES to non-destructively probe the TiW/Cu interface, as well as a SXPS depth profile to obtain an elemental distribution map across the TiW/Cu layers, and assess the chemical stability of this interface.\par

\subsection{Across the Metallisation}\label{DepthP}
Quantification with either SXPS or HAXPES at synchrotron facilities is challenging due to the lack of critical parameters necessary to obtain reliable quantification, such as relative atomic sensitivity factors (RASFs) and a precise characterisation of the transmission function of the analyser. However, for laboratory-based SXPS systems these critical parameters are known. Therefore, to quantitatively determine the distribution of titanium across the metallisation and the depletion within the bulk TiW, a depth profile was conducted using a Kratos Axis Supra laboratory-based SXPS instrument.\par

\begin{figure*}[ht!]
\centering
    \includegraphics[keepaspectratio, width=\textwidth]{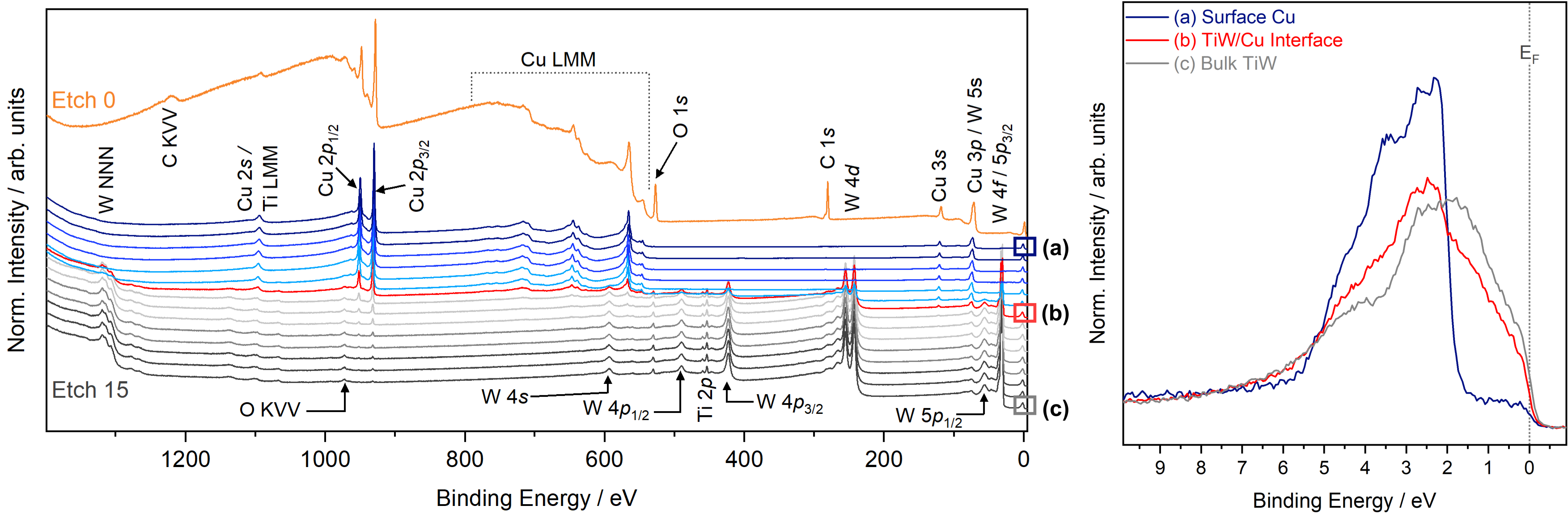}
    \caption{Survey spectra collected after each etch cycle on the 0.5~h annealed sample, going from the as-received measurement ``Etch 0'' (top) to the last etch cycle ``Etch 15'' (bottom). The red line indicates the point at which the atomic percentage ratio of W:Cu first becomes W majority and so is termed the ``interface''. The valence band spectra provide a good measure for the depth profile, with the three key valence band spectra presented on the right-hand side. The valence band collected after \textbf{(a)} the initial etch cycle (Etch 1), \textbf{(b)} the 7\textsuperscript{th} etch cycle (i.e.~``interface''), and \textbf{(c)} the final etch cycle in the TiW (Etch 15), which can be considered a bulk TiW measurement. Clear changes occur in the valence band as we go from a metallic Cu rich to metallic TiW rich system.}
    \label{fig:Surv_dp}
\end{figure*}

The depth profile was conducted via in-situ Ar\textsuperscript{+} sputtering, whereby the copper capping layer was gradually removed until the TiW was reached. Fig.~\ref{fig:Surv_dp} displays the survey spectra collected after each sputter cycle, highlighting the removal of the copper overlayer and the subsequent increase in intensity of the Ti and W signals. To provide quantitative elemental information, the collected core level spectra after each sputter cycle were peak fitted, with the areas corrected using the appropriate RASFs to determine a relative atomic percentage ratio between Cu, W, and Ti across the bilayer (see Supplementary Information IV for further information regarding the peak fit analysis).\par

Peak fit analysis of the W~4\textit{f} core level is particularly challenging due to the close proximity of the W~4\textit{f} core lines to the W~5\textit{p}\textsubscript{3/2} core line.~\cite{kalha2021lifetime} In the present case another complication that needs addressing is the overlap of the Ti~3\textit{p} core level located at 32.6~eV for titanium metal with the W~4\textit{f}\textsubscript{5/2} core level peak.~\cite{Barth_1985} In this system, the composition is heavily tungsten dominated and one would expect a considerably lower Ti~3\textit{p} peak intensity relative to the W~4\textit{f} peaks. Moreover, the W~4\textit{f}\textsubscript{7/2} photoionisation cross section at the Al~K$\alpha$ excitation energy is an order of magnitude greater than the Ti~3\textit{p} cross section.~\cite{Scofield1973TheoreticalKeV, JJackson2018Galore:Spectroscopy, Kalha20} In terms of line shapes, the Ti~3\textit{p}\textsubscript{3/2} core line has a theoretical life time width of 1.2~eV according to measurements by Campbell~\textit{et al.}, therefore, if the peak was present in a significant intensity it should be observable as the W~4\textit{f} peaks have a full width at half maximum (FWHM) of approximately 0.5~eV.~\cite{Campbell2001WidthsLevels} Lastly, the W~4\textit{f} core level collected within the bulk of the TiW (i.e.\ last etch cycle), displayed in Supplementary Information IV, shows no indication of a prominent Ti~3\textit{p} core line piercing the W~4\textit{f}\textsubscript{5/2} core line or affecting its line shape. Therefore, if the Ti~3\textit{p} core line was present with significant intensity it should affect the asymmetry and tailing of the W~4\textit{f}\textsubscript{5/2} peak. This is not the case, with the line shape of the W~4\textit{f} doublet peaks appearing very similar. Extensive peak fit analysis was conducted on the core level region and it was determined that if the W~4\textit{f} doublet peaks are constrained to have the same FWHM and line shape, the resultant area ratio is 0.75, matching with the expected ratio as determined from the degeneracy of the \textit{f} spin states. Therefore the Ti~3\textit{p} core line is negligible with respect to the W~4\textit{f} intensity and does not interfere with the W~4\textit{f} region. Consequently it does not need not be considered in peak-fit analysis for quantification.\par

The elemental distribution profiles of the metals present for all samples are displayed in Fig.~\ref{fig:Depth_Profile}. The complete depth profile containing the O and C signals can be found in Supplementary Information V. The profiles displayed in Fig.~\ref{fig:Depth_Profile} show very little difference between the as-deposited and annealed samples, with the Cu and TiW signals well separated in agreement with the results from the TEM/EDS measurements. At the copper surface of the 5~h annealed sample, prior to etching (i.e.~etch 0), the atomic ratio of Ti:Cu is 4.01:95.99, lowering to 1.29:98.71 after the first etch cycle, and 0.04:99.96 after the second cycle before Ti being undetected in the third cycle, indicating that the titanium distribution is skewed to the surface and not uniformly distributed in the copper bulk. Titanium accumulated at the copper surface is able to undergo complete oxidation as the concentration is relatively small and so is not at risk of saturation, which is why no metallic features are retained in the as-received Ti~2\textit{p} core level spectra in Fig.~\ref{fig:Ti2p_surface}(b).\par

\begin{figure}[ht!]
\centering
    \includegraphics[keepaspectratio, width=\linewidth]{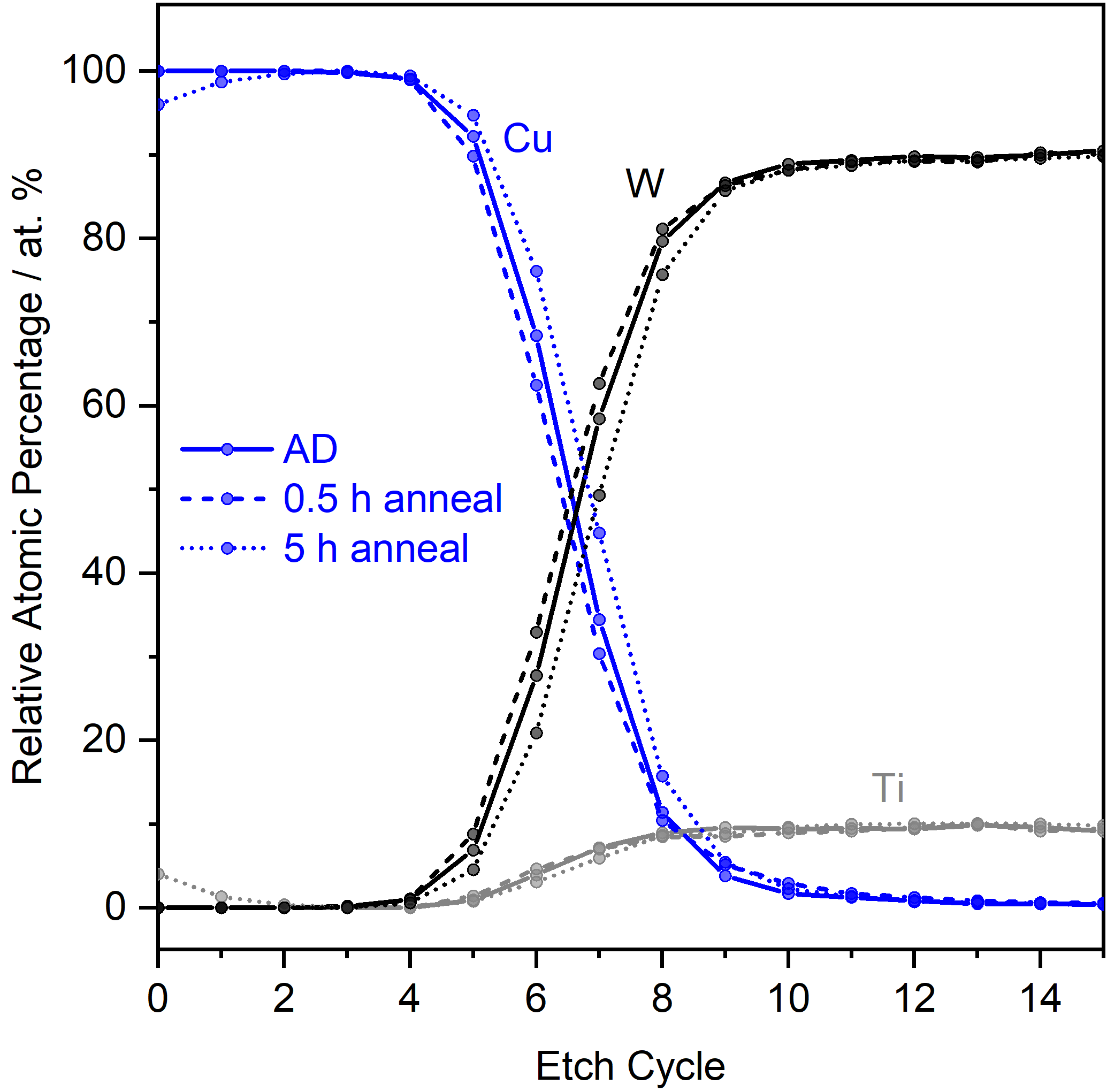}
    \caption{Elemental depth profiles collected for all samples. Etch 0 refers to measurements on the as-received sample, prior to etching.}
    \label{fig:Depth_Profile}
\end{figure}

The depth profiles suggest that the interface does not broaden or migrate significantly with annealing, confirming its stability upon thermal treatment at 400{\textdegree}C for prolonged timescales. The point of cross-over between the Cu and W signals changes only subtly and the slight variations observed are attributed to different sputter removal rates and the thickness of the adventitious carbon overlayer for each sample. Annealing at 400{\textdegree}C will affect the crystal structure and density of the copper film due to extensive recrystallisation, affecting the sputter removal rate.~\cite{Harper_1997}\par

The average Ti:W ratio (determined by comparing the RASFs corrected total Ti~2\textit{p} area and W~4\textit{f} area) across the last five etch steps (etch cycle 11-15) of the depth profile (i.e.\ bulk TiW) across all samples is close to 10:90. The error associated with the Ti:W values is $\pm$0.4~at.\% due to the complexity of peak fitting the core lines, making the individual values (9.5:90.5, 9.6:90.4, and 10.0:90.0 for the as-deposited, 0.5~h and 5~h samples) essentially indistinguishable. As the out-diffusion of Ti is minimal the small loss of Ti cannot be robustly observed in the bulk TiW. This further confirms the stability of the diffusion barrier at 400{\textdegree}C, aligning well with past TiW/Cu studies.~\cite{Wang_1993,Chiou_1995, FUGGER20142487}\par

As mentioned in Section~\ref{sec:Method}, a target consisting of 30~at.\% Ti was used for the deposition of the TiW. However, the quantified concentration of the resultant film determined from SXPS clearly deviates from this as detailed above. Others have reported similar discrepancies when depositing TiW in Ar discharges,~\cite{Bergstrom_1997, Plappert_2012, FUGGER20142487} and attribute these to the inherent issue of sputter depositing two metals with very different atomic weights, with titanium deposited and then preferentially sputtered by backscattered Ar\textsuperscript{+} ions, increasing the tungsten concentration. Bergstrom~\textit{et al.} show that depositing TiW in xenon (Xe) rather than Ar provides a film composition equal to that of the target as the heavier mass of Xe compared to Ar eliminates backscattering.~\cite{Bergstrom_1997}.\par

The reason for the good stability of the TiW diffusion barrier in response to annealing may be attributed to the Ti:W ratio present. Cunningham~\textit{et al.} state a solubility limit of Ti in W of 10~at.\% at 600{\textdegree}C and from the depth profiles values close to 10~at.\% Ti are found in the present samples.~\cite{Cunningham_1970} Therefore, it is not expected that any significant amount of more mobile, excess titanium is present, but rather that almost all Ti is in a stable solid solution with W. This can explain the low level of out-diffusion observed.\par

\subsection{TiW/Cu Interface}\label{TiW_Interface}

\subsubsection{Core Levels}

Only two reported detailed studies have explored the TiW layer using XPS, the first from Alay~\textit{et al.}, who focused on conducting an extensive chemical and structural characterisation of TiW films using not just XPS but also X-ray diffraction (XRD), RBS, electron probe microanalysis (EPMA), and AES,~\cite{Alay1991} and the second being recent work by the authors of the present work exploring the effects of both post-deposition annealing and titanium concentration on the oxidation behaviour of TiW.~\cite{Kalha_2021_TiWO} However, in both cases the TiW was exposed to the ambient environment, thereby forming a native and extensive surface oxide layer, not representative of the true device state. Several studies using AES have been conducted on TiW/Al~\cite{Olowolafe_1985, Furlan1991} and TiW/Cu~\cite{Wang_1993, OLOWOLAFE199337} bilayer samples, however, all rely on using in-situ sputtering to obtain elemental distribution depth profiles across the heterostructure. Although, depth profiling has the clear advantage of providing elemental quantification, which is the reason we have also employed it in the previous section, the destructive nature of the sputtering process compromises the assessment of the chemical states across the material. This is particularly prevalent for metal oxide systems whereby upon sputtering, the oxides are typically reduced into a continuum of lower oxide and metal states. Of most interest to industry is the assessment of the TiW and TiW/Cu interface in its true state, which formed the main motivation to thin the copper overlayer in-situ and study the preserved interface with synchrotron-based SXPS ($h\nu$ = 1.6~keV) and the unaffected bulk TiW with synchrotron-based HAXPES ($h\nu$ = 5.9~keV), exploiting the higher probing depth of  HAXPES.\par

\begin{figure*}
\centering
    \includegraphics[keepaspectratio, width= \textwidth]{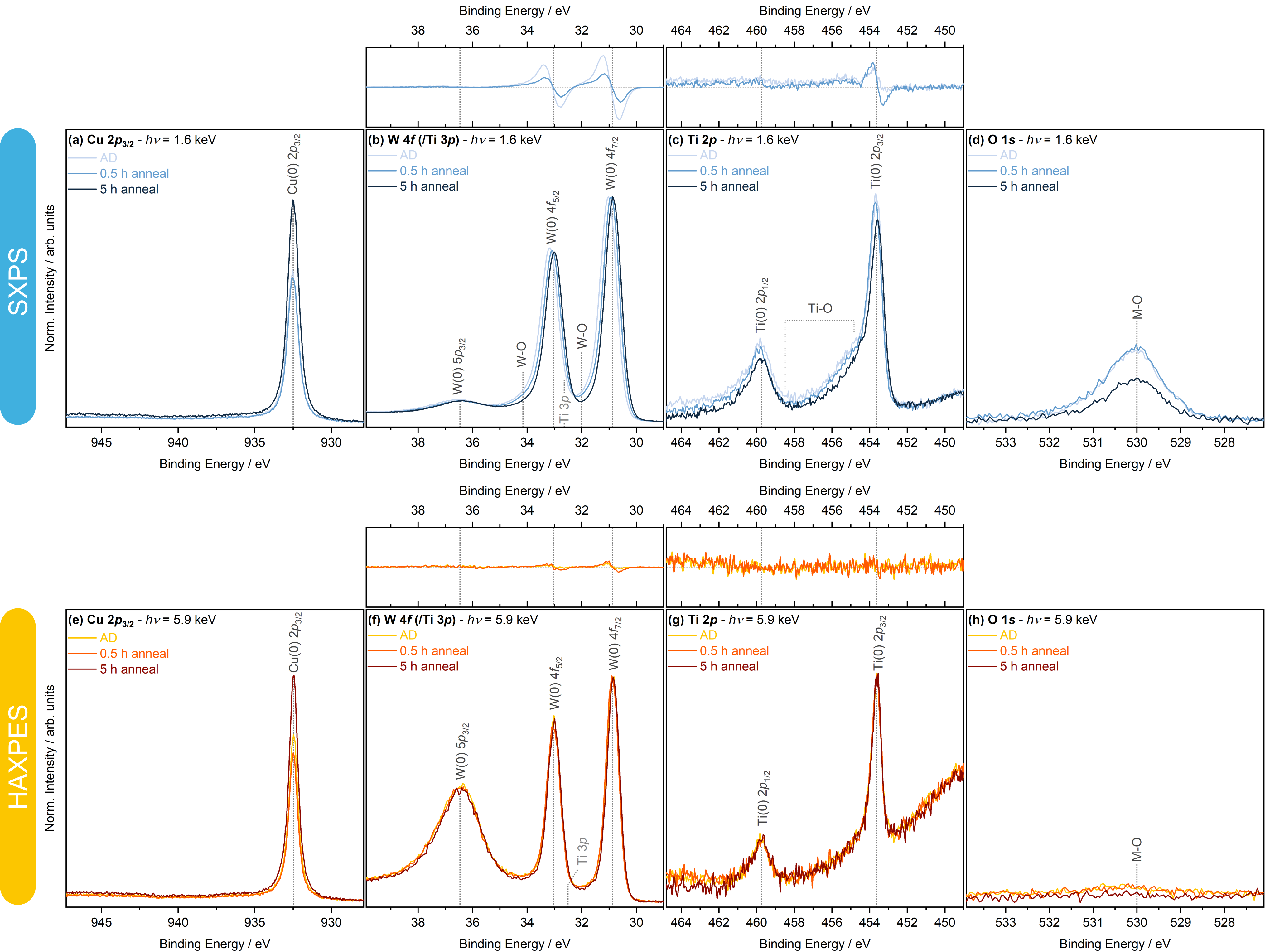}
    \caption{Core level spectra collected at the TiW/Cu interface (after the in-situ thinning of the Cu) with SXPS (top row) and HAXPES (bottom row), including from left to right, Cu~2\textit{p}\textsubscript{3/2}, W~4\textit{f}, Ti~2\textit{p}, and O~1\textit{s}. Spectra were normalised to the maximum intensity of the measured W~4\textit{f}\textsubscript{7/2} core level after the removal of a constant linear background. The O~1s core level y-axis was scaled to match the signal-to-noise ratio of the Ti~2\textit{p} core level. Above the W~4\textit{f} and Ti~2\textit{p} core level spectra are difference plots showing the difference in signal intensity of the AD and 0.5~h anneal sample relative to the 5~h anneal sample. The difference plots are plotted on the same relative intensity y-axis scale for each core level.}
    \label{fig:Interface_TiW}
\end{figure*}

Survey spectra collected with both synchrotron-based SXPS and HAXPES at the TiW/Cu interface after the in-situ thinning of the copper can be found in Supplementary Information VI. The SXPS and HAXPES recorded survey spectra display Cu, Ti, W and O signals, whereas the signal from carbon is very small and is only identifiable in the SXP spectra. The key core levels measured with both techniques at the TiW/Cu interface are displayed and compared in Fig.~\ref{fig:Interface_TiW}, with the BE positions of the Ti~2\textit{p}\textsubscript{3/2} and W~4\textit{f}\textsubscript{7/2} core level peaks listed in Tab.~\ref{BE}. The IMFP of W~4\textit{f} electrons in copper metal calculated using the TPP-2M formula is 2.2 and 6.3~nm, when using the 1.6 and 5.9~keV photon energies, respectively. Based on the statements made in Section~\ref{sec:Method}, the probing depth of the W~4\textit{f} electrons in copper metal is approximately 6.5 and 19.0~nm (i.e. 3$\lambda$), with SXPS and HAXPES, respectively. Considering a strong tungsten signal is observable with SXPS, the maximum copper thickness after in-situ thinning is estimated to be 6.5~nm.\par

\begin{table}[h!]
     \caption{\label{BE}BE positions of the Ti~2\textit{p}\textsubscript{3/2} and W~4\textit{f}\textsubscript{7/2} core lines taken from SXPS and HAXPES measurements. The errors associated with the BE values are $\pm$195~meV for the SXPS and $\pm$120~meV for the HAXPES measurements, respectively, based on their individual energy resolution.}
     \begin{ruledtabular}
    \begin{tabular}{ccccc}

       & \multicolumn{2}{c}{SXPS / eV} & \multicolumn{2}{c}{HAXPES / eV} \\
       \hline
Sample & Ti 2\textit{p}\textsubscript{3/2}        & W 4\textit{f}\textsubscript{7/2}      & Ti 2\textit{p}\textsubscript{3/2}         & W 4\textit{f}\textsubscript{7/2}        \\
   \hline
AD     & 453.6          & 31.0         & 453.6          & 30.9           \\
0.5 h  & 453.7          & 30.9         & 453.6          & 30.9           \\
5 h    & 453.5          & 30.8         & 453.6          & 30.9          
\end{tabular}
     \end{ruledtabular}
\end{table}

The Cu~2\textit{p}\textsubscript{3/2} core level spectra collected with both SXPS and HAXPES are displayed in Fig.~\ref{fig:Interface_TiW}(a) and (e), respectively. Copper is clearly detected with significant intensity confirming that the interface has not been reached and the TiW is not exposed in the analysis chamber, but rather the copper has been thinned sufficiently to allow for SXPS measurements across the TiW/Cu interface and HAXPES measurements into the bulk TiW. The spectra display no satellite features toward the higher BE side of the main photoelectron peak and the Cu~2\textit{p}\textsubscript{3/2} peak position is located at 932.5~eV, confirming that copper is in its metallic state. Additionally, no BE shifts or changes to the line shape (i.e.~asymmetry or FWHM) occur with annealing, suggesting that no reaction between the copper and TiW occurs. The Cu~2\textit{p}\textsubscript{3/2} signal intensity varies between samples relative to the W~4\textit{f} main signal intensity. The AD and 0.5~h annealed sample spectra show similar Cu~2\textit{p}\textsubscript{3/2} intensities, whereas for the 5~h sample spectrum the Cu~2\textit{p}\textsubscript{3/2} signal intensity is approximately 50\% greater. The differences in the remaining copper thickness across the samples is attributed to the repeatability of the sputtering process as well as differences in sample composition and density.\par 

Fig.~\ref{fig:Interface_TiW}(b) and (f) display the W~4\textit{f}/5\textit{p}\textsubscript{3/2} core level region measured with SXPS and HAXPES, respectively. With increasing photon energy the intensity of the W~5\textit{p}\textsubscript{3/2} peak increases relative to the W~4\textit{f} peak intensities as a consequence of the differences in decay rate of the photoionisation cross sections between the two orbitals when moving from the soft to the hard X-ray regime (see Supplementary Information VII for the photoionisation cross section decay profiles as a function of photon kinetic energy).~\cite{HAXPES_Big_Boy} The BE positions and asymmetric line shape of the W~4\textit{f} core level in both the soft and hard XP spectra correlate well with tungsten in its metallic state.~\cite{Barr1978, Engelhard2000ThirdSpectroscopy} However, the binding energy position of the W~4\textit{f}\textsubscript{7/2} and W~4\textit{f}\textsubscript{5/2} are located on average at 30.9~eV and 33.0~eV, respectively, which are approximately 0.5~eV lower compared to pure metallic tungsten.~\cite{kalha2021lifetime} This negative shift in BE position away from the pure tungsten metal BE position can be attributed to an alloying effect in the presence of titanium. Similar shifts have also been observed for Ti-Ni,~\cite{SENKOVSKIY2012190} Ti-Cu~\cite{COCKE1990119} and Ti-Pd~\cite{TANAKA1990429} alloys when varying the alloying concentration and are a consequence of unique electronic environments that develop when two metals are alloyed together.\par

Subtle spectral changes are observed in the SXPS W~4\textit{f} spectra, highlighted in the difference plot, whereas the changes in the HAXPES spectra are much smaller. Fig.~\ref{fig:Interface_TiW}(b) shows that with increasing annealing duration the BE position of the W~4\textit{f} peaks shifts to lower energies, with a -0.2~eV shift observed between the AD and 5~h annealed sample. Moreover, the line shape of the core level appears to change between each sample, with the AD and 0.5~h samples showing similar shapes, but the spectrum of the 5~h sample displays a lower intensity of the valley region between the W~4\textit{f} doublet (see Supplementary Information VIII for a relative BE scale plot of these spectra, further emphasising these differences). Given that the depth profiles discussed in the previous section show no significant change in the TiW concentration with annealing, the observed spectral differences cannot be explained on the basis of changes in the Ti:W ratio. Instead, these spectral differences are evidence of subtle changes to the chemical states across the interface, namely the formation of W-O environments. The SXP O~1\textit{s} spectra displayed in Fig.~\ref{fig:Interface_TiW}(d) show that relative to the W~4\textit{f} core line intensity, the 5~h sample has the lowest oxygen intensity. The main peak intensity of the O~1\textit{s} core line is at approximately 530.0~eV, in agreement with a metal oxide (M-O) environment.~\cite{DIMITROV2002100} This can explain the observed differences in the SXPS W~4\textit{f} spectrum. The AD and 0.5~h annealed samples show the highest level of oxygen, which in turn leads to a positive shift in the BE position of the W~4\textit{f} peaks and change in line shape, both of which represent the formation of a local chemical environment of W-O bonds.~\cite{Kalha_2021_TiWO} However, due to the low concentration of oxygen, a full, extended bulk interfacial oxide layer cannot form, evidenced by the lack of distinct chemically shifted core level peaks. As mentioned earlier, the Ti~3\textit{p} core line for metallic titanium is expected to be positioned at 32.6~eV, overlapping with the W~4\textit{f}\textsubscript{5/2} core line, and so is marked with a guide-line in Fig.~\ref{fig:Interface_TiW}, however, given that it has been established that the intensity of this peak is negligible with respect to the W~4\textit{f} peaks, the observed changes in line shapes can only be attributed to the presence of an interfacial oxide and not the presence of the Ti~3\textit{p} peak. Lastly, it is difficult to comment on whether annealing has an effect on this interfacial oxide as the copper overlayer thickness between the samples varies but it can be confirmed that regardless of the annealing duration, metal-oxide states are present at the TiW/Cu interface.\par

The origin of the inclusion of oxygen into this system and its accumulation at the interface will be further discussed in Section~\ref{Interfacial_Discussion}. However, to summarise, it is assumed that given that the deposition process occurs under high vacuum, residual concentrations of oxygen will be present. Therefore, prior to the deposition of copper, the TiW face will be exposed to the chamber environment gettering oxygen and leading to the development of M-O environments. Due to the presence of titanium, oxidation of the TiW interface even at low residual oxygen concentrations is favourable as titanium has excellent gettering properties,~\cite{Beggs} and a high chemical affinity towards oxygen.~\cite{Liu1988, Kalha_2021_TiWO} \par

The SXPS Ti~2\textit{p} core level spectra are displayed in Fig.~\ref{fig:Interface_TiW}(c). The spectra display core line peaks with asymmetric tails and at average BE positions of 453.7~eV (Ti~2\textit{p}\textsubscript{3/2}) and 459.9~eV (Ti~2\textit{p}\textsubscript{1/2}), giving a SOS distance of 6.2~eV, confirming that titanium within the TiW is present in its metallic state.~\cite{Doniach1969Many-electronMetals,COCKE1990119, SENKOVSKIY2012190} 
The reported binding energy positions of pure titanium metal in the literature are less well-defined and more varied than those of tungsten metal, making it difficult to truly confirm if the Ti~2\textit{p} core level of TiW also deviates from the BE position of pure metallic titanium, similar to the observation of the W~4\textit{f} shifts, alluding to an alloying effect. Bert{\'{o}}ti~\textit{et al.} compiled literature values of the Ti~2\textit{p}\textsubscript{3/2} core level peak of titanium metal citing a range of 453.9-454.3~eV and reporting a 453.8$\pm$0.2~eV value from their own work.~\cite{BERTOTI1995357} Therefore, based on this, it appears that the BE positions of the Ti~2\textit{p} core level peaks measured for our TiW system also show a slight negative shift relative to the pure metal, similar to the W~4\textit{f} binding energy shifts. Aside from BE shifts, the Ti~2\textit{p} SXPS spectra also show variations in signal intensity but this is most likely due to the remaining copper thickness differences between samples.\par

The HAXPES Ti~2\textit{p} core level displayed in Fig.~\ref{fig:Interface_TiW}(g) show no change in the signal intensity or peak position as a function of annealing duration. Compared to the SXPS spectrum, the background is considerably higher in the Ti~2\textit{p}\textsubscript{3/2} region. This is due to the plasmon from the close lying W~4\textit{p}\textsubscript{3/2} peak and the intensity of this plasmon is enhanced with HAXPES due to photoionisation-cross section effects. Additionally, differences in line shapes between the SXPS and HAXPES recorded Ti~2\textit{p} spectra are observed, especially on the higher BE side of the Ti~2\textit{p}\textsubscript{3/2} between 454.5-458.5~eV (a comparison of the SXPS and HAXPES core levels after the subtraction of a background, highlighting this spectral change is included in Supplementary Information IX). The change in intensity across this region clearly exceeds the difference in energy resolution between the SXPS and HAXPES experiments, suggesting that this is intrinsic to the sample interface and an indication of differences in chemical states, rather than a consequence of energy resolution. Carley~\textit{et al.} share a similar observation during oxygen absorption experiments on titanium and other metal foils using XPS.~\cite{Carley1987} The authors determined that under low oxygen exposure conditions the main peak BE positions remained consistent with metallic species. However, with increasing oxygen exposure, a noticeable “filling in” of the region between doublet peaks or an increased tail height toward higher binding energies (i.e. above the Ti~2\textit{p}\textsubscript{1/2} core line) was observed, correlating to the formation and development of Ti-O environments at the TiW/Cu interface.~\cite{Carley1987,Puglia1995,Kuznetsov1992} Therefore, much like the W~4\textit{f} spectra, the Ti~2\textit{p} spectra give evidence of the formation of a Ti-O environment at the interface.\par

The O~1\textit{s} spectra collected with HAXPES, displayed in Fig.~\ref{fig:Interface_TiW}(h) are considerably lower in intensity compared to the SXPS collected spectra, suggesting that the bulk TiW contains trace quantities of oxygen, whereas the concentration of oxygen at the interface is far greater. It is well known that the photoionisation cross section (and therefore signal intensity) of core levels decay with increasing photon energy, with each core level having different decay rates.~\cite{HAXPES_Big_Boy} However, comparing the Scofield photoionisation cross sections of the Ti~2\textit{p}\textsubscript{3/2} and O~1\textit{s} (see Supplementary Information VII), it is obvious that their decay rates are similar. Furthermore, when comparing the W~4\textit{f}\textsubscript{7/2} cross section to the O~1\textit{s} an overlap is observed at 3.6~keV, with the O~1\textit{s} cross section exceeding the W~4\textit{f}\textsubscript{7/2} when using a 5.9~keV photon energy. Therefore, we can be confident that the observed low intensity of the HAXPES recorded O~1\textit{s} spectrum is in fact due to the small concentration of oxygen in the bulk rather than a photoionisation cross section effect. The W~4\textit{f} and Ti~2\textit{p} spectra collected with HAXPES also reflect this observation as they show almost negligible changes in the line shape and BE positions across the samples and do not appear to have the same hallmarks of M-O formation as observed with SXPS.\par

Overall, it is clear that by comparing the SXPS and HAXPES W~4\textit{f} and Ti~2\textit{p} spectra, an interfacial oxide consisting of a local co-ordination of Ti-O and W-O states are present at the TiW/Cu interface but not a bulk oxide layer. This assumption is further corroborated by the fact that titanium is observed to diffuse during annealing and, if the interface was fully oxidised, titanium oxide would be precipitated and immobile to diffuse through the copper metallisation.\par

\subsubsection{Electronic Structure}

The electronic structure of occupied states across the TiW/Cu interface can be probed through collection of the valence band (VB) spectra. The VB is the highest measurable kinetic energy region (lowest BE region) in XPS, and therefore electrons originating from the VB have the largest IMFP and the signal is representative of the maximum probing depth.\par

\begin{figure}[ht!]
\centering
    \includegraphics[keepaspectratio, width= \linewidth]{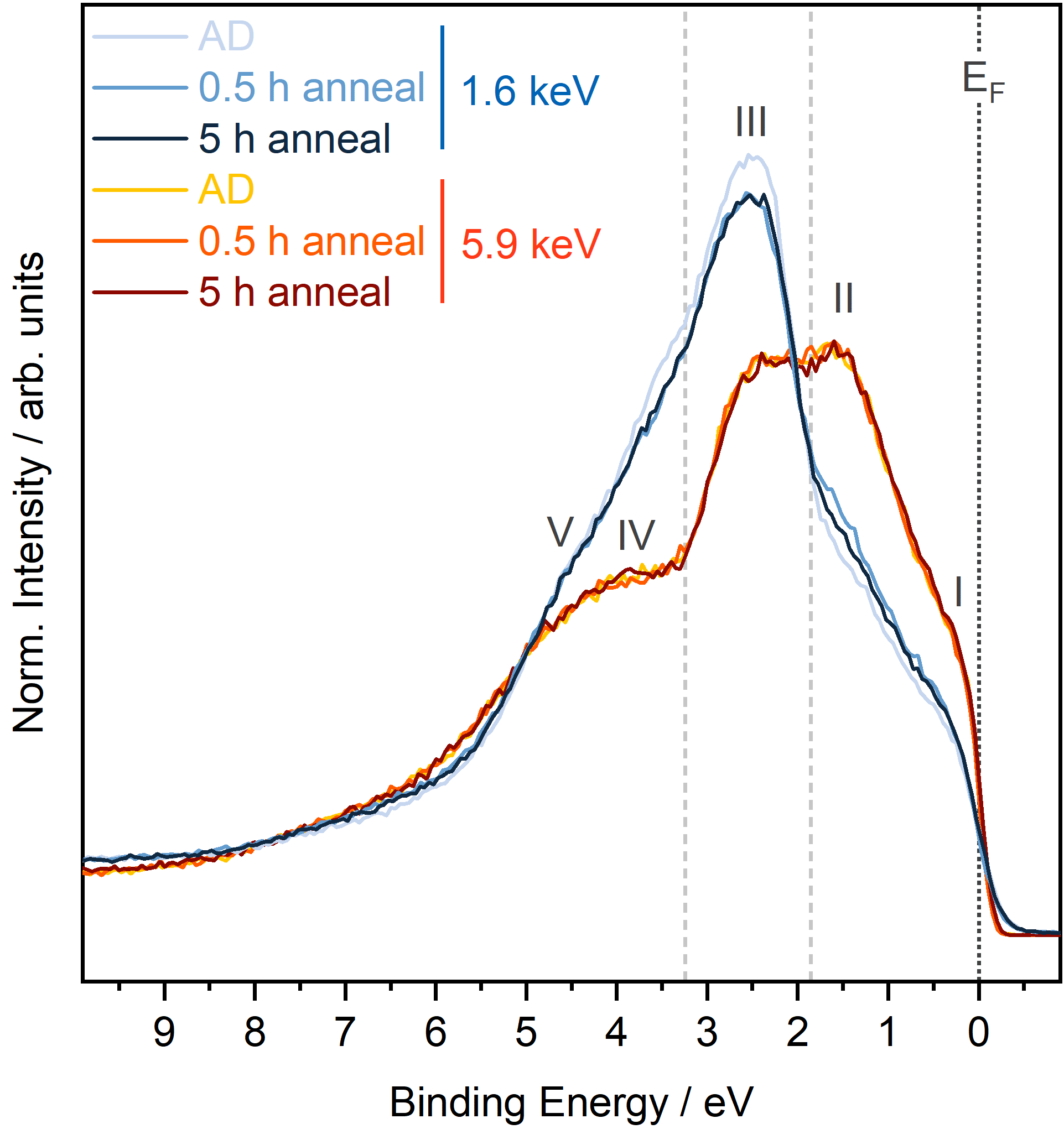}
    \caption{Valence band spectra acquired at the TiW/Cu interface after in-situ thinning of the copper metallisation for all samples, collected with two photon excitation energies - 1.6~keV (SXPS) and 5.9~keV (HAXPES). Two guide lines are given at 1.9 and 3.2~eV marking key line shape changes between the SXPS and HAXPES measured valence band spectra. Spectra are normalised to their respective areas and aligned to their intrinsic Fermi edge (E\textsubscript{F}).}
    \label{fig:VB}
\end{figure}

Fig.~\ref{fig:VB} displays the VB spectra collected with both SXPS and HAXPES across the TiW/Cu interface after the in-situ thinning of the copper. A stark difference is observed between the valence band spectra collected with soft and hard X-rays. The main features are labelled with Roman numerals (I-V). These features are located at approximately 0.2~eV (I), 1.6~eV (II), 2.5~eV (III), 3.8~eV (IV) and, 4.7~eV (V). Due to the difference in probing depth of SXPS relative to HAXPES, the soft X-ray collected valence band is more heavily influenced by the remaining copper metallisation and closely resembles the valence band of copper reported by others.~\cite{Hufner_1973, Copper_VB_1993} In contrast, the HAXPES measurements probe predominately the bulk TiW and closely resemble the VB of tungsten metal.~\cite{Engelhard2000ThirdSpectroscopy, kalha2021lifetime}\par

The SXP spectra show differences between the samples. The gradient of the region between 0-2~eV appears to fluctuate between the samples. However, no systematic shift is observed. Whereas, the intensity of feature III (attributed to mainly Cu~3\textit{d} states) appears identical for the AD and 0.5~h annealed sample, but increases for the 5~h sample. These differences are attributed to the differences in remaining Cu overlayer thickness between the samples after in-situ sputtering, which will influence the the ratio of Cu and W states in the VB, affecting the VB shape.\par

The HAXPES spectra show negligible change with increasing annealing duration. The titanium diffusion mechanism and changes to the concentration within the bulk of the TiW is too small to warrant significant change in the VB, reflecting the core level spectra. Additionally, the one-electron photoionisation cross section of Ti~3\textit{d} at 5.9~keV is 3.1$\times$10\textsuperscript{-25}~cm\textsuperscript{2}, compared to 5.1$\times$10\textsuperscript{-23}~cm\textsuperscript{2} for the W~5\textit{d} orbital and given the significantly higher concentration of tungsten in the TiW layer, the tungsten overshadows the contributions of titanium to the VB. Therefore, even if the depletion was significant, it would be difficult to infer this from VB spectra.~\cite{Scofield1973TheoreticalKeV, Kalha20} The lack of change in electronic structure across the TiW bulk further corroborates the findings from the depth profiles and HAXPES core level analysis. According to Tanaka~\textit{et al.} the changing of the alloying concentration of a binary alloy will affect the electronic structure of the alloy, consequently leading to significant changes in the line shape (i.e. asymmetry) and BE position of the core levels of the two metal components.~\cite{TANAKA1990429} However, the HAXPES recorded Ti~2\textit{p} and W~4\textit{f} core levels show an almost negligible change in the line shape and BE position, and the depth profile results show that the depletion of Ti from the bulk TiW after annealing is not significant. Therefore, these observations explain why the HAXPES recorded VB shows negligible change between the as-deposited and annealed samples.\par

\section{Discussion}

\subsection{Intermetallic Compound Formation}

Several studies have explored the metallurgical stability of Ti/Cu and TiW/Cu systems during thermal treatments, tracking the temperature of failure and assessing any occurring chemical reactions.~\cite{Liotard_1985, Jian_1992, OLOWOLAFE199337, Takeyama_1996, Ramarotafika, FUGGER20142487} According to the Ti-Cu phase diagram,~\cite{Murray1983,Okamoto2002} Ti and Cu exhibit trace-solubilities in one another, with titanium showing a maximum solubility in copper of approximately 8~at.\% at 855{\textdegree}C, otherwise they are subject to forming intermetallic phases. Murray~\textit{et al.} do not report phase transitions below 500{\textdegree}C,~\cite{Murray1983} whereas Okomoto~\textit{et al.} provide an updated phase diagram going down to 200{\textdegree}C suggesting intermetallic Cu-Ti phases can form below 500{\textdegree}C.~\cite{Okamoto2002}.\par

Copper and tungsten are completely immiscible,~\cite{VULLERS2015213} therefore, in TiW/Cu systems the only potential reactions are Cu-Ti reactions at the interface. However, from the core level analysis at the interface no BE shifts or changes in line-shape occur that would suggest the formation of such a compound. If a reaction was to occur one would expect to observe a significant change when comparing the AD sample, where the Ti-Cu reaction should not occur and the 5~h annealed sample, where the conditions favour intermetallic phase formation. Cocke~\textit{et al.} report that for Ti\textsubscript{36}Cu\textsubscript{64} alloys, the Ti~2\textit{p}\textsubscript{3/2} core level should shift by +0.5~eV relative to the pure titanium metal core level BE, whereas the Cu~2\textit{p}\textsubscript{3/2} peak should shift by -0.2~eV relative to the pure copper metal core level BE position.~\cite{COCKE1990119} In the present case, a negative shift in BE of the Ti~2\textit{p} core line with annealing is observed, which as discussed earlier, falls more in-line with the changes to the O~1\textit{s} core line and the development of Ti-O states, and no BE shift or even FWHM change in the Cu~2\textit{p}\textsubscript{3/2} core line is notable.\par 

In Ti/Cu systems, several studies have detected the formation of intermetallic compounds during annealing as low as 350{\textdegree}C.~\cite{Liotard_1985, Jian_1992, OLOWOLAFE199337} In such systems, copper and titanium interdiffusion proceeded, leading to the consumption of titanium and the growth of the TiCu\textsubscript{3} phase at the expense of TiCu.~\cite{Liotard_1985, Jian_1992} This ultimately led to the demise of titanium as a suitable power semiconductor diffusion barrier candidate on its own. However, in terms of TiW/Cu studies, the formation of Cu-Ti intermetallic phases at temperatures below 500{\textdegree}C is not readily observed. Olowolafe~\textit{et al.} studied Si/SiO\textsubscript{2}/TiW(50~nm)/Cu(100~nm)/TiW(50~nm) multi-layer stacks. The structures were annealed at 500{\textdegree}C for 1-2~h and from AES and XRD measurements, no sign of interdiffusion or intermetallic compound formation was observed.~\cite{OLOWOLAFE199337} Similarly, Wang~\textit{et al.} report no reaction or CuTi formation between Cu and TiW up to 875{\textdegree}C after 30s of rapid thermal annealing (RTA), although at temperatures above 800{\textdegree}C slight titanium interdiffusion into copper was observed.~\cite{Wang_1993} However, in this case the RTA may be too quick to allow for a significant reaction or sufficient titanium diffusion and therefore is unlikely to be comparable to our study. Fugger~\textit{et al.} explored annealing TiW(50~nm)/Cu(1000~nm) systems at 400{\textdegree}C for 8~h, and observe no Cu-Ti intermetallic phase formation.~\cite{FUGGER20142487} Based on these studies and the TEM/SXPS/HAXPES data presented in this work, it is clear that the TiW/Cu interface is stable during annealing at 400{\textdegree}C for as long as 5~h and no Cu-Ti intermetallic compounds form.\par

\subsection{Interfacial Oxide Formation}\label{Interfacial_Discussion}

From the acquired SXPS data at the interface, it is clear that there is an accumulation of oxygen at the interface leading to the formation of a thin interfacial layer of Ti-O and W-O states. Given that the TiW interface is not subjected to excessively high temperatures and the concentration of oxygen is insufficient to form an extended bulk oxide, it is expected that the TiW oxidation behaviour will not follow the parabolic law outlined by the classic Wagner theory but rather obey the Cabrera-Mott theory which is applicable low temperature oxidation and the formation of thin oxide films.~\cite{COCKE1990119, FUJITA2013347} The surface reaction at the interface will occur step-wise, first with the atomic dissociation of oxygen and physisorption to the TiW surface, followed by chemisorption into the lattice and the subsequent migration of oxygen ions through the metal surface after which a metal/metal oxide and metal/oxide/gas interface forms. The growth of the oxide layer then proceeds via the movement of ions and electrons/holes across the oxide layer. The theory states that the metal ion diffusion across the developing metal oxide layer is governed by the electric field generated between the adsorbed oxygen on the surface and the underlying metal, which leads to a very fast initial rate of oxidation before a slowing down.~\cite{FUJITA2013347}\par

In a binary alloy like TiW, the metal component with the largest enthalpy of formation for its oxide will oxidise first (i.e.~titanium) and be the dominant component to diffuse across the metal oxide layer.~\cite{Kalha_2021_TiWO} The second metal (i.e.~tungsten) will oxidise only, if the temperature is sufficiently high or the oxygen partial pressure is high enough to induce oxidation. From the O~1\textit{s} spectra alone it is difficult to confirm the ratio of titanium to tungsten oxides as the chemical shift between the two is difficult to resolve (TiO\textsubscript{2} = 529.7~eV and WO\textsubscript{3} = 530.2~eV).~\cite{DIMITROV2002100} However, from the SXPS core levels, the Ti~2\textit{p} spectra show the most change in line shape relative to the bulk ``metallic'' HAXPES spectra, suggesting preferential oxidation of the Ti does occur.\par

It is difficult to confirm the exact origin of the oxygen that has accumulated at the TiW interface, but three potential routes can be identified: \textbf{(a)} the diffusion of oxygen through the 25~nm-thick copper layer (whether that be during deposition, annealing or when the sample is removed from the deposition chamber) and/or \textbf{(b)} the gettering of residual oxygen within the deposition chamber by the bare TiW surface prior to the deposition of copper and/or \textbf{(c)} oxygen within the copper deposit accumulating at the TiW interface. Considering route (a), the SXPS measurements on the as-received samples show that significant oxidation of the copper surface does not occur as only a native Cu(I) oxide is present rather than Cu(II) oxide. Leon~\textit{et al.} conducted polarised reflectometry measurements on sputter deposited 32~nm-thick copper films exposed to room temperature (300~K), ambient pressure conditions (0.1~MPa), and a relative humidity of 87~\% after 1 and 253 days of exposure. From their measurements, the native oxide thickness was extracted to be 1.81~nm after 1~day, growing to 6.59~nm after 253 days.~\cite{Leon_2016} This suggests that the oxygen diffusion path length in sputtered deposited copper is less than 10~nm when under standard conditions. Therefore, based on this it can be ruled out that the interfacial TiW oxide develops via the room temperature diffusion of oxygen through the 25~nm copper layer when the sample is exposed to air. Alternatively, residual oxygen is expected to be present in the deposition chamber as it operates only under high vacuum conditions (10\textsuperscript{-8}-10\textsuperscript{-7}~mbar). The oxidation of copper is known to enhance with higher temperatures~\cite{Rhodin_1950} and annealing timescales,~\cite{Stiedl_2019} and therefore oxygen may be able to diffuse through the copper, towards the TiW interface during annealing. However, in this case significant differences in the chemical state of the copper surface and TiW interface would be expected between the AD and 5~h annealed sample. However, this is not the case, ruling out this pathway.\par

Alternatively, TiW may getter residual oxygen within the deposition chamber following route (b). As assumed above, residual quantities of oxygen are expected in the deposition chamber. Therefore, when the TiW is deposited its surface will be exposed to this atmosphere prior to the deposition of copper, providing the perfect opportunity for oxygen to accumulate at the TiW/Cu interface and partially oxidise. The third potential route, (c), involves the accumulation of oxygen from within the copper deposit at the TiW/Cu interface. The complete depth profiles displayed in Supplementary Information V show oxygen is present within the copper layer. This can be attributed to the purity of the target and sputtering conditions as residual oxygen in the chamber may be incorporated into the copper film. Therefore, the source of oxygen at the TiW/Cu interface may be due to residual oxygen at the base of the copper film being drawn to the TiW interface. Overall, routes (b) and (c) appear most likely and the source of oxygen can be assumed to be largely from the deposition chamber environment.\par

The inclusion of oxygen at the interface may be of benefit to the system and improve the stability of the TiW/Cu interface and effectiveness of the TiW barrier. Oparowski~\textit{et al.} showed that the performance of TiW diffusion barriers against Au or Al interdiffusion could be improved by varying the sputter gas composition during deposition.~\cite{OPAROWSKI1987313} Depositing TiW in a 95:5 Ar:O gas mixture compared to a pure Ar gas mixture was shown to increase the barrier breakdown temperature from 300 to 400{\textdegree}C, however, the average adhesion structure was shown to be significantly compromised with the addition of oxygen to the gas mixture. This was attributed to the formation of TiO\textsubscript{2} and WO\textsubscript{3} at the interface prior to metallisation deposition.\par

Wang~\textit{et al.} explored this further by studying a Si/Ti\textsubscript{27}W\textsubscript{73}/Cu sample configuration with and without a venting break between the TiW and Cu deposition steps. The venting break involved opening the deposition chamber and exposing the TiW face to air for 30~minutes.~\cite{Wang_1993} The authors showed that the vented samples possessed a higher TiW breakdown temperature of 875{\textdegree}C compared to 825{\textdegree}C for the no vent sample. The explanation for this improvement is that exposure to air leads to a ``decoration'' of oxides at the TiW grain boundaries, reducing the mobility of titanium out of the barrier and subsequently retarding the interdiffusion of Si or metal atoms through the TiW and/or suppressing the reaction of TiW with either the Si or metal interfaces, both of which improve the barrier performance. Therefore, the presence of oxide species at the interface in the present samples may explain the stability of the diffusion barrier.\par

\section{Conclusions}

This work probes the chemical states present in both TiW and Cu as well as across their interface for the first time. The Ti concentration close to the solubility limit in W as well as the unintentional presence of oxide species at the interface can explain the exceptional stability of the TiW diffusion barriers. The combined use of SXPS and HAXPES coupled with depth-profiling allows the exploration of the depth dependence of elemental distributions and chemical states, enabling the characterisation of the titanium diffusion mechanism that occurs during annealing.\par

Titanium diffuses out of the barrier and accumulates at the copper surface when annealed at 400{\textdegree}C, with prolonged annealing timescales resulting in a greater out-diffusion of titanium from the TiW. Accumulated titanium at the copper surface undergoes oxidation to the predominately Ti(IV) oxide state with a small contribution from Ti(III) states. Annealing at 400{\textdegree}C for 0.5~h and 5~h did not lead to the failure of the TiW barrier, indicated by the lack of Cu-Si and Cu-Ti intermetallic formation. Furthermore, TEM images show that after 5~h of annealing the TiW remains stable and copper is not observed to diffuse downward but is rather constrained above the TiW. HAXPES measurements were able to characterise the TiW layer in its true state, showing that the TiW is representative of a binary metal alloy system, with trace quantities of oxygen attributed to the gettering of residual oxygen by the TiW face prior to the deposition of copper. SXPS measurements across the interface detected low concentrations of metal-oxygen states (W-O and Ti-O), suggesting that a thin interfacial oxide forms between the TiW and Cu layers. Despite the depletion of titanium from the barrier layer, the electronic structure of bulk TiW does not change with annealing, suggesting that the extent of titanium depletion is too small to have any observable influence.\par

Results obtained for the specific metallisation technology explored in this work, exemplify the usefulness of the employed characterisation protocol to study not only the spacial distributions of elements, but importantly the chemical states present. In addition, the in-situ partial decapping of the Cu overlayer makes it possible to  study the true state of the interface in-situ. This approach holds promise for the exploration of other metallisation schemes and their buried interfaces, where similar open questions regarding stability and breakdown remain.\par

\section*{Supplementary Material}

The Supplementary Information contains TEM images collected for the as-deposited sample, the complete EDS dataset including W and Cu signals for both the as-deposited and 5~h anneal samples, survey spectra collected with synchrotron-based SXPS and HAXPES, peak-fit analysis of the Ti~2\textit{p} core level spectra collected at the copper surface of as-received samples, the complete SXPS depth profile including the C and O signals, photoionisation cross section profiles for the core levels analysed across the measured photon energy range and additional figures aiding with the assessment of the core level spectra collected at the interface.

\begin{acknowledgments}
CK acknowledges the support from the Department of Chemistry, UCL. AR acknowledges the support from the Analytical Chemistry Trust Fund for her CAMS-UK Fellowship. These experiments were conducted with the support of the Diamond Light Source, beamline I09 (proposal SI24248). The authors would like to thank Dave McCue, I09 beamline technician, for his support of the experiments. The SXPS depth profile measurements were performed at the EPSRC National Facility for XPS (“HarwellXPS”) operated by Cardiff University and UCL, under Contract No. PR16195.
\end{acknowledgments}

\section*{Data availability statement}
The majority of the data that supports the findings of this study are available within the article and its Supplementary Material.

\section*{Disclosures}
The authors declare no conflicts of interest.

\section{References}

\bibliography{References}

\begin{thebibliography}{84}%
\makeatletter
\providecommand \@ifxundefined [1]{%
 \@ifx{#1\undefined}
}%
\providecommand \@ifnum [1]{%
 \ifnum #1\expandafter \@firstoftwo
 \else \expandafter \@secondoftwo
 \fi
}%
\providecommand \@ifx [1]{%
 \ifx #1\expandafter \@firstoftwo
 \else \expandafter \@secondoftwo
 \fi
}%
\providecommand \natexlab [1]{#1}%
\providecommand \enquote  [1]{``#1''}%
\providecommand \bibnamefont  [1]{#1}%
\providecommand \bibfnamefont [1]{#1}%
\providecommand \citenamefont [1]{#1}%
\providecommand \href@noop [0]{\@secondoftwo}%
\providecommand \href [0]{\begingroup \@sanitize@url \@href}%
\providecommand \@href[1]{\@@startlink{#1}\@@href}%
\providecommand \@@href[1]{\endgroup#1\@@endlink}%
\providecommand \@sanitize@url [0]{\catcode `\\12\catcode `\$12\catcode
  `\&12\catcode `\#12\catcode `\^12\catcode `\_12\catcode `\%12\relax}%
\providecommand \@@startlink[1]{}%
\providecommand \@@endlink[0]{}%
\providecommand \url  [0]{\begingroup\@sanitize@url \@url }%
\providecommand \@url [1]{\endgroup\@href {#1}{\urlprefix }}%
\providecommand \urlprefix  [0]{URL }%
\providecommand \Eprint [0]{\href }%
\providecommand \doibase [0]{http://dx.doi.org/}%
\providecommand \selectlanguage [0]{\@gobble}%
\providecommand \bibinfo  [0]{\@secondoftwo}%
\providecommand \bibfield  [0]{\@secondoftwo}%
\providecommand \translation [1]{[#1]}%
\providecommand \BibitemOpen [0]{}%
\providecommand \bibitemStop [0]{}%
\providecommand \bibitemNoStop [0]{.\EOS\space}%
\providecommand \EOS [0]{\spacefactor3000\relax}%
\providecommand \BibitemShut  [1]{\csname bibitem#1\endcsname}%
\let\auto@bib@innerbib\@empty
\bibitem [{\citenamefont {Shen}\ and\ \citenamefont {Omura}(2007)}]{Shen_2007}%
  \BibitemOpen
  \bibfield  {author} {\bibinfo {author} {\bibfnamefont {Z.~J.}\ \bibnamefont
  {Shen}}\ and\ \bibinfo {author} {\bibfnamefont {I.}~\bibnamefont {Omura}},\
  }\href {\doibase 10.1109/JPROC.2006.890118} {\bibfield  {journal} {\bibinfo
  {journal} {Proceedings of the IEEE}\ }\textbf {\bibinfo {volume} {95}},\
  \bibinfo {pages} {778} (\bibinfo {year} {2007})}\BibitemShut {NoStop}%
\bibitem [{\citenamefont {Do}\ \emph {et~al.}(2020)\citenamefont {Do},
  \citenamefont {Li}, \citenamefont {Trovão},\ and\ \citenamefont
  {Boulon}}]{2020_Do}%
  \BibitemOpen
  \bibfield  {author} {\bibinfo {author} {\bibfnamefont {T.~V.}\ \bibnamefont
  {Do}}, \bibinfo {author} {\bibfnamefont {K.}~\bibnamefont {Li}}, \bibinfo
  {author} {\bibfnamefont {J.~P.}\ \bibnamefont {Trovão}}, \ and\ \bibinfo
  {author} {\bibfnamefont {L.}~\bibnamefont {Boulon}},\ }in\ \href {\doibase
  10.1109/VPPC49601.2020.9330854} {\emph {\bibinfo {booktitle} {2020 IEEE
  Vehicle Power and Propulsion Conference (VPPC)}}}\ (\bibinfo {year} {2020})\
  pp.\ \bibinfo {pages} {1--6}\BibitemShut {NoStop}%
\bibitem [{\citenamefont {Iannaccone}\ \emph {et~al.}(2021)\citenamefont
  {Iannaccone}, \citenamefont {Sbrana}, \citenamefont {Morelli},\ and\
  \citenamefont {Strangio}}]{Iannaccone_2021}%
  \BibitemOpen
  \bibfield  {author} {\bibinfo {author} {\bibfnamefont {G.}~\bibnamefont
  {Iannaccone}}, \bibinfo {author} {\bibfnamefont {C.}~\bibnamefont {Sbrana}},
  \bibinfo {author} {\bibfnamefont {I.}~\bibnamefont {Morelli}}, \ and\
  \bibinfo {author} {\bibfnamefont {S.}~\bibnamefont {Strangio}},\ }\href
  {\doibase 10.1109/ACCESS.2021.3118897} {\bibfield  {journal} {\bibinfo
  {journal} {IEEE Access}\ }\textbf {\bibinfo {volume} {9}},\ \bibinfo {pages}
  {139446} (\bibinfo {year} {2021})}\BibitemShut {NoStop}%
\bibitem [{\citenamefont {Wong}\ \emph {et~al.}(1998)\citenamefont {Wong},
  \citenamefont {Kaja},\ and\ \citenamefont {DeHaven}}]{IBM}%
  \BibitemOpen
  \bibfield  {author} {\bibinfo {author} {\bibfnamefont {K.~K.~H.}\
  \bibnamefont {Wong}}, \bibinfo {author} {\bibfnamefont {S.}~\bibnamefont
  {Kaja}}, \ and\ \bibinfo {author} {\bibfnamefont {P.~W.}\ \bibnamefont
  {DeHaven}},\ }\href {\doibase 10.1147/rd.425.0587} {\bibfield  {journal}
  {\bibinfo  {journal} {IBM Journal of Research and Development}\ }\textbf
  {\bibinfo {volume} {42}},\ \bibinfo {pages} {587–596} (\bibinfo {year}
  {1998})}\BibitemShut {NoStop}%
\bibitem [{\citenamefont {Behrens}(2013)}]{Behrens_2013}%
  \BibitemOpen
  \bibfield  {author} {\bibinfo {author} {\bibfnamefont {T.}~\bibnamefont
  {Behrens}},\ }in\ \href {\doibase 10.1109/EPE.2013.6634478} {\emph {\bibinfo
  {booktitle} {2013 15th European Conference on Power Electronics and
  Applications (EPE)}}}\ (\bibinfo {year} {2013})\ pp.\ \bibinfo {pages}
  {1--10}\BibitemShut {NoStop}%
\bibitem [{\citenamefont {Souli}\ \emph {et~al.}(2017)\citenamefont {Souli},
  \citenamefont {Terziyska}, \citenamefont {Keckes}, \citenamefont {Robl},
  \citenamefont {Zechner},\ and\ \citenamefont {Mitterer}}]{Souli_2017}%
  \BibitemOpen
  \bibfield  {author} {\bibinfo {author} {\bibfnamefont {I.}~\bibnamefont
  {Souli}}, \bibinfo {author} {\bibfnamefont {V.~L.}\ \bibnamefont
  {Terziyska}}, \bibinfo {author} {\bibfnamefont {J.}~\bibnamefont {Keckes}},
  \bibinfo {author} {\bibfnamefont {W.}~\bibnamefont {Robl}}, \bibinfo {author}
  {\bibfnamefont {J.}~\bibnamefont {Zechner}}, \ and\ \bibinfo {author}
  {\bibfnamefont {C.}~\bibnamefont {Mitterer}},\ }\href {\doibase
  10.1116/1.4975805} {\bibfield  {journal} {\bibinfo  {journal} {Journal of
  Vacuum Science \& Technology B}\ }\textbf {\bibinfo {volume} {35}},\ \bibinfo
  {pages} {022201} (\bibinfo {year} {2017})}\BibitemShut {NoStop}%
\bibitem [{\citenamefont {Stolt}\ \emph {et~al.}(1991)\citenamefont {Stolt},
  \citenamefont {D'Heurle},\ and\ \citenamefont {Harper}}]{STOLT1991147}%
  \BibitemOpen
  \bibfield  {author} {\bibinfo {author} {\bibfnamefont {L.}~\bibnamefont
  {Stolt}}, \bibinfo {author} {\bibfnamefont {F.}~\bibnamefont {D'Heurle}}, \
  and\ \bibinfo {author} {\bibfnamefont {J.}~\bibnamefont {Harper}},\ }\href
  {\doibase https://doi.org/10.1016/0040-6090(91)90037-X} {\bibfield  {journal}
  {\bibinfo  {journal} {Thin Solid Films}\ }\textbf {\bibinfo {volume} {200}},\
  \bibinfo {pages} {147 } (\bibinfo {year} {1991})}\BibitemShut {NoStop}%
\bibitem [{\citenamefont {Shacham-Diamand}\ \emph {et~al.}(1993)\citenamefont
  {Shacham-Diamand}, \citenamefont {Dedhia}, \citenamefont {Hoffstetter},\ and\
  \citenamefont {Oldham}}]{Shacham_Diamand_1993}%
  \BibitemOpen
  \bibfield  {author} {\bibinfo {author} {\bibfnamefont {Y.}~\bibnamefont
  {Shacham-Diamand}}, \bibinfo {author} {\bibfnamefont {A.}~\bibnamefont
  {Dedhia}}, \bibinfo {author} {\bibfnamefont {D.}~\bibnamefont {Hoffstetter}},
  \ and\ \bibinfo {author} {\bibfnamefont {W.~G.}\ \bibnamefont {Oldham}},\
  }\href {\doibase 10.1149/1.2220837} {\bibfield  {journal} {\bibinfo
  {journal} {Journal of The Electrochemical Society}\ }\textbf {\bibinfo
  {volume} {140}},\ \bibinfo {pages} {2427} (\bibinfo {year}
  {1993})}\BibitemShut {NoStop}%
\bibitem [{\citenamefont {Sachdeva}\ \emph {et~al.}(2001)\citenamefont
  {Sachdeva}, \citenamefont {Istratov},\ and\ \citenamefont
  {Weber}}]{Sachdeva_2001}%
  \BibitemOpen
  \bibfield  {author} {\bibinfo {author} {\bibfnamefont {R.}~\bibnamefont
  {Sachdeva}}, \bibinfo {author} {\bibfnamefont {A.~A.}\ \bibnamefont
  {Istratov}}, \ and\ \bibinfo {author} {\bibfnamefont {E.~R.}\ \bibnamefont
  {Weber}},\ }\href {\doibase 10.1063/1.1415350} {\bibfield  {journal}
  {\bibinfo  {journal} {Applied Physics Letters}\ }\textbf {\bibinfo {volume}
  {79}},\ \bibinfo {pages} {2937} (\bibinfo {year} {2001})}\BibitemShut
  {NoStop}%
\bibitem [{\citenamefont {{Cunningham}}\ \emph {et~al.}(1970)\citenamefont
  {{Cunningham}}, \citenamefont {{Fuller}},\ and\ \citenamefont
  {{Haywood}}}]{Cunningham_1970}%
  \BibitemOpen
  \bibfield  {author} {\bibinfo {author} {\bibfnamefont {J.~A.}\ \bibnamefont
  {{Cunningham}}}, \bibinfo {author} {\bibfnamefont {C.~R.}\ \bibnamefont
  {{Fuller}}}, \ and\ \bibinfo {author} {\bibfnamefont {C.~T.}\ \bibnamefont
  {{Haywood}}},\ }\href {\doibase 10.1109/TR.1970.5216441} {\bibfield
  {journal} {\bibinfo  {journal} {IEEE Transactions on Reliability}\ }\textbf
  {\bibinfo {volume} {R-19}},\ \bibinfo {pages} {182} (\bibinfo {year}
  {1970})}\BibitemShut {NoStop}%
\bibitem [{\citenamefont {Shen}\ \emph {et~al.}(1986)\citenamefont {Shen},
  \citenamefont {Smith}, \citenamefont {Anthony},\ and\ \citenamefont
  {Matyi}}]{Shen_1986}%
  \BibitemOpen
  \bibfield  {author} {\bibinfo {author} {\bibfnamefont {B.~W.}\ \bibnamefont
  {Shen}}, \bibinfo {author} {\bibfnamefont {G.~C.}\ \bibnamefont {Smith}},
  \bibinfo {author} {\bibfnamefont {J.~M.}\ \bibnamefont {Anthony}}, \ and\
  \bibinfo {author} {\bibfnamefont {R.~J.}\ \bibnamefont {Matyi}},\ }\href
  {\doibase 10.1116/1.583460} {\bibfield  {journal} {\bibinfo  {journal}
  {Journal of Vacuum Science \& Technology B: Microelectronics Processing and
  Phenomena}\ }\textbf {\bibinfo {volume} {4}},\ \bibinfo {pages} {1369}
  (\bibinfo {year} {1986})}\BibitemShut {NoStop}%
\bibitem [{\citenamefont {Olowolafe}\ \emph {et~al.}(1993)\citenamefont
  {Olowolafe}, \citenamefont {Mogab},\ and\ \citenamefont
  {Gregory}}]{OLOWOLAFE199337}%
  \BibitemOpen
  \bibfield  {author} {\bibinfo {author} {\bibfnamefont {J.}~\bibnamefont
  {Olowolafe}}, \bibinfo {author} {\bibfnamefont {C.}~\bibnamefont {Mogab}}, \
  and\ \bibinfo {author} {\bibfnamefont {R.}~\bibnamefont {Gregory}},\ }\href
  {\doibase https://doi.org/10.1016/0040-6090(93)90184-Q} {\bibfield  {journal}
  {\bibinfo  {journal} {Thin Solid Films}\ }\textbf {\bibinfo {volume} {227}},\
  \bibinfo {pages} {37} (\bibinfo {year} {1993})}\BibitemShut {NoStop}%
\bibitem [{\citenamefont {Day}\ \emph {et~al.}(1977)\citenamefont {Day},
  \citenamefont {Christou},\ and\ \citenamefont {Macpherson}}]{Christou_1977}%
  \BibitemOpen
  \bibfield  {author} {\bibinfo {author} {\bibfnamefont {H.~M.}\ \bibnamefont
  {Day}}, \bibinfo {author} {\bibfnamefont {A.}~\bibnamefont {Christou}}, \
  and\ \bibinfo {author} {\bibfnamefont {A.~C.}\ \bibnamefont {Macpherson}},\
  }\href {\doibase 10.1116/1.569396} {\bibfield  {journal} {\bibinfo  {journal}
  {Journal of Vacuum Science and Technology}\ }\textbf {\bibinfo {volume}
  {14}},\ \bibinfo {pages} {939} (\bibinfo {year} {1977})}\BibitemShut
  {NoStop}%
\bibitem [{\citenamefont {Nicolet}(1978)}]{NICOLET_1978}%
  \BibitemOpen
  \bibfield  {author} {\bibinfo {author} {\bibfnamefont {M.-A.}\ \bibnamefont
  {Nicolet}},\ }\href {\doibase https://doi.org/10.1016/0040-6090(78)90184-0}
  {\bibfield  {journal} {\bibinfo  {journal} {Thin Solid Films}\ }\textbf
  {\bibinfo {volume} {52}},\ \bibinfo {pages} {415} (\bibinfo {year}
  {1978})}\BibitemShut {NoStop}%
\bibitem [{\citenamefont {Ghate}\ \emph {et~al.}(1978)\citenamefont {Ghate},
  \citenamefont {Blair}, \citenamefont {Fuller},\ and\ \citenamefont
  {McGuire}}]{GHATE1978117}%
  \BibitemOpen
  \bibfield  {author} {\bibinfo {author} {\bibfnamefont {P.}~\bibnamefont
  {Ghate}}, \bibinfo {author} {\bibfnamefont {J.}~\bibnamefont {Blair}},
  \bibinfo {author} {\bibfnamefont {C.}~\bibnamefont {Fuller}}, \ and\ \bibinfo
  {author} {\bibfnamefont {G.}~\bibnamefont {McGuire}},\ }\href {\doibase
  https://doi.org/10.1016/0040-6090(78)90024-X} {\bibfield  {journal} {\bibinfo
   {journal} {Thin Solid Films}\ }\textbf {\bibinfo {volume} {53}},\ \bibinfo
  {pages} {117 } (\bibinfo {year} {1978})}\BibitemShut {NoStop}%
\bibitem [{\citenamefont {Canali}\ \emph {et~al.}(1982)\citenamefont {Canali},
  \citenamefont {Celotti}, \citenamefont {Fantini},\ and\ \citenamefont
  {Zanoni}}]{CANALI19829}%
  \BibitemOpen
  \bibfield  {author} {\bibinfo {author} {\bibfnamefont {C.}~\bibnamefont
  {Canali}}, \bibinfo {author} {\bibfnamefont {G.}~\bibnamefont {Celotti}},
  \bibinfo {author} {\bibfnamefont {F.}~\bibnamefont {Fantini}}, \ and\
  \bibinfo {author} {\bibfnamefont {E.}~\bibnamefont {Zanoni}},\ }\href
  {\doibase https://doi.org/10.1016/0040-6090(82)90345-5} {\bibfield  {journal}
  {\bibinfo  {journal} {Thin Solid Films}\ }\textbf {\bibinfo {volume} {88}},\
  \bibinfo {pages} {9} (\bibinfo {year} {1982})}\BibitemShut {NoStop}%
\bibitem [{\citenamefont {Babcock}\ and\ \citenamefont
  {Tu}(1982)}]{Babcoock_1982}%
  \BibitemOpen
  \bibfield  {author} {\bibinfo {author} {\bibfnamefont {S.~E.}\ \bibnamefont
  {Babcock}}\ and\ \bibinfo {author} {\bibfnamefont {K.~N.}\ \bibnamefont
  {Tu}},\ }\href {\doibase 10.1063/1.330031} {\bibfield  {journal} {\bibinfo
  {journal} {Journal of Applied Physics}\ }\textbf {\bibinfo {volume} {53}},\
  \bibinfo {pages} {6898} (\bibinfo {year} {1982})}\BibitemShut {NoStop}%
\bibitem [{\citenamefont {Olowolafe}\ \emph {et~al.}(1985)\citenamefont
  {Olowolafe}, \citenamefont {Palmstro/m}, \citenamefont {Colgan},\ and\
  \citenamefont {Mayer}}]{Olowolafe_1985}%
  \BibitemOpen
  \bibfield  {author} {\bibinfo {author} {\bibfnamefont {J.~O.}\ \bibnamefont
  {Olowolafe}}, \bibinfo {author} {\bibfnamefont {C.~J.}\ \bibnamefont
  {Palmstro/m}}, \bibinfo {author} {\bibfnamefont {E.~G.}\ \bibnamefont
  {Colgan}}, \ and\ \bibinfo {author} {\bibfnamefont {J.~W.}\ \bibnamefont
  {Mayer}},\ }\href {\doibase 10.1063/1.335764} {\bibfield  {journal} {\bibinfo
   {journal} {Journal of Applied Physics}\ }\textbf {\bibinfo {volume} {58}},\
  \bibinfo {pages} {3440} (\bibinfo {year} {1985})}\BibitemShut {NoStop}%
\bibitem [{\citenamefont {Babcock}\ and\ \citenamefont
  {Tu}(1986)}]{Babcock_1986}%
  \BibitemOpen
  \bibfield  {author} {\bibinfo {author} {\bibfnamefont {S.~E.}\ \bibnamefont
  {Babcock}}\ and\ \bibinfo {author} {\bibfnamefont {K.~N.}\ \bibnamefont
  {Tu}},\ }\href {\doibase 10.1063/1.336470} {\bibfield  {journal} {\bibinfo
  {journal} {Journal of Applied Physics}\ }\textbf {\bibinfo {volume} {59}},\
  \bibinfo {pages} {1599} (\bibinfo {year} {1986})}\BibitemShut {NoStop}%
\bibitem [{\citenamefont {Krautz}\ \emph {et~al.}(1988)\citenamefont {Krautz},
  \citenamefont {Wenzel}, \citenamefont {Bornkessel},\ and\ \citenamefont
  {Blasek}}]{Krautz_1988}%
  \BibitemOpen
  \bibfield  {author} {\bibinfo {author} {\bibfnamefont {H.}~\bibnamefont
  {Krautz}}, \bibinfo {author} {\bibfnamefont {C.}~\bibnamefont {Wenzel}},
  \bibinfo {author} {\bibfnamefont {K.}~\bibnamefont {Bornkessel}}, \ and\
  \bibinfo {author} {\bibfnamefont {G.}~\bibnamefont {Blasek}},\ }\href@noop {}
  {\bibfield  {journal} {\bibinfo  {journal} {Physica Status Solidi (a)}\
  }\textbf {\bibinfo {volume} {110}},\ \bibinfo {pages} {K77} (\bibinfo {year}
  {1988})}\BibitemShut {NoStop}%
\bibitem [{\citenamefont {Ashkenazi}\ \emph {et~al.}(1993)\citenamefont
  {Ashkenazi}, \citenamefont {Komen},\ and\ \citenamefont
  {Lerner}}]{ASHKENAZI1993746}%
  \BibitemOpen
  \bibfield  {author} {\bibinfo {author} {\bibfnamefont {A.}~\bibnamefont
  {Ashkenazi}}, \bibinfo {author} {\bibfnamefont {Y.}~\bibnamefont {Komen}}, \
  and\ \bibinfo {author} {\bibfnamefont {I.}~\bibnamefont {Lerner}},\ }\href
  {\doibase https://doi.org/10.1016/0169-4332(93)90749-2} {\bibfield  {journal}
  {\bibinfo  {journal} {Applied Surface Science}\ }\textbf {\bibinfo {volume}
  {65-66}},\ \bibinfo {pages} {746} (\bibinfo {year} {1993})}\BibitemShut
  {NoStop}%
\bibitem [{\citenamefont {Wang}\ \emph {et~al.}(1993)\citenamefont {Wang},
  \citenamefont {Suthar}, \citenamefont {Hoeflich},\ and\ \citenamefont
  {Burrow}}]{Wang_1993}%
  \BibitemOpen
  \bibfield  {author} {\bibinfo {author} {\bibfnamefont {S.}~\bibnamefont
  {Wang}}, \bibinfo {author} {\bibfnamefont {S.}~\bibnamefont {Suthar}},
  \bibinfo {author} {\bibfnamefont {C.}~\bibnamefont {Hoeflich}}, \ and\
  \bibinfo {author} {\bibfnamefont {B.~J.}\ \bibnamefont {Burrow}},\ }\href
  {\doibase 10.1063/1.353135} {\bibfield  {journal} {\bibinfo  {journal}
  {Journal of Applied Physics}\ }\textbf {\bibinfo {volume} {73}},\ \bibinfo
  {pages} {2301} (\bibinfo {year} {1993})}\BibitemShut {NoStop}%
\bibitem [{\citenamefont {Chiou}\ \emph {et~al.}(1995)\citenamefont {Chiou},
  \citenamefont {Juang},\ and\ \citenamefont {Chen}}]{Chiou_1995}%
  \BibitemOpen
  \bibfield  {author} {\bibinfo {author} {\bibfnamefont {J.-C.}\ \bibnamefont
  {Chiou}}, \bibinfo {author} {\bibfnamefont {K.-C.}\ \bibnamefont {Juang}}, \
  and\ \bibinfo {author} {\bibfnamefont {M.-C.}\ \bibnamefont {Chen}},\ }\href
  {\doibase 10.1149/1.2044295} {\bibfield  {journal} {\bibinfo  {journal}
  {Journal of The Electrochemical Society}\ }\textbf {\bibinfo {volume}
  {142}},\ \bibinfo {pages} {2326} (\bibinfo {year} {1995})}\BibitemShut
  {NoStop}%
\bibitem [{\citenamefont {Chang}\ \emph {et~al.}(2000)\citenamefont {Chang},
  \citenamefont {Wu}, \citenamefont {Huang}, \citenamefont {Shih},\ and\
  \citenamefont {Chao}}]{Chang_2000}%
  \BibitemOpen
  \bibfield  {author} {\bibinfo {author} {\bibfnamefont {C.~S.}\ \bibnamefont
  {Chang}}, \bibinfo {author} {\bibfnamefont {T.~B.}\ \bibnamefont {Wu}},
  \bibinfo {author} {\bibfnamefont {C.~K.}\ \bibnamefont {Huang}}, \bibinfo
  {author} {\bibfnamefont {W.~C.}\ \bibnamefont {Shih}}, \ and\ \bibinfo
  {author} {\bibfnamefont {L.~L.}\ \bibnamefont {Chao}},\ }\href {\doibase
  10.1143/jjap.39.6413} {\bibfield  {journal} {\bibinfo  {journal} {Japanese
  Journal of Applied Physics}\ }\textbf {\bibinfo {volume} {39}},\ \bibinfo
  {pages} {6413} (\bibinfo {year} {2000})}\BibitemShut {NoStop}%
\bibitem [{\citenamefont {Bhagat}\ \emph {et~al.}(2006)\citenamefont {Bhagat},
  \citenamefont {Han},\ and\ \citenamefont {Alford}}]{BHAGAT20061998}%
  \BibitemOpen
  \bibfield  {author} {\bibinfo {author} {\bibfnamefont {S.}~\bibnamefont
  {Bhagat}}, \bibinfo {author} {\bibfnamefont {H.}~\bibnamefont {Han}}, \ and\
  \bibinfo {author} {\bibfnamefont {T.}~\bibnamefont {Alford}},\ }\href
  {\doibase https://doi.org/10.1016/j.tsf.2006.03.049} {\bibfield  {journal}
  {\bibinfo  {journal} {Thin Solid Films}\ }\textbf {\bibinfo {volume} {515}},\
  \bibinfo {pages} {1998} (\bibinfo {year} {2006})}\BibitemShut {NoStop}%
\bibitem [{\citenamefont {Petrović}\ \emph {et~al.}(2010)\citenamefont
  {Petrović}, \citenamefont {Peruško}, \citenamefont {Gaković},
  \citenamefont {Mitrić}, \citenamefont {Kovač}, \citenamefont {Zalar},
  \citenamefont {Milinović}, \citenamefont {Bogdanović-Radović},\ and\
  \citenamefont {Milosavljević}}]{PETROVIC20102099}%
  \BibitemOpen
  \bibfield  {author} {\bibinfo {author} {\bibfnamefont {S.}~\bibnamefont
  {Petrović}}, \bibinfo {author} {\bibfnamefont {D.}~\bibnamefont {Peruško}},
  \bibinfo {author} {\bibfnamefont {B.}~\bibnamefont {Gaković}}, \bibinfo
  {author} {\bibfnamefont {M.}~\bibnamefont {Mitrić}}, \bibinfo {author}
  {\bibfnamefont {J.}~\bibnamefont {Kovač}}, \bibinfo {author} {\bibfnamefont
  {A.}~\bibnamefont {Zalar}}, \bibinfo {author} {\bibfnamefont
  {V.}~\bibnamefont {Milinović}}, \bibinfo {author} {\bibfnamefont
  {I.}~\bibnamefont {Bogdanović-Radović}}, \ and\ \bibinfo {author}
  {\bibfnamefont {M.}~\bibnamefont {Milosavljević}},\ }\href {\doibase
  https://doi.org/10.1016/j.surfcoat.2009.09.048} {\bibfield  {journal}
  {\bibinfo  {journal} {Surface and Coatings Technology}\ }\textbf {\bibinfo
  {volume} {204}},\ \bibinfo {pages} {2099} (\bibinfo {year} {2010})},\
  \bibinfo {note} {proceedings of the European Materials Research Socierty
  (E-MRS)Spring Meeting 2009}\BibitemShut {NoStop}%
\bibitem [{\citenamefont {Wang}\ and\ \citenamefont
  {Liang}(2011)}]{WANG2011979}%
  \BibitemOpen
  \bibfield  {author} {\bibinfo {author} {\bibfnamefont {Q.}~\bibnamefont
  {Wang}}\ and\ \bibinfo {author} {\bibfnamefont {S.}~\bibnamefont {Liang}},\
  }\href {\doibase https://doi.org/10.1016/j.vacuum.2010.12.012} {\bibfield
  {journal} {\bibinfo  {journal} {Vacuum}\ }\textbf {\bibinfo {volume} {85}},\
  \bibinfo {pages} {979} (\bibinfo {year} {2011})}\BibitemShut {NoStop}%
\bibitem [{\citenamefont {Plappert}\ \emph {et~al.}(2012)\citenamefont
  {Plappert}, \citenamefont {Humbel}, \citenamefont {Koprowski},\ and\
  \citenamefont {Nowottnick}}]{Plappert_2012}%
  \BibitemOpen
  \bibfield  {author} {\bibinfo {author} {\bibfnamefont {M.}~\bibnamefont
  {Plappert}}, \bibinfo {author} {\bibfnamefont {O.}~\bibnamefont {Humbel}},
  \bibinfo {author} {\bibfnamefont {A.}~\bibnamefont {Koprowski}}, \ and\
  \bibinfo {author} {\bibfnamefont {M.}~\bibnamefont {Nowottnick}},\ }\href
  {\doibase https://doi.org/10.1016/j.microrel.2012.06.066} {\bibfield
  {journal} {\bibinfo  {journal} {Microelectronics Reliability}\ }\textbf
  {\bibinfo {volume} {52}},\ \bibinfo {pages} {1993} (\bibinfo {year}
  {2012})}\BibitemShut {NoStop}%
\bibitem [{\citenamefont {Fugger}\ \emph {et~al.}(2014)\citenamefont {Fugger},
  \citenamefont {Plappert}, \citenamefont {Schäffer}, \citenamefont {Humbel},
  \citenamefont {Hutter}, \citenamefont {Danninger},\ and\ \citenamefont
  {Nowottnick}}]{FUGGER20142487}%
  \BibitemOpen
  \bibfield  {author} {\bibinfo {author} {\bibfnamefont {M.}~\bibnamefont
  {Fugger}}, \bibinfo {author} {\bibfnamefont {M.}~\bibnamefont {Plappert}},
  \bibinfo {author} {\bibfnamefont {C.}~\bibnamefont {Schäffer}}, \bibinfo
  {author} {\bibfnamefont {O.}~\bibnamefont {Humbel}}, \bibinfo {author}
  {\bibfnamefont {H.}~\bibnamefont {Hutter}}, \bibinfo {author} {\bibfnamefont
  {H.}~\bibnamefont {Danninger}}, \ and\ \bibinfo {author} {\bibfnamefont
  {M.}~\bibnamefont {Nowottnick}},\ }\href {\doibase
  https://doi.org/10.1016/j.microrel.2014.04.016} {\bibfield  {journal}
  {\bibinfo  {journal} {Microelectronics Reliability}\ }\textbf {\bibinfo
  {volume} {54}},\ \bibinfo {pages} {2487} (\bibinfo {year}
  {2014})}\BibitemShut {NoStop}%
\bibitem [{\citenamefont {Woicik}(2016)}]{Woicik2016}%
  \BibitemOpen
  \bibfield  {author} {\bibinfo {author} {\bibfnamefont {J.}~\bibnamefont
  {Woicik}},\ }\href {\doibase 10.1007/978-3-319-24043-5} {\emph {\bibinfo
  {title} {{Hard X-ray Photoelectron Spectroscopy (HAXPES)}}}}\ (\bibinfo
  {publisher} {Springer},\ \bibinfo {year} {2016})\BibitemShut {NoStop}%
\bibitem [{\citenamefont {Kalha}\ \emph
  {et~al.}(2021{\natexlab{a}})\citenamefont {Kalha}, \citenamefont {Fernando},
  \citenamefont {Bhatt}, \citenamefont {Johansson}, \citenamefont {Linblad},
  \citenamefont {Rensmo}, \citenamefont {Mendina}, \citenamefont {Lindblad},
  \citenamefont {Siol}, \citenamefont {Jeurgens}, \citenamefont {Cancellieri},
  \citenamefont {Rossnagel}, \citenamefont {Medjanik}, \citenamefont
  {Schoenhense}, \citenamefont {Simon}, \citenamefont {alex Gray},
  \citenamefont {Nemsak}, \citenamefont {Lömker}, \citenamefont {Schlueter},\
  and\ \citenamefont {Regoutz}}]{HAXPES_Big_Boy}%
  \BibitemOpen
  \bibfield  {author} {\bibinfo {author} {\bibfnamefont {C.}~\bibnamefont
  {Kalha}}, \bibinfo {author} {\bibfnamefont {N.~K.}\ \bibnamefont {Fernando}},
  \bibinfo {author} {\bibfnamefont {P.}~\bibnamefont {Bhatt}}, \bibinfo
  {author} {\bibfnamefont {F.}~\bibnamefont {Johansson}}, \bibinfo {author}
  {\bibfnamefont {A.}~\bibnamefont {Linblad}}, \bibinfo {author} {\bibfnamefont
  {H.}~\bibnamefont {Rensmo}}, \bibinfo {author} {\bibfnamefont {L.~Z.}\
  \bibnamefont {Mendina}}, \bibinfo {author} {\bibfnamefont {R.}~\bibnamefont
  {Lindblad}}, \bibinfo {author} {\bibfnamefont {S.}~\bibnamefont {Siol}},
  \bibinfo {author} {\bibfnamefont {L.}~\bibnamefont {Jeurgens}}, \bibinfo
  {author} {\bibfnamefont {C.}~\bibnamefont {Cancellieri}}, \bibinfo {author}
  {\bibfnamefont {K.}~\bibnamefont {Rossnagel}}, \bibinfo {author}
  {\bibfnamefont {K.}~\bibnamefont {Medjanik}}, \bibinfo {author}
  {\bibfnamefont {G.}~\bibnamefont {Schoenhense}}, \bibinfo {author}
  {\bibfnamefont {M.}~\bibnamefont {Simon}}, \bibinfo {author} {\bibnamefont
  {alex Gray}}, \bibinfo {author} {\bibfnamefont {S.}~\bibnamefont {Nemsak}},
  \bibinfo {author} {\bibfnamefont {P.}~\bibnamefont {Lömker}}, \bibinfo
  {author} {\bibfnamefont {C.}~\bibnamefont {Schlueter}}, \ and\ \bibinfo
  {author} {\bibfnamefont {A.}~\bibnamefont {Regoutz}},\ }\href
  {http://iopscience.iop.org/article/10.1088/1361-648X/abeacd} {\bibfield
  {journal} {\bibinfo  {journal} {Journal of Physics: Condensed Matter}\ }
  (\bibinfo {year} {2021}{\natexlab{a}})}\BibitemShut {NoStop}%
\bibitem [{\citenamefont {Saghaeian}\ \emph {et~al.}(2019)\citenamefont
  {Saghaeian}, \citenamefont {Keckes}, \citenamefont {Woehlert}, \citenamefont
  {Rosenthal}, \citenamefont {Reisinger},\ and\ \citenamefont
  {Todt}}]{SAGHAEIAN2019137576}%
  \BibitemOpen
  \bibfield  {author} {\bibinfo {author} {\bibfnamefont {F.}~\bibnamefont
  {Saghaeian}}, \bibinfo {author} {\bibfnamefont {J.}~\bibnamefont {Keckes}},
  \bibinfo {author} {\bibfnamefont {S.}~\bibnamefont {Woehlert}}, \bibinfo
  {author} {\bibfnamefont {M.}~\bibnamefont {Rosenthal}}, \bibinfo {author}
  {\bibfnamefont {M.}~\bibnamefont {Reisinger}}, \ and\ \bibinfo {author}
  {\bibfnamefont {J.}~\bibnamefont {Todt}},\ }\href {\doibase
  https://doi.org/10.1016/j.tsf.2019.137576} {\bibfield  {journal} {\bibinfo
  {journal} {Thin Solid Films}\ }\textbf {\bibinfo {volume} {691}},\ \bibinfo
  {pages} {137576} (\bibinfo {year} {2019})}\BibitemShut {NoStop}%
\bibitem [{\citenamefont {Lee}\ and\ \citenamefont {Duncan}(2018)}]{Duncan_18}%
  \BibitemOpen
  \bibfield  {author} {\bibinfo {author} {\bibfnamefont {T.-L.}\ \bibnamefont
  {Lee}}\ and\ \bibinfo {author} {\bibfnamefont {D.~A.}\ \bibnamefont
  {Duncan}},\ }\href {\doibase 10.1080/08940886.2018.1483653} {\bibfield
  {journal} {\bibinfo  {journal} {Synchrotron Radiation News}\ }\textbf
  {\bibinfo {volume} {31}},\ \bibinfo {pages} {16} (\bibinfo {year}
  {2018})}\BibitemShut {NoStop}%
\bibitem [{\citenamefont {Tanuma}\ \emph {et~al.}(1994)\citenamefont {Tanuma},
  \citenamefont {Powell},\ and\ \citenamefont {Penn}}]{TPP-2M_QUASES}%
  \BibitemOpen
  \bibfield  {author} {\bibinfo {author} {\bibfnamefont {S.}~\bibnamefont
  {Tanuma}}, \bibinfo {author} {\bibfnamefont {C.~J.}\ \bibnamefont {Powell}},
  \ and\ \bibinfo {author} {\bibfnamefont {D.~R.}\ \bibnamefont {Penn}},\
  }\href {\doibase https://doi.org/10.1002/sia.740210302} {\bibfield  {journal}
  {\bibinfo  {journal} {Surface and Interface Analysis}\ }\textbf {\bibinfo
  {volume} {21}},\ \bibinfo {pages} {165} (\bibinfo {year} {1994})}\BibitemShut
  {NoStop}%
\bibitem [{\citenamefont {Walton}\ \emph {et~al.}(2010)\citenamefont {Walton},
  \citenamefont {Wincott}, \citenamefont {Fairley},\ and\ \citenamefont
  {Carrick}}]{CASA}%
  \BibitemOpen
  \bibfield  {author} {\bibinfo {author} {\bibfnamefont {J.}~\bibnamefont
  {Walton}}, \bibinfo {author} {\bibfnamefont {P.}~\bibnamefont {Wincott}},
  \bibinfo {author} {\bibfnamefont {N.}~\bibnamefont {Fairley}}, \ and\
  \bibinfo {author} {\bibfnamefont {A.}~\bibnamefont {Carrick}},\ }\href@noop
  {} {\emph {\bibinfo {title} {Peak Fitting with CasaXPS: A Casa Pocket
  Book}}}\ (\bibinfo  {publisher} {Accolyte Science},\ \bibinfo {address}
  {United Kingdom},\ \bibinfo {year} {2010})\BibitemShut {NoStop}%
\bibitem [{\citenamefont {Harper}\ and\ \citenamefont
  {Rodbell}(1997)}]{Harper_1997}%
  \BibitemOpen
  \bibfield  {author} {\bibinfo {author} {\bibfnamefont {J.~M.~E.}\
  \bibnamefont {Harper}}\ and\ \bibinfo {author} {\bibfnamefont {K.~P.}\
  \bibnamefont {Rodbell}},\ }\href {\doibase 10.1116/1.589407} {\bibfield
  {journal} {\bibinfo  {journal} {Journal of Vacuum Science \& Technology B:
  Microelectronics and Nanometer Structures Processing, Measurement, and
  Phenomena}\ }\textbf {\bibinfo {volume} {15}},\ \bibinfo {pages} {763}
  (\bibinfo {year} {1997})}\BibitemShut {NoStop}%
\bibitem [{\citenamefont {Dirks}\ \emph {et~al.}(1990)\citenamefont {Dirks},
  \citenamefont {Wolters},\ and\ \citenamefont {Nellissen}}]{DIRKS1990201}%
  \BibitemOpen
  \bibfield  {author} {\bibinfo {author} {\bibfnamefont {A.}~\bibnamefont
  {Dirks}}, \bibinfo {author} {\bibfnamefont {R.}~\bibnamefont {Wolters}}, \
  and\ \bibinfo {author} {\bibfnamefont {A.}~\bibnamefont {Nellissen}},\ }\href
  {\doibase https://doi.org/10.1016/S0040-6090(05)80028-8} {\bibfield
  {journal} {\bibinfo  {journal} {Thin Solid Films}\ }\textbf {\bibinfo
  {volume} {193-194}},\ \bibinfo {pages} {201 } (\bibinfo {year}
  {1990})}\BibitemShut {NoStop}%
\bibitem [{\citenamefont {Karlsson}\ \emph {et~al.}(1992)\citenamefont
  {Karlsson}, \citenamefont {Gunnarsson},\ and\ \citenamefont
  {Jepsen}}]{Karlsson_1992}%
  \BibitemOpen
  \bibfield  {author} {\bibinfo {author} {\bibfnamefont {K.}~\bibnamefont
  {Karlsson}}, \bibinfo {author} {\bibfnamefont {O.}~\bibnamefont
  {Gunnarsson}}, \ and\ \bibinfo {author} {\bibfnamefont {O.}~\bibnamefont
  {Jepsen}},\ }\href {\doibase 10.1088/0953-8984/4/11/009} {\bibfield
  {journal} {\bibinfo  {journal} {Journal of Physics: Condensed Matter}\
  }\textbf {\bibinfo {volume} {4}},\ \bibinfo {pages} {2801} (\bibinfo {year}
  {1992})}\BibitemShut {NoStop}%
\bibitem [{\citenamefont {Schön}(1973)}]{SCHON197396}%
  \BibitemOpen
  \bibfield  {author} {\bibinfo {author} {\bibfnamefont {G.}~\bibnamefont
  {Schön}},\ }\href {\doibase https://doi.org/10.1016/0039-6028(73)90206-9}
  {\bibfield  {journal} {\bibinfo  {journal} {Surface Science}\ }\textbf
  {\bibinfo {volume} {35}},\ \bibinfo {pages} {96} (\bibinfo {year}
  {1973})}\BibitemShut {NoStop}%
\bibitem [{\citenamefont {Kaushik}(1989)}]{KAUSHIK1989581}%
  \BibitemOpen
  \bibfield  {author} {\bibinfo {author} {\bibfnamefont {V.~K.}\ \bibnamefont
  {Kaushik}},\ }\href {\doibase https://doi.org/10.1016/0584-8547(89)80137-5}
  {\bibfield  {journal} {\bibinfo  {journal} {Spectrochimica Acta Part B:
  Atomic Spectroscopy}\ }\textbf {\bibinfo {volume} {44}},\ \bibinfo {pages}
  {581} (\bibinfo {year} {1989})}\BibitemShut {NoStop}%
\bibitem [{\citenamefont {Miller}\ and\ \citenamefont
  {Simmons}(1993{\natexlab{a}})}]{Cu_Miller}%
  \BibitemOpen
  \bibfield  {author} {\bibinfo {author} {\bibfnamefont {A.~C.}\ \bibnamefont
  {Miller}}\ and\ \bibinfo {author} {\bibfnamefont {G.~W.}\ \bibnamefont
  {Simmons}},\ }\href {\doibase 10.1116/1.1247725} {\bibfield  {journal}
  {\bibinfo  {journal} {Surface Science Spectra}\ }\textbf {\bibinfo {volume}
  {2}},\ \bibinfo {pages} {55} (\bibinfo {year}
  {1993}{\natexlab{a}})}\BibitemShut {NoStop}%
\bibitem [{\citenamefont {Vasquez}(1998)}]{Vasquez_1998}%
  \BibitemOpen
  \bibfield  {author} {\bibinfo {author} {\bibfnamefont {R.~P.}\ \bibnamefont
  {Vasquez}},\ }\href {\doibase 10.1116/1.1247881} {\bibfield  {journal}
  {\bibinfo  {journal} {Surface Science Spectra}\ }\textbf {\bibinfo {volume}
  {5}},\ \bibinfo {pages} {257} (\bibinfo {year} {1998})}\BibitemShut {NoStop}%
\bibitem [{\citenamefont {Tahir}\ and\ \citenamefont
  {Tougaard}(2012)}]{Tahir_2012}%
  \BibitemOpen
  \bibfield  {author} {\bibinfo {author} {\bibfnamefont {D.}~\bibnamefont
  {Tahir}}\ and\ \bibinfo {author} {\bibfnamefont {S.}~\bibnamefont
  {Tougaard}},\ }\href {\doibase 10.1088/0953-8984/24/17/175002} {\bibfield
  {journal} {\bibinfo  {journal} {Journal of Physics: Condensed Matter}\
  }\textbf {\bibinfo {volume} {24}},\ \bibinfo {pages} {175002} (\bibinfo
  {year} {2012})}\BibitemShut {NoStop}%
\bibitem [{\citenamefont {Sen}\ \emph {et~al.}(1976)\citenamefont {Sen},
  \citenamefont {Riga},\ and\ \citenamefont {Verbist}}]{SEN1976560}%
  \BibitemOpen
  \bibfield  {author} {\bibinfo {author} {\bibfnamefont {S.}~\bibnamefont
  {Sen}}, \bibinfo {author} {\bibfnamefont {J.}~\bibnamefont {Riga}}, \ and\
  \bibinfo {author} {\bibfnamefont {J.}~\bibnamefont {Verbist}},\ }\href
  {\doibase https://doi.org/10.1016/0009-2614(76)80329-6} {\bibfield  {journal}
  {\bibinfo  {journal} {Chemical Physics Letters}\ }\textbf {\bibinfo {volume}
  {39}},\ \bibinfo {pages} {560} (\bibinfo {year} {1976})}\BibitemShut
  {NoStop}%
\bibitem [{\citenamefont {Sayers}\ and\ \citenamefont
  {Armstrong}(1978)}]{SAYERS1978301}%
  \BibitemOpen
  \bibfield  {author} {\bibinfo {author} {\bibfnamefont {C.~N.}\ \bibnamefont
  {Sayers}}\ and\ \bibinfo {author} {\bibfnamefont {N.~R.}\ \bibnamefont
  {Armstrong}},\ }\href {\doibase https://doi.org/10.1016/0039-6028(78)90008-0}
  {\bibfield  {journal} {\bibinfo  {journal} {Surface Science}\ }\textbf
  {\bibinfo {volume} {77}},\ \bibinfo {pages} {301} (\bibinfo {year}
  {1978})}\BibitemShut {NoStop}%
\bibitem [{\citenamefont {Diebold}\ and\ \citenamefont
  {Madey}(1996)}]{Diebold_1996}%
  \BibitemOpen
  \bibfield  {author} {\bibinfo {author} {\bibfnamefont {U.}~\bibnamefont
  {Diebold}}\ and\ \bibinfo {author} {\bibfnamefont {T.~E.}\ \bibnamefont
  {Madey}},\ }\href {\doibase 10.1116/1.1247794} {\bibfield  {journal}
  {\bibinfo  {journal} {Surface Science Spectra}\ }\textbf {\bibinfo {volume}
  {4}},\ \bibinfo {pages} {227} (\bibinfo {year} {1996})}\BibitemShut {NoStop}%
\bibitem [{\citenamefont {Kurtz}\ and\ \citenamefont
  {Henrich}(1998)}]{Kurtz_1998}%
  \BibitemOpen
  \bibfield  {author} {\bibinfo {author} {\bibfnamefont {R.~L.}\ \bibnamefont
  {Kurtz}}\ and\ \bibinfo {author} {\bibfnamefont {V.~E.}\ \bibnamefont
  {Henrich}},\ }\href {\doibase 10.1116/1.1247874} {\bibfield  {journal}
  {\bibinfo  {journal} {Surface Science Spectra}\ }\textbf {\bibinfo {volume}
  {5}},\ \bibinfo {pages} {179} (\bibinfo {year} {1998})}\BibitemShut {NoStop}%
\bibitem [{\citenamefont {Bertóti}\ \emph {et~al.}(1995)\citenamefont
  {Bertóti}, \citenamefont {Mohai}, \citenamefont {Sullivan},\ and\
  \citenamefont {Saied}}]{BERTOTI1995357}%
  \BibitemOpen
  \bibfield  {author} {\bibinfo {author} {\bibfnamefont {I.}~\bibnamefont
  {Bertóti}}, \bibinfo {author} {\bibfnamefont {M.}~\bibnamefont {Mohai}},
  \bibinfo {author} {\bibfnamefont {J.}~\bibnamefont {Sullivan}}, \ and\
  \bibinfo {author} {\bibfnamefont {S.}~\bibnamefont {Saied}},\ }\href
  {\doibase https://doi.org/10.1016/0169-4332(94)00545-1} {\bibfield  {journal}
  {\bibinfo  {journal} {Applied Surface Science}\ }\textbf {\bibinfo {volume}
  {84}},\ \bibinfo {pages} {357} (\bibinfo {year} {1995})}\BibitemShut
  {NoStop}%
\bibitem [{\citenamefont {Kalha}\ \emph
  {et~al.}(2021{\natexlab{b}})\citenamefont {Kalha}, \citenamefont {Ratcliff},
  \citenamefont {Moreno}, \citenamefont {Mohr}, \citenamefont {Mantsinen},
  \citenamefont {Fernando}, \citenamefont {Thakur}, \citenamefont {Lee},
  \citenamefont {Tseng}, \citenamefont {Nunney}, \citenamefont {Kahk},
  \citenamefont {Lischner},\ and\ \citenamefont {Regoutz}}]{kalha2021lifetime}%
  \BibitemOpen
  \bibfield  {author} {\bibinfo {author} {\bibfnamefont {C.}~\bibnamefont
  {Kalha}}, \bibinfo {author} {\bibfnamefont {L.~E.}\ \bibnamefont {Ratcliff}},
  \bibinfo {author} {\bibfnamefont {J.~J.~G.}\ \bibnamefont {Moreno}}, \bibinfo
  {author} {\bibfnamefont {S.}~\bibnamefont {Mohr}}, \bibinfo {author}
  {\bibfnamefont {M.}~\bibnamefont {Mantsinen}}, \bibinfo {author}
  {\bibfnamefont {N.~K.}\ \bibnamefont {Fernando}}, \bibinfo {author}
  {\bibfnamefont {P.~K.}\ \bibnamefont {Thakur}}, \bibinfo {author}
  {\bibfnamefont {T.-L.}\ \bibnamefont {Lee}}, \bibinfo {author} {\bibfnamefont
  {H.-H.}\ \bibnamefont {Tseng}}, \bibinfo {author} {\bibfnamefont {T.~S.}\
  \bibnamefont {Nunney}}, \bibinfo {author} {\bibfnamefont {J.~M.}\
  \bibnamefont {Kahk}}, \bibinfo {author} {\bibfnamefont {J.}~\bibnamefont
  {Lischner}}, \ and\ \bibinfo {author} {\bibfnamefont {A.}~\bibnamefont
  {Regoutz}},\ }\href@noop {} {\bibfield  {journal} {\bibinfo  {journal}
  {arXiv}\ } (\bibinfo {year} {2021}{\natexlab{b}})},\ \Eprint
  {http://arxiv.org/abs/2109.04761} {2109.04761} \BibitemShut {NoStop}%
\bibitem [{\citenamefont {Barth}\ \emph {et~al.}(1985)\citenamefont {Barth},
  \citenamefont {Gerken},\ and\ \citenamefont {Kunz}}]{Barth_1985}%
  \BibitemOpen
  \bibfield  {author} {\bibinfo {author} {\bibfnamefont {J.}~\bibnamefont
  {Barth}}, \bibinfo {author} {\bibfnamefont {F.}~\bibnamefont {Gerken}}, \
  and\ \bibinfo {author} {\bibfnamefont {C.}~\bibnamefont {Kunz}},\ }\href
  {\doibase 10.1103/PhysRevB.31.2022} {\bibfield  {journal} {\bibinfo
  {journal} {Phys. Rev. B}\ }\textbf {\bibinfo {volume} {31}},\ \bibinfo
  {pages} {2022} (\bibinfo {year} {1985})}\BibitemShut {NoStop}%
\bibitem [{\citenamefont {Scofield}(1973)}]{Scofield1973TheoreticalKeV}%
  \BibitemOpen
  \bibfield  {author} {\bibinfo {author} {\bibfnamefont {J.~H.}\ \bibnamefont
  {Scofield}},\ }\href@noop {} {\emph {\bibinfo {title} {{Theoretical
  Photoionization Cross Sections from 1 to 1500 keV}}}},\ \bibinfo {type}
  {Tech. Rep.}\ (\bibinfo  {institution} {Lawrence Livermore Laboratory},\
  \bibinfo {year} {1973})\BibitemShut {NoStop}%
\bibitem [{\citenamefont {J~Jackson}\ \emph {et~al.}(2018)\citenamefont
  {J~Jackson}, \citenamefont {M~Ganose}, \citenamefont {Regoutz}, \citenamefont
  {G.~Egdell},\ and\ \citenamefont
  {O~Scanlon}}]{JJackson2018Galore:Spectroscopy}%
  \BibitemOpen
  \bibfield  {author} {\bibinfo {author} {\bibfnamefont {A.}~\bibnamefont
  {J~Jackson}}, \bibinfo {author} {\bibfnamefont {A.}~\bibnamefont {M~Ganose}},
  \bibinfo {author} {\bibfnamefont {A.}~\bibnamefont {Regoutz}}, \bibinfo
  {author} {\bibfnamefont {R.}~\bibnamefont {G.~Egdell}}, \ and\ \bibinfo
  {author} {\bibfnamefont {D.}~\bibnamefont {O~Scanlon}},\ }\href {\doibase
  10.21105/joss.00773} {\bibfield  {journal} {\bibinfo  {journal} {Journal of
  Open Source Software}\ }\textbf {\bibinfo {volume} {3}},\ \bibinfo {pages}
  {773} (\bibinfo {year} {2018})}\BibitemShut {NoStop}%
\bibitem [{\citenamefont {Kalha}\ \emph {et~al.}(2020)\citenamefont {Kalha},
  \citenamefont {Fernando},\ and\ \citenamefont {Regoutz}}]{Kalha20}%
  \BibitemOpen
  \bibfield  {author} {\bibinfo {author} {\bibfnamefont {C.}~\bibnamefont
  {Kalha}}, \bibinfo {author} {\bibfnamefont {N.}~\bibnamefont {Fernando}}, \
  and\ \bibinfo {author} {\bibfnamefont {A.}~\bibnamefont {Regoutz}},\ }\href
  {\doibase 10.6084/m9.figshare.12967079.v1} {\enquote {\bibinfo {title}
  {{Digitisation of Scofield Photoionisation Cross Section Tabulated Data}},}\
  } (\bibinfo {year} {2020})\BibitemShut {NoStop}%
\bibitem [{\citenamefont {Campbell}\ and\ \citenamefont
  {Papp}(2001)}]{Campbell2001WidthsLevels}%
  \BibitemOpen
  \bibfield  {author} {\bibinfo {author} {\bibfnamefont {J.~L.}\ \bibnamefont
  {Campbell}}\ and\ \bibinfo {author} {\bibfnamefont {T.}~\bibnamefont
  {Papp}},\ }\href {\doibase 10.1006/adnd.2000.0848} {\bibfield  {journal}
  {\bibinfo  {journal} {Atomic Data and Nuclear Data Tables}\ }\textbf
  {\bibinfo {volume} {77}},\ \bibinfo {pages} {1} (\bibinfo {year}
  {2001})}\BibitemShut {NoStop}%
\bibitem [{\citenamefont {Bergstrom}\ \emph {et~al.}(1997)\citenamefont
  {Bergstrom}, \citenamefont {Petrov},\ and\ \citenamefont
  {Greene}}]{Bergstrom_1997}%
  \BibitemOpen
  \bibfield  {author} {\bibinfo {author} {\bibfnamefont {D.~B.}\ \bibnamefont
  {Bergstrom}}, \bibinfo {author} {\bibfnamefont {I.}~\bibnamefont {Petrov}}, \
  and\ \bibinfo {author} {\bibfnamefont {J.~E.}\ \bibnamefont {Greene}},\
  }\href {\doibase 10.1063/1.366039} {\bibfield  {journal} {\bibinfo  {journal}
  {Journal of Applied Physics}\ }\textbf {\bibinfo {volume} {82}},\ \bibinfo
  {pages} {2312} (\bibinfo {year} {1997})}\BibitemShut {NoStop}%
\bibitem [{\citenamefont {Alay}\ \emph {et~al.}(1991)\citenamefont {Alay},
  \citenamefont {Bender}, \citenamefont {Brijs}, \citenamefont {Demesmaeker},\
  and\ \citenamefont {Vandervorst}}]{Alay1991}%
  \BibitemOpen
  \bibfield  {author} {\bibinfo {author} {\bibfnamefont {J.~L.}\ \bibnamefont
  {Alay}}, \bibinfo {author} {\bibfnamefont {H.}~\bibnamefont {Bender}},
  \bibinfo {author} {\bibfnamefont {G.}~\bibnamefont {Brijs}}, \bibinfo
  {author} {\bibfnamefont {A.}~\bibnamefont {Demesmaeker}}, \ and\ \bibinfo
  {author} {\bibfnamefont {W.}~\bibnamefont {Vandervorst}},\ }\href {\doibase
  10.1002/sia.740170613} {\bibfield  {journal} {\bibinfo  {journal} {Surface
  and Interface Analysis}\ }\textbf {\bibinfo {volume} {17}},\ \bibinfo {pages}
  {373} (\bibinfo {year} {1991})}\BibitemShut {NoStop}%
\bibitem [{\citenamefont {Kalha}\ \emph
  {et~al.}(2021{\natexlab{c}})\citenamefont {Kalha}, \citenamefont
  {Bichelmaier}, \citenamefont {Fernando}, \citenamefont {Berens},
  \citenamefont {Thakur}, \citenamefont {Lee}, \citenamefont
  {Gutiérrez~Moreno}, \citenamefont {Mohr}, \citenamefont {Ratcliff},
  \citenamefont {Reisinger}, \citenamefont {Zechner}, \citenamefont
  {Nelhiebel},\ and\ \citenamefont {Regoutz}}]{Kalha_2021_TiWO}%
  \BibitemOpen
  \bibfield  {author} {\bibinfo {author} {\bibfnamefont {C.}~\bibnamefont
  {Kalha}}, \bibinfo {author} {\bibfnamefont {S.}~\bibnamefont {Bichelmaier}},
  \bibinfo {author} {\bibfnamefont {N.~K.}\ \bibnamefont {Fernando}}, \bibinfo
  {author} {\bibfnamefont {J.~V.}\ \bibnamefont {Berens}}, \bibinfo {author}
  {\bibfnamefont {P.~K.}\ \bibnamefont {Thakur}}, \bibinfo {author}
  {\bibfnamefont {T.-L.}\ \bibnamefont {Lee}}, \bibinfo {author} {\bibfnamefont
  {J.~J.}\ \bibnamefont {Gutiérrez~Moreno}}, \bibinfo {author} {\bibfnamefont
  {S.}~\bibnamefont {Mohr}}, \bibinfo {author} {\bibfnamefont {L.~E.}\
  \bibnamefont {Ratcliff}}, \bibinfo {author} {\bibfnamefont {M.}~\bibnamefont
  {Reisinger}}, \bibinfo {author} {\bibfnamefont {J.}~\bibnamefont {Zechner}},
  \bibinfo {author} {\bibfnamefont {M.}~\bibnamefont {Nelhiebel}}, \ and\
  \bibinfo {author} {\bibfnamefont {A.}~\bibnamefont {Regoutz}},\ }\href
  {\doibase 10.1063/5.0048304} {\bibfield  {journal} {\bibinfo  {journal}
  {Journal of Applied Physics}\ }\textbf {\bibinfo {volume} {129}},\ \bibinfo
  {pages} {195302} (\bibinfo {year} {2021}{\natexlab{c}})}\BibitemShut
  {NoStop}%
\bibitem [{\citenamefont {Furlan}\ \emph {et~al.}(1991)\citenamefont {Furlan},
  \citenamefont {Van~der Spiegel},\ and\ \citenamefont {Swart}}]{Furlan1991}%
  \BibitemOpen
  \bibfield  {author} {\bibinfo {author} {\bibfnamefont {R.}~\bibnamefont
  {Furlan}}, \bibinfo {author} {\bibfnamefont {J.}~\bibnamefont {Van~der
  Spiegel}}, \ and\ \bibinfo {author} {\bibfnamefont {J.~W.}\ \bibnamefont
  {Swart}},\ }\href@noop {} {\bibfield  {journal} {\bibinfo  {journal} {Journal
  of the Electrochemical Society}\ }\textbf {\bibinfo {volume} {138}},\
  \bibinfo {pages} {2377} (\bibinfo {year} {1991})}\BibitemShut {NoStop}%
\bibitem [{\citenamefont {Barr}(1978)}]{Barr1978}%
  \BibitemOpen
  \bibfield  {author} {\bibinfo {author} {\bibfnamefont {T.~L.}\ \bibnamefont
  {Barr}},\ }\href {\doibase 10.1021/j100505a006} {\bibfield  {journal}
  {\bibinfo  {journal} {The Journal of Physical Chemistry}\ }\textbf {\bibinfo
  {volume} {82}},\ \bibinfo {pages} {1801} (\bibinfo {year}
  {1978})}\BibitemShut {NoStop}%
\bibitem [{\citenamefont {Engelhard}\ and\ \citenamefont
  {Baer}(2000)}]{Engelhard2000ThirdSpectroscopy}%
  \BibitemOpen
  \bibfield  {author} {\bibinfo {author} {\bibfnamefont {M.}~\bibnamefont
  {Engelhard}}\ and\ \bibinfo {author} {\bibfnamefont {D.}~\bibnamefont
  {Baer}},\ }\href {\doibase 10.1116/1.1311915} {\bibfield  {journal} {\bibinfo
   {journal} {Surface Science Spectra}\ }\textbf {\bibinfo {volume} {7}},\
  \bibinfo {pages} {1} (\bibinfo {year} {2000})}\BibitemShut {NoStop}%
\bibitem [{\citenamefont {Senkovskiy}\ \emph {et~al.}(2012)\citenamefont
  {Senkovskiy}, \citenamefont {Usachov}, \citenamefont {Fedorov}, \citenamefont
  {Vilkov}, \citenamefont {Shelyakov},\ and\ \citenamefont
  {Adamchuk}}]{SENKOVSKIY2012190}%
  \BibitemOpen
  \bibfield  {author} {\bibinfo {author} {\bibfnamefont {B.}~\bibnamefont
  {Senkovskiy}}, \bibinfo {author} {\bibfnamefont {D.}~\bibnamefont {Usachov}},
  \bibinfo {author} {\bibfnamefont {A.}~\bibnamefont {Fedorov}}, \bibinfo
  {author} {\bibfnamefont {O.}~\bibnamefont {Vilkov}}, \bibinfo {author}
  {\bibfnamefont {A.}~\bibnamefont {Shelyakov}}, \ and\ \bibinfo {author}
  {\bibfnamefont {V.}~\bibnamefont {Adamchuk}},\ }\href {\doibase
  https://doi.org/10.1016/j.jallcom.2012.05.059} {\bibfield  {journal}
  {\bibinfo  {journal} {Journal of Alloys and Compounds}\ }\textbf {\bibinfo
  {volume} {537}},\ \bibinfo {pages} {190} (\bibinfo {year}
  {2012})}\BibitemShut {NoStop}%
\bibitem [{\citenamefont {Cocke}\ \emph {et~al.}(1990)\citenamefont {Cocke},
  \citenamefont {Hess}, \citenamefont {Mebrahtu}, \citenamefont {Mencer},\ and\
  \citenamefont {Naugle}}]{COCKE1990119}%
  \BibitemOpen
  \bibfield  {author} {\bibinfo {author} {\bibfnamefont {D.}~\bibnamefont
  {Cocke}}, \bibinfo {author} {\bibfnamefont {T.}~\bibnamefont {Hess}},
  \bibinfo {author} {\bibfnamefont {T.}~\bibnamefont {Mebrahtu}}, \bibinfo
  {author} {\bibfnamefont {D.}~\bibnamefont {Mencer}}, \ and\ \bibinfo {author}
  {\bibfnamefont {D.}~\bibnamefont {Naugle}},\ }\href {\doibase
  https://doi.org/10.1016/0167-2738(90)90478-A} {\bibfield  {journal} {\bibinfo
   {journal} {Solid State Ionics}\ }\textbf {\bibinfo {volume} {43}},\ \bibinfo
  {pages} {119} (\bibinfo {year} {1990})}\BibitemShut {NoStop}%
\bibitem [{\citenamefont {Tanaka}\ \emph {et~al.}(1990)\citenamefont {Tanaka},
  \citenamefont {Ushida}, \citenamefont {Sumiyama},\ and\ \citenamefont
  {Nakamura}}]{TANAKA1990429}%
  \BibitemOpen
  \bibfield  {author} {\bibinfo {author} {\bibfnamefont {K.}~\bibnamefont
  {Tanaka}}, \bibinfo {author} {\bibfnamefont {M.}~\bibnamefont {Ushida}},
  \bibinfo {author} {\bibfnamefont {K.}~\bibnamefont {Sumiyama}}, \ and\
  \bibinfo {author} {\bibfnamefont {Y.}~\bibnamefont {Nakamura}},\ }\href@noop
  {} {\bibfield  {journal} {\bibinfo  {journal} {Journal of Non-Crystalline
  Solids}\ }\textbf {\bibinfo {volume} {117-118}},\ \bibinfo {pages} {429}
  (\bibinfo {year} {1990})},\ \bibinfo {note} {proceedings of the Seventh
  International Conference on Liquid and Amorphous Metals}\BibitemShut
  {NoStop}%
\bibitem [{\citenamefont {Dimitrov}\ and\ \citenamefont
  {Komatsu}(2002)}]{DIMITROV2002100}%
  \BibitemOpen
  \bibfield  {author} {\bibinfo {author} {\bibfnamefont {V.}~\bibnamefont
  {Dimitrov}}\ and\ \bibinfo {author} {\bibfnamefont {T.}~\bibnamefont
  {Komatsu}},\ }\href {\doibase https://doi.org/10.1006/jssc.2001.9378}
  {\bibfield  {journal} {\bibinfo  {journal} {Journal of Solid State
  Chemistry}\ }\textbf {\bibinfo {volume} {163}},\ \bibinfo {pages} {100}
  (\bibinfo {year} {2002})}\BibitemShut {NoStop}%
\bibitem [{\citenamefont {{Beggs}}(1956)}]{Beggs}%
  \BibitemOpen
  \bibfield  {author} {\bibinfo {author} {\bibfnamefont {J.~E.}\ \bibnamefont
  {{Beggs}}},\ }\href {\doibase 10.1109/T-ED.1956.14110} {\bibfield  {journal}
  {\bibinfo  {journal} {IRE Transactions on Electron Devices}\ }\textbf
  {\bibinfo {volume} {3}},\ \bibinfo {pages} {93} (\bibinfo {year}
  {1956})}\BibitemShut {NoStop}%
\bibitem [{\citenamefont {Liu}\ and\ \citenamefont {Welsch}(1988)}]{Liu1988}%
  \BibitemOpen
  \bibfield  {author} {\bibinfo {author} {\bibfnamefont {Z.}~\bibnamefont
  {Liu}}\ and\ \bibinfo {author} {\bibfnamefont {G.}~\bibnamefont {Welsch}},\
  }\href {\doibase 10.1007/BF02628396} {\bibfield  {journal} {\bibinfo
  {journal} {Metallurgical Transactions A}\ }\textbf {\bibinfo {volume} {19}},\
  \bibinfo {pages} {1121} (\bibinfo {year} {1988})}\BibitemShut {NoStop}%
\bibitem [{\citenamefont {Doniach}\ and\ \citenamefont
  {{\v{S}}unji{\'{c}}}(1969)}]{Doniach1969Many-electronMetals}%
  \BibitemOpen
  \bibfield  {author} {\bibinfo {author} {\bibfnamefont {S.}~\bibnamefont
  {Doniach}}\ and\ \bibinfo {author} {\bibfnamefont {M.}~\bibnamefont
  {{\v{S}}unji{\'{c}}}},\ }\href@noop {} {\bibfield  {journal} {\bibinfo
  {journal} {Journal of Physics C Solid State Physics}\ }\textbf {\bibinfo
  {volume} {3}},\ \bibinfo {pages} {285} (\bibinfo {year} {1969})}\BibitemShut
  {NoStop}%
\bibitem [{\citenamefont {Carley}\ \emph {et~al.}(1987)\citenamefont {Carley},
  \citenamefont {Chalker}, \citenamefont {Riviere},\ and\ \citenamefont
  {Roberts}}]{Carley1987}%
  \BibitemOpen
  \bibfield  {author} {\bibinfo {author} {\bibfnamefont {A.~F.}\ \bibnamefont
  {Carley}}, \bibinfo {author} {\bibfnamefont {P.~R.}\ \bibnamefont {Chalker}},
  \bibinfo {author} {\bibfnamefont {J.~C.}\ \bibnamefont {Riviere}}, \ and\
  \bibinfo {author} {\bibfnamefont {M.~W.}\ \bibnamefont {Roberts}},\ }\href
  {\doibase 10.1039/F19878300351} {\bibfield  {journal} {\bibinfo  {journal}
  {J. Chem. Soc.{,} Faraday Trans. 1}\ }\textbf {\bibinfo {volume} {83}},\
  \bibinfo {pages} {351} (\bibinfo {year} {1987})}\BibitemShut {NoStop}%
\bibitem [{\citenamefont {Puglia}\ \emph {et~al.}(1995)\citenamefont {Puglia},
  \citenamefont {Nilsson}, \citenamefont {Hernnäs}, \citenamefont {Karis},
  \citenamefont {Bennich},\ and\ \citenamefont {Mårtensson}}]{Puglia1995}%
  \BibitemOpen
  \bibfield  {author} {\bibinfo {author} {\bibfnamefont {C.}~\bibnamefont
  {Puglia}}, \bibinfo {author} {\bibfnamefont {A.}~\bibnamefont {Nilsson}},
  \bibinfo {author} {\bibfnamefont {B.}~\bibnamefont {Hernnäs}}, \bibinfo
  {author} {\bibfnamefont {O.}~\bibnamefont {Karis}}, \bibinfo {author}
  {\bibfnamefont {P.}~\bibnamefont {Bennich}}, \ and\ \bibinfo {author}
  {\bibfnamefont {N.}~\bibnamefont {Mårtensson}},\ }\href {\doibase
  https://doi.org/10.1016/0039-6028(95)00798-9} {\bibfield  {journal} {\bibinfo
   {journal} {Surface Science}\ }\textbf {\bibinfo {volume} {342}},\ \bibinfo
  {pages} {119 } (\bibinfo {year} {1995})}\BibitemShut {NoStop}%
\bibitem [{\citenamefont {Kuznetsov}\ \emph {et~al.}(1992)\citenamefont
  {Kuznetsov}, \citenamefont {Zhuravlev},\ and\ \citenamefont
  {Gubanov}}]{Kuznetsov1992}%
  \BibitemOpen
  \bibfield  {author} {\bibinfo {author} {\bibfnamefont {M.}~\bibnamefont
  {Kuznetsov}}, \bibinfo {author} {\bibfnamefont {J.}~\bibnamefont
  {Zhuravlev}}, \ and\ \bibinfo {author} {\bibfnamefont {V.}~\bibnamefont
  {Gubanov}},\ }\href {\doibase https://doi.org/10.1016/0368-2048(92)80016-2}
  {\bibfield  {journal} {\bibinfo  {journal} {Journal of Electron Spectroscopy
  and Related Phenomena}\ }\textbf {\bibinfo {volume} {58}},\ \bibinfo {pages}
  {169 } (\bibinfo {year} {1992})}\BibitemShut {NoStop}%
\bibitem [{\citenamefont {H\"ufner}\ \emph {et~al.}(1973)\citenamefont
  {H\"ufner}, \citenamefont {Wertheim},\ and\ \citenamefont
  {Wernick}}]{Hufner_1973}%
  \BibitemOpen
  \bibfield  {author} {\bibinfo {author} {\bibfnamefont {S.}~\bibnamefont
  {H\"ufner}}, \bibinfo {author} {\bibfnamefont {G.~K.}\ \bibnamefont
  {Wertheim}}, \ and\ \bibinfo {author} {\bibfnamefont {J.~H.}\ \bibnamefont
  {Wernick}},\ }\href {\doibase 10.1103/PhysRevB.8.4511} {\bibfield  {journal}
  {\bibinfo  {journal} {Phys. Rev. B}\ }\textbf {\bibinfo {volume} {8}},\
  \bibinfo {pages} {4511} (\bibinfo {year} {1973})}\BibitemShut {NoStop}%
\bibitem [{\citenamefont {Miller}\ and\ \citenamefont
  {Simmons}(1993{\natexlab{b}})}]{Copper_VB_1993}%
  \BibitemOpen
  \bibfield  {author} {\bibinfo {author} {\bibfnamefont {A.~C.}\ \bibnamefont
  {Miller}}\ and\ \bibinfo {author} {\bibfnamefont {G.~W.}\ \bibnamefont
  {Simmons}},\ }\href {\doibase 10.1116/1.1247725} {\bibfield  {journal}
  {\bibinfo  {journal} {Surface Science Spectra}\ }\textbf {\bibinfo {volume}
  {2}},\ \bibinfo {pages} {55} (\bibinfo {year}
  {1993}{\natexlab{b}})}\BibitemShut {NoStop}%
\bibitem [{\citenamefont {Liotard}\ \emph {et~al.}(1985)\citenamefont
  {Liotard}, \citenamefont {Gupta}, \citenamefont {Psaras},\ and\ \citenamefont
  {Ho}}]{Liotard_1985}%
  \BibitemOpen
  \bibfield  {author} {\bibinfo {author} {\bibfnamefont {J.~L.}\ \bibnamefont
  {Liotard}}, \bibinfo {author} {\bibfnamefont {D.}~\bibnamefont {Gupta}},
  \bibinfo {author} {\bibfnamefont {P.~A.}\ \bibnamefont {Psaras}}, \ and\
  \bibinfo {author} {\bibfnamefont {P.~S.}\ \bibnamefont {Ho}},\ }\href
  {\doibase 10.1063/1.334422} {\bibfield  {journal} {\bibinfo  {journal}
  {Journal of Applied Physics}\ }\textbf {\bibinfo {volume} {57}},\ \bibinfo
  {pages} {1895} (\bibinfo {year} {1985})}\BibitemShut {NoStop}%
\bibitem [{\citenamefont {Li}\ \emph {et~al.}(1992)\citenamefont {Li},
  \citenamefont {Strane}, \citenamefont {Russell}, \citenamefont {Hong},
  \citenamefont {Mayer}, \citenamefont {Marais}, \citenamefont {Theron},\ and\
  \citenamefont {Pretorius}}]{Jian_1992}%
  \BibitemOpen
  \bibfield  {author} {\bibinfo {author} {\bibfnamefont {J.}~\bibnamefont
  {Li}}, \bibinfo {author} {\bibfnamefont {J.~W.}\ \bibnamefont {Strane}},
  \bibinfo {author} {\bibfnamefont {S.~W.}\ \bibnamefont {Russell}}, \bibinfo
  {author} {\bibfnamefont {S.~Q.}\ \bibnamefont {Hong}}, \bibinfo {author}
  {\bibfnamefont {J.~W.}\ \bibnamefont {Mayer}}, \bibinfo {author}
  {\bibfnamefont {T.~K.}\ \bibnamefont {Marais}}, \bibinfo {author}
  {\bibfnamefont {C.~C.}\ \bibnamefont {Theron}}, \ and\ \bibinfo {author}
  {\bibfnamefont {R.}~\bibnamefont {Pretorius}},\ }\href {\doibase
  10.1063/1.351533} {\bibfield  {journal} {\bibinfo  {journal} {Journal of
  Applied Physics}\ }\textbf {\bibinfo {volume} {72}},\ \bibinfo {pages} {2810}
  (\bibinfo {year} {1992})}\BibitemShut {NoStop}%
\bibitem [{\citenamefont {Takeyama}\ \emph {et~al.}(1996)\citenamefont
  {Takeyama}, \citenamefont {Noya}, \citenamefont {Sakanishi}, \citenamefont
  {Seki},\ and\ \citenamefont {Sasaki}}]{Takeyama_1996}%
  \BibitemOpen
  \bibfield  {author} {\bibinfo {author} {\bibfnamefont {M.}~\bibnamefont
  {Takeyama}}, \bibinfo {author} {\bibfnamefont {A.}~\bibnamefont {Noya}},
  \bibinfo {author} {\bibfnamefont {K.}~\bibnamefont {Sakanishi}}, \bibinfo
  {author} {\bibfnamefont {H.}~\bibnamefont {Seki}}, \ and\ \bibinfo {author}
  {\bibfnamefont {K.}~\bibnamefont {Sasaki}},\ }\href {\doibase
  10.1143/jjap.35.4027} {\bibfield  {journal} {\bibinfo  {journal} {Japanese
  Journal of Applied Physics}\ }\textbf {\bibinfo {volume} {35}},\ \bibinfo
  {pages} {4027} (\bibinfo {year} {1996})}\BibitemShut {NoStop}%
\bibitem [{\citenamefont {{Ramarotafika}}\ and\ \citenamefont
  {{Lemperiere}}(1997)}]{Ramarotafika}%
  \BibitemOpen
  \bibfield  {author} {\bibinfo {author} {\bibfnamefont {H.}~\bibnamefont
  {{Ramarotafika}}}\ and\ \bibinfo {author} {\bibfnamefont {G.}~\bibnamefont
  {{Lemperiere}}},\ }in\ \href {\doibase 10.1109/MAM.1997.621082} {\emph
  {\bibinfo {booktitle} {European Workshop Materials for Advanced
  Metallization,}}}\ (\bibinfo {year} {1997})\ pp.\ \bibinfo {pages}
  {122--123}\BibitemShut {NoStop}%
\bibitem [{\citenamefont {Murray}(1983)}]{Murray1983}%
  \BibitemOpen
  \bibfield  {author} {\bibinfo {author} {\bibfnamefont {J.~L.}\ \bibnamefont
  {Murray}},\ }\href {\doibase 10.1007/BF02880329} {\bibfield  {journal}
  {\bibinfo  {journal} {Bulletin of Alloy Phase Diagrams}\ }\textbf {\bibinfo
  {volume} {4}},\ \bibinfo {pages} {81} (\bibinfo {year} {1983})}\BibitemShut
  {NoStop}%
\bibitem [{\citenamefont {Okamoto}(2002)}]{Okamoto2002}%
  \BibitemOpen
  \bibfield  {author} {\bibinfo {author} {\bibfnamefont {H.}~\bibnamefont
  {Okamoto}},\ }\href {\doibase 10.1361/105497102770331307} {\bibfield
  {journal} {\bibinfo  {journal} {Journal of Phase Equilibria}\ }\textbf
  {\bibinfo {volume} {23}},\ \bibinfo {pages} {549} (\bibinfo {year}
  {2002})}\BibitemShut {NoStop}%
\bibitem [{\citenamefont {Vüllers}\ and\ \citenamefont
  {Spolenak}(2015)}]{VULLERS2015213}%
  \BibitemOpen
  \bibfield  {author} {\bibinfo {author} {\bibfnamefont {F.}~\bibnamefont
  {Vüllers}}\ and\ \bibinfo {author} {\bibfnamefont {R.}~\bibnamefont
  {Spolenak}},\ }\href {\doibase https://doi.org/10.1016/j.actamat.2015.07.050}
  {\bibfield  {journal} {\bibinfo  {journal} {Acta Materialia}\ }\textbf
  {\bibinfo {volume} {99}},\ \bibinfo {pages} {213} (\bibinfo {year}
  {2015})}\BibitemShut {NoStop}%
\bibitem [{\citenamefont {Fujita}\ \emph {et~al.}(2013)\citenamefont {Fujita},
  \citenamefont {Ando}, \citenamefont {Uchikoshi}, \citenamefont {Mimura},\
  and\ \citenamefont {Isshiki}}]{FUJITA2013347}%
  \BibitemOpen
  \bibfield  {author} {\bibinfo {author} {\bibfnamefont {K.}~\bibnamefont
  {Fujita}}, \bibinfo {author} {\bibfnamefont {D.}~\bibnamefont {Ando}},
  \bibinfo {author} {\bibfnamefont {M.}~\bibnamefont {Uchikoshi}}, \bibinfo
  {author} {\bibfnamefont {K.}~\bibnamefont {Mimura}}, \ and\ \bibinfo {author}
  {\bibfnamefont {M.}~\bibnamefont {Isshiki}},\ }\href {\doibase
  https://doi.org/10.1016/j.apsusc.2013.03.096} {\bibfield  {journal} {\bibinfo
   {journal} {Applied Surface Science}\ }\textbf {\bibinfo {volume} {276}},\
  \bibinfo {pages} {347} (\bibinfo {year} {2013})}\BibitemShut {NoStop}%
\bibitem [{\citenamefont {Diaz~Leon}\ \emph {et~al.}(2016)\citenamefont
  {Diaz~Leon}, \citenamefont {Fryauf}, \citenamefont {Cormia}, \citenamefont
  {Zhang}, \citenamefont {Samuels}, \citenamefont {Williams},\ and\
  \citenamefont {Kobayashi}}]{Leon_2016}%
  \BibitemOpen
  \bibfield  {author} {\bibinfo {author} {\bibfnamefont {J.~J.}\ \bibnamefont
  {Diaz~Leon}}, \bibinfo {author} {\bibfnamefont {D.~M.}\ \bibnamefont
  {Fryauf}}, \bibinfo {author} {\bibfnamefont {R.~D.}\ \bibnamefont {Cormia}},
  \bibinfo {author} {\bibfnamefont {M.-X.~M.}\ \bibnamefont {Zhang}}, \bibinfo
  {author} {\bibfnamefont {K.}~\bibnamefont {Samuels}}, \bibinfo {author}
  {\bibfnamefont {R.~S.}\ \bibnamefont {Williams}}, \ and\ \bibinfo {author}
  {\bibfnamefont {N.~P.}\ \bibnamefont {Kobayashi}},\ }\href {\doibase
  10.1021/acsami.6b06626} {\bibfield  {journal} {\bibinfo  {journal} {ACS
  Applied Materials \& Interfaces}\ }\textbf {\bibinfo {volume} {8}},\ \bibinfo
  {pages} {22337} (\bibinfo {year} {2016})}\BibitemShut {NoStop}%
\bibitem [{\citenamefont {Rhodin}(1950)}]{Rhodin_1950}%
  \BibitemOpen
  \bibfield  {author} {\bibinfo {author} {\bibfnamefont {T.~N.}\ \bibnamefont
  {Rhodin}},\ }\href {\doibase 10.1021/ja01167a079} {\bibfield  {journal}
  {\bibinfo  {journal} {Journal of the American Chemical Society}\ }\textbf
  {\bibinfo {volume} {72}},\ \bibinfo {pages} {5102} (\bibinfo {year}
  {1950})}\BibitemShut {NoStop}%
\bibitem [{\citenamefont {Stiedl}\ \emph {et~al.}(2019)\citenamefont {Stiedl},
  \citenamefont {Green}, \citenamefont {Chassé},\ and\ \citenamefont
  {Rebner}}]{Stiedl_2019}%
  \BibitemOpen
  \bibfield  {author} {\bibinfo {author} {\bibfnamefont {J.}~\bibnamefont
  {Stiedl}}, \bibinfo {author} {\bibfnamefont {S.}~\bibnamefont {Green}},
  \bibinfo {author} {\bibfnamefont {T.}~\bibnamefont {Chassé}}, \ and\
  \bibinfo {author} {\bibfnamefont {K.}~\bibnamefont {Rebner}},\ }\href
  {\doibase 10.1177/0003702818797959} {\bibfield  {journal} {\bibinfo
  {journal} {Applied Spectroscopy}\ }\textbf {\bibinfo {volume} {73}},\
  \bibinfo {pages} {59} (\bibinfo {year} {2019})}\BibitemShut {NoStop}%
\bibitem [{\citenamefont {Oparowski}\ \emph {et~al.}(1987)\citenamefont
  {Oparowski}, \citenamefont {Sisson},\ and\ \citenamefont
  {Biederman}}]{OPAROWSKI1987313}%
  \BibitemOpen
  \bibfield  {author} {\bibinfo {author} {\bibfnamefont {J.}~\bibnamefont
  {Oparowski}}, \bibinfo {author} {\bibfnamefont {R.}~\bibnamefont {Sisson}}, \
  and\ \bibinfo {author} {\bibfnamefont {R.}~\bibnamefont {Biederman}},\ }\href
  {\doibase https://doi.org/10.1016/0040-6090(87)90192-1} {\bibfield  {journal}
  {\bibinfo  {journal} {Thin Solid Films}\ }\textbf {\bibinfo {volume} {153}},\
  \bibinfo {pages} {313} (\bibinfo {year} {1987})}\BibitemShut {NoStop}%
\end{thebibliography}%


\begin{thebibliography}{2}%
\makeatletter
\providecommand \@ifxundefined [1]{%
 \@ifx{#1\undefined}
}%
\providecommand \@ifnum [1]{%
 \ifnum #1\expandafter \@firstoftwo
 \else \expandafter \@secondoftwo
 \fi
}%
\providecommand \@ifx [1]{%
 \ifx #1\expandafter \@firstoftwo
 \else \expandafter \@secondoftwo
 \fi
}%
\providecommand \natexlab [1]{#1}%
\providecommand \enquote  [1]{``#1''}%
\providecommand \bibnamefont  [1]{#1}%
\providecommand \bibfnamefont [1]{#1}%
\providecommand \citenamefont [1]{#1}%
\providecommand \href@noop [0]{\@secondoftwo}%
\providecommand \href [0]{\begingroup \@sanitize@url \@href}%
\providecommand \@href[1]{\@@startlink{#1}\@@href}%
\providecommand \@@href[1]{\endgroup#1\@@endlink}%
\providecommand \@sanitize@url [0]{\catcode `\\12\catcode `\$12\catcode
  `\&12\catcode `\#12\catcode `\^12\catcode `\_12\catcode `\%12\relax}%
\providecommand \@@startlink[1]{}%
\providecommand \@@endlink[0]{}%
\providecommand \url  [0]{\begingroup\@sanitize@url \@url }%
\providecommand \@url [1]{\endgroup\@href {#1}{\urlprefix }}%
\providecommand \urlprefix  [0]{URL }%
\providecommand \Eprint [0]{\href }%
\providecommand \doibase [0]{http://dx.doi.org/}%
\providecommand \selectlanguage [0]{\@gobble}%
\providecommand \bibinfo  [0]{\@secondoftwo}%
\providecommand \bibfield  [0]{\@secondoftwo}%
\providecommand \translation [1]{[#1]}%
\providecommand \BibitemOpen [0]{}%
\providecommand \bibitemStop [0]{}%
\providecommand \bibitemNoStop [0]{.\EOS\space}%
\providecommand \EOS [0]{\spacefactor3000\relax}%
\providecommand \BibitemShut  [1]{\csname bibitem#1\endcsname}%
\let\auto@bib@innerbib\@empty
\bibitem [{\citenamefont {Scofield}(1973)}]{Scofield1973TheoreticalKeV}%
  \BibitemOpen
  \bibfield  {author} {\bibinfo {author} {\bibfnamefont {J.~H.}\ \bibnamefont
  {Scofield}},\ }\href@noop {} {\emph {\bibinfo {title} {{Theoretical
  Photoionization Cross Sections from 1 to 1500 keV}}}},\ \bibinfo {type}
  {Tech. Rep.}\ (\bibinfo  {institution} {Lawrence Livermore Laboratory},\
  \bibinfo {year} {1973})\BibitemShut {NoStop}%
\bibitem [{\citenamefont {Kalha}\ \emph {et~al.}(2020)\citenamefont {Kalha},
  \citenamefont {Fernando},\ and\ \citenamefont {Regoutz}}]{Kalha20}%
  \BibitemOpen
  \bibfield  {author} {\bibinfo {author} {\bibfnamefont {C.}~\bibnamefont
  {Kalha}}, \bibinfo {author} {\bibfnamefont {N.}~\bibnamefont {Fernando}}, \
  and\ \bibinfo {author} {\bibfnamefont {A.}~\bibnamefont {Regoutz}},\ }\href
  {\doibase 10.6084/m9.figshare.12967079.v1} {\enquote {\bibinfo {title}
  {{Digitisation of Scofield Photoionisation Cross Section Tabulated Data}},}\
  } (\bibinfo {year} {2020})\BibitemShut {NoStop}%
\end{thebibliography}%
\bibliographystyle{apsrev4-1.bst}

\end{document}


\title{Evaluation of the Thermal Stability of TiW/Cu Heterojunctions Using a Combined SXPS and HAXPES Approach: Supplementary Information}

\author{C.~Kalha}
\affiliation{Department of Chemistry, University College London, 20 Gordon Street, London, WC1H~0AJ, United Kingdom.}

\author{M.~Reisinger}
\affiliation{Kompetenzzentrum Automobil- und Industrie-Elektronik GmbH, Europastraße 8, 9524 Villach, Austria.}

\author{P.~K.~Thakur}%
\author{T.~-L.~Lee}%
\affiliation{Diamond Light Source Ltd., Harwell Science and Innovation Campus, Didcot, Oxfordshire, OX1 3QR, United Kingdom.}

\author{S. Venkatesan}
\affiliation{Infineon Technologies AG, Am Campeon 1-15, 85579 Neubiberg, Germany.}

\author{M.~Isaacs}%
\affiliation{Harwell XPS, Research Complex at Harwell (RCaH), Didcot, OX11 0FA}
\affiliation{Department of Chemistry, University College London, 20 Gordon Street, London, WC1H~0AJ, United Kingdom.}

\author{R.~G.~Palgrave}
\affiliation{Department of Chemistry, University College London, 20 Gordon Street, London, WC1H~0AJ, United Kingdom.}

\author{J.~Zechner}
\author{M.~Nelhiebel}
\affiliation{Kompetenzzentrum Automobil- und Industrie-Elektronik GmbH, Europastraße 8, 9524 Villach, Austria.}

\author{A.~Regoutz}
 \email{a.regoutz@ucl.ac.uk}
\affiliation{Department of Chemistry, University College London, 20 Gordon Street, London, WC1H~0AJ, United Kingdom.}

\date{\today}

\maketitle

\newpage

\tableofcontents

\cleardoublepage

\section{\label{sec:HRTEM}As-deposited TEM images}

\begin{figure}[H]
\centering
    \includegraphics[keepaspectratio, width= 0.6\textwidth]{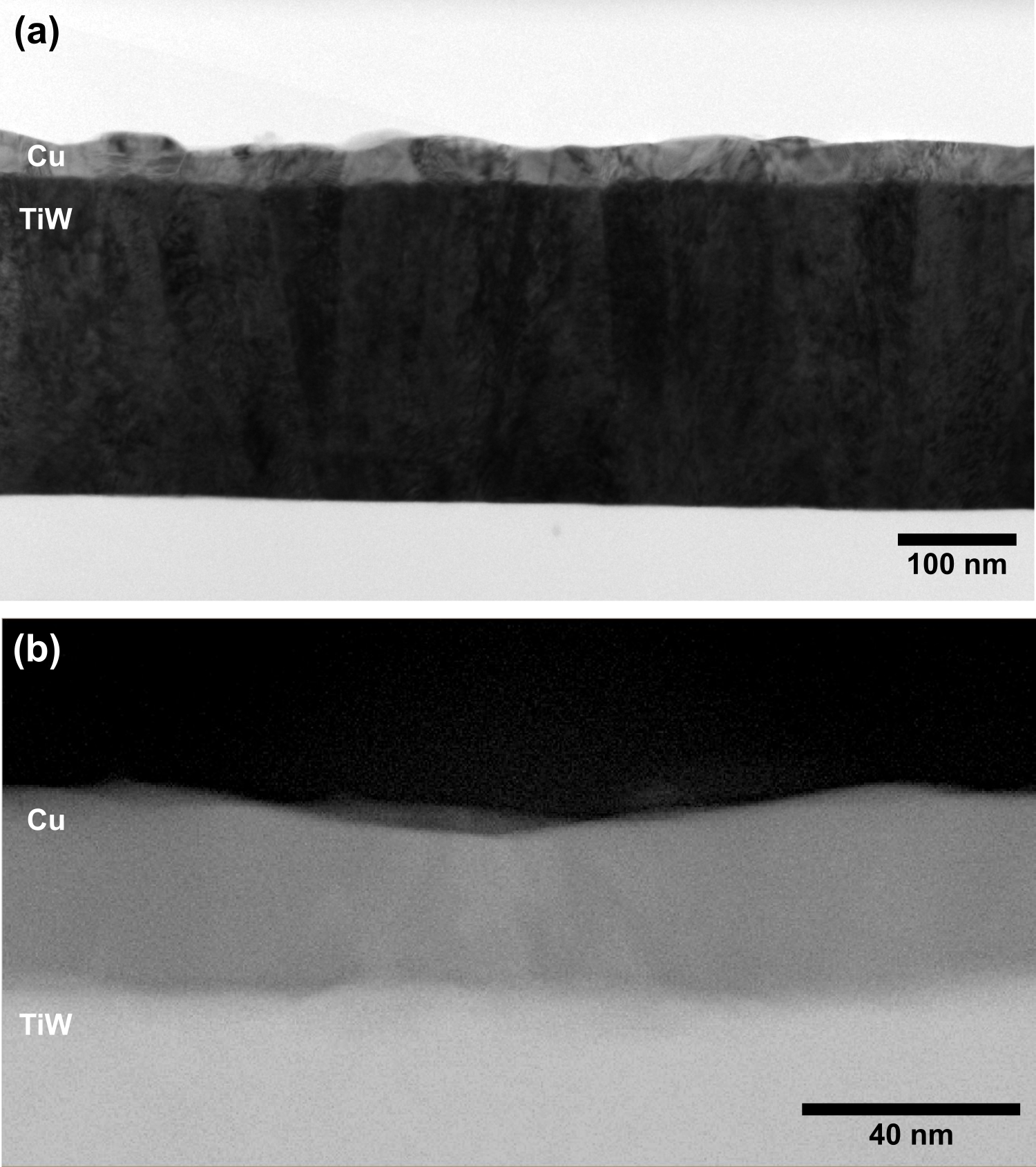}
    \caption{TEM images of the AD sample, including (a) a bright-field TEM image and (b) HAADF image.}
    \label{fig:HRTEM}
\end{figure}

Fig.~\ref{fig:HRTEM} displays high-resolution TEM cross-sectional images collected on an as-deposited (AD) Si/SiO\textsubscript{2}/TiW(300~nm)/Cu(25~nm) sample.

\newpage

    \section{\label{sec:Survey_Spectra_Surface}Survey Spectra Collected at Copper Surface}

\begin{figure}[H]
\centering
    \includegraphics[keepaspectratio, width= 0.8\textwidth]{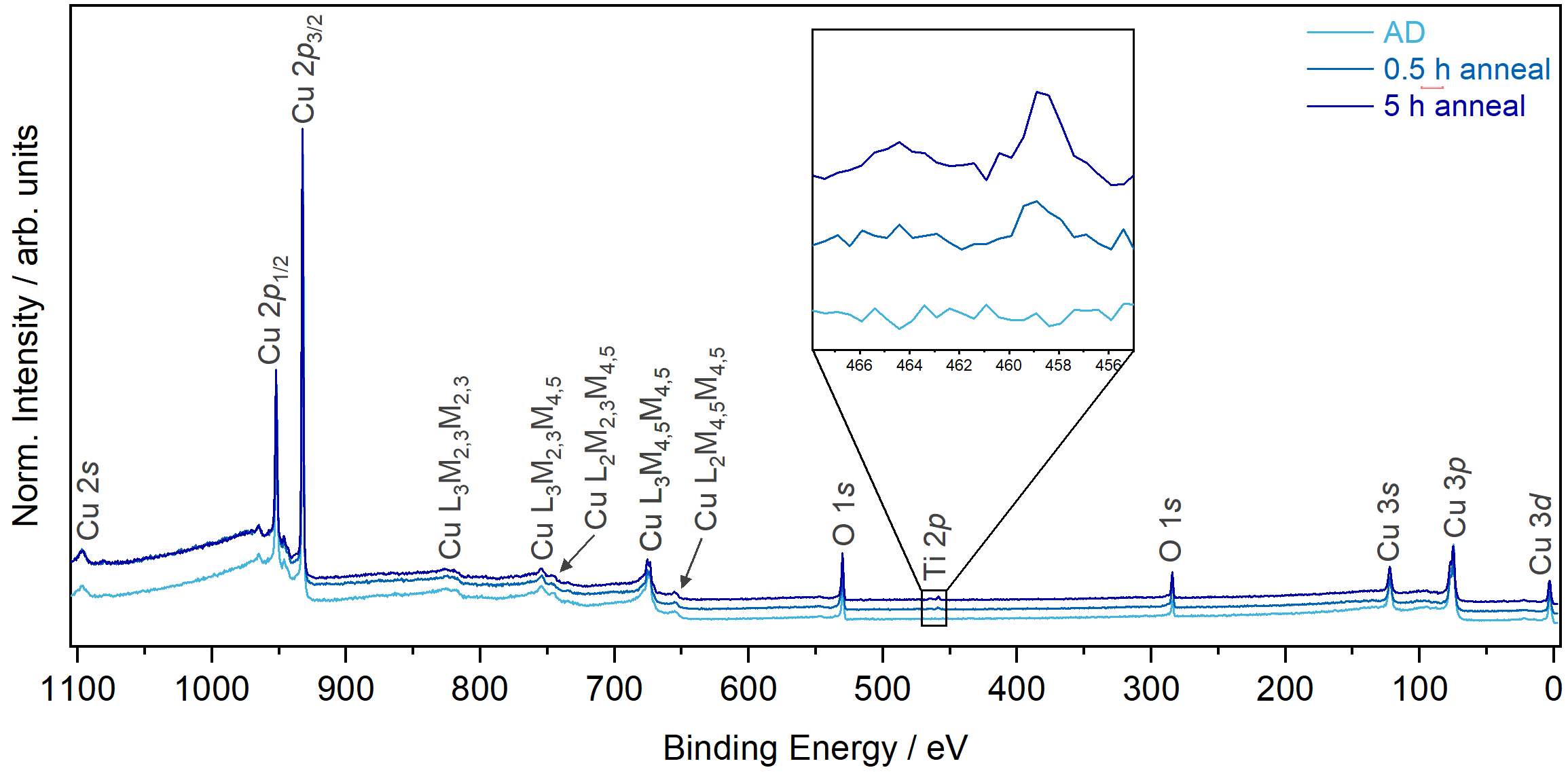}
    \caption{Synchrotron-based SXPS survey spectra collected on the as-received samples (i.e.\ no etch). Spectra are vertically offset and normalised to the maximum intensity peak to aid with comparison.}
    \label{fig:Cu_Survey}
\end{figure}

Fig.~\ref{fig:Cu_Survey} displays the as-received sample survey spectra collected using synchrotron-based SXPS (\textit{h$\nu$} = 1.6~keV). All observed core and Auger lines are indicated.

\newpage

\section{\label{sec:EDS}Complete EDS maps}

\begin{figure}[ht!]
\centering
    \includegraphics[keepaspectratio, width= 0.7\textwidth]{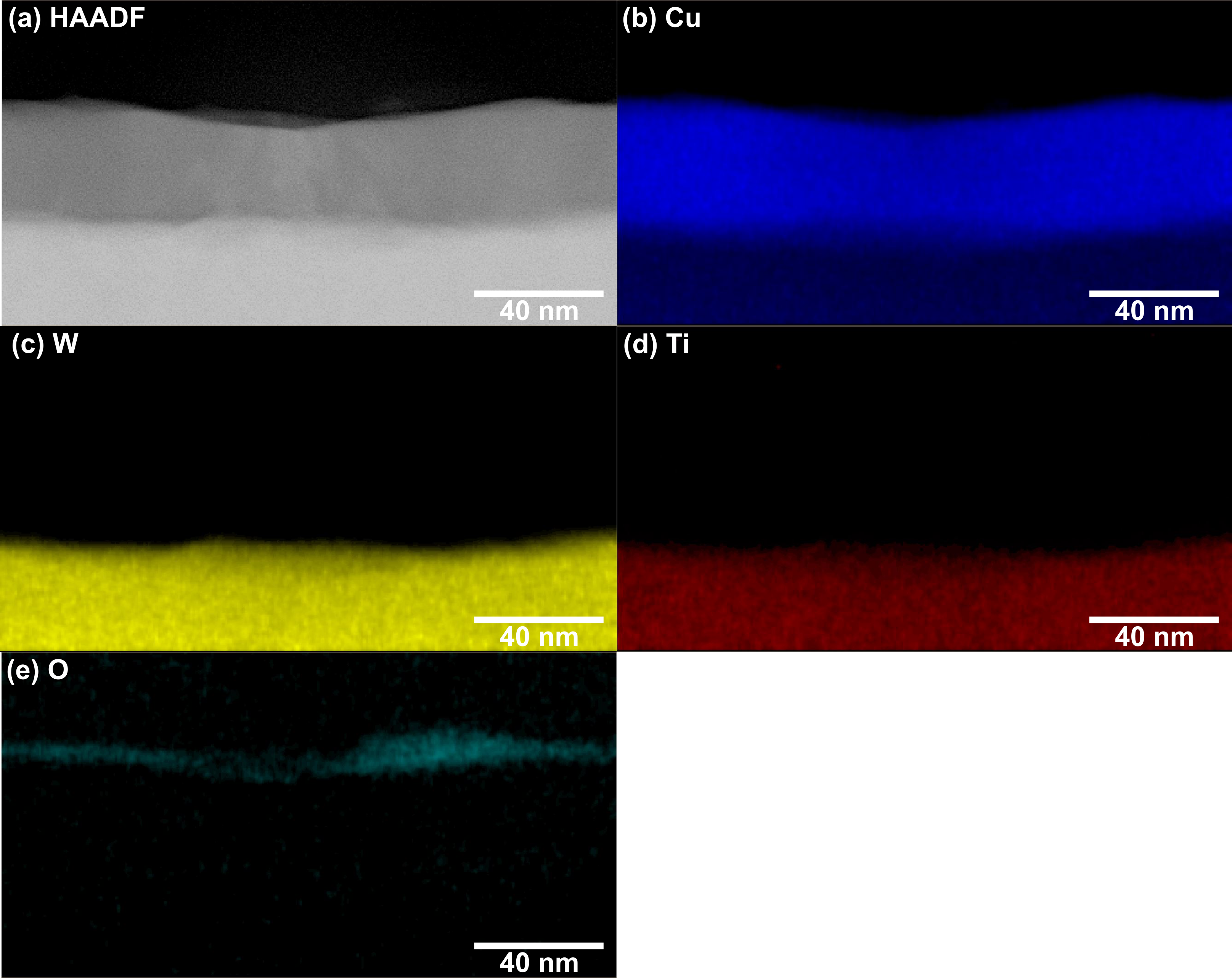}
    \caption{EDS elemental maps taken on a cross-sectional image of the AD sample, including a (a) HAADF image, and (b) Cu, (c) W, (d) Ti, and (e) O EDS maps.}
    \label{fig:EDS_AD}
\end{figure}

\begin{figure}[ht!]
\centering
    \includegraphics[keepaspectratio, width= 0.8\textwidth]{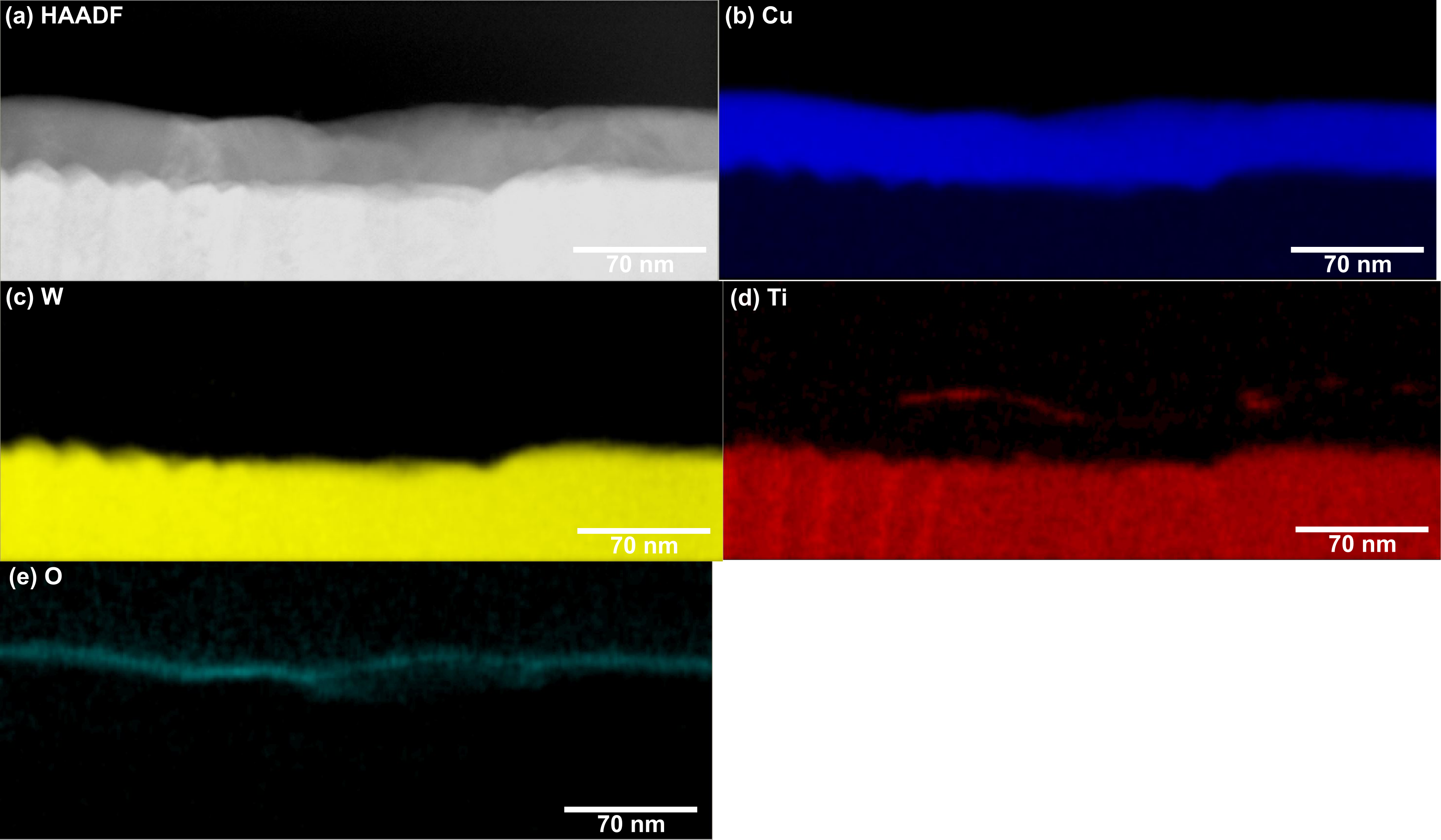}
    \caption{EDS elemental maps taken on a cross-sectional image of the 5~h annealed sample, including include a (a) HAADF image, and (b) Cu, (c) W, (d) Ti, and (e) O EDS maps.}
    \label{fig:EDS_5h}
\end{figure}

Fig.~\ref{fig:EDS_AD} and Fig~\ref{fig:EDS_5h} display the complete set of EDS maps taken on cross sections of the AD and 5~h anneal samples, respectively.

\newpage

\section{\label{PF}Peak Fit Analysis of Depth Profile Data}

\begin{figure}[H]
\centering
    \includegraphics[keepaspectratio, width=0.8\textwidth]{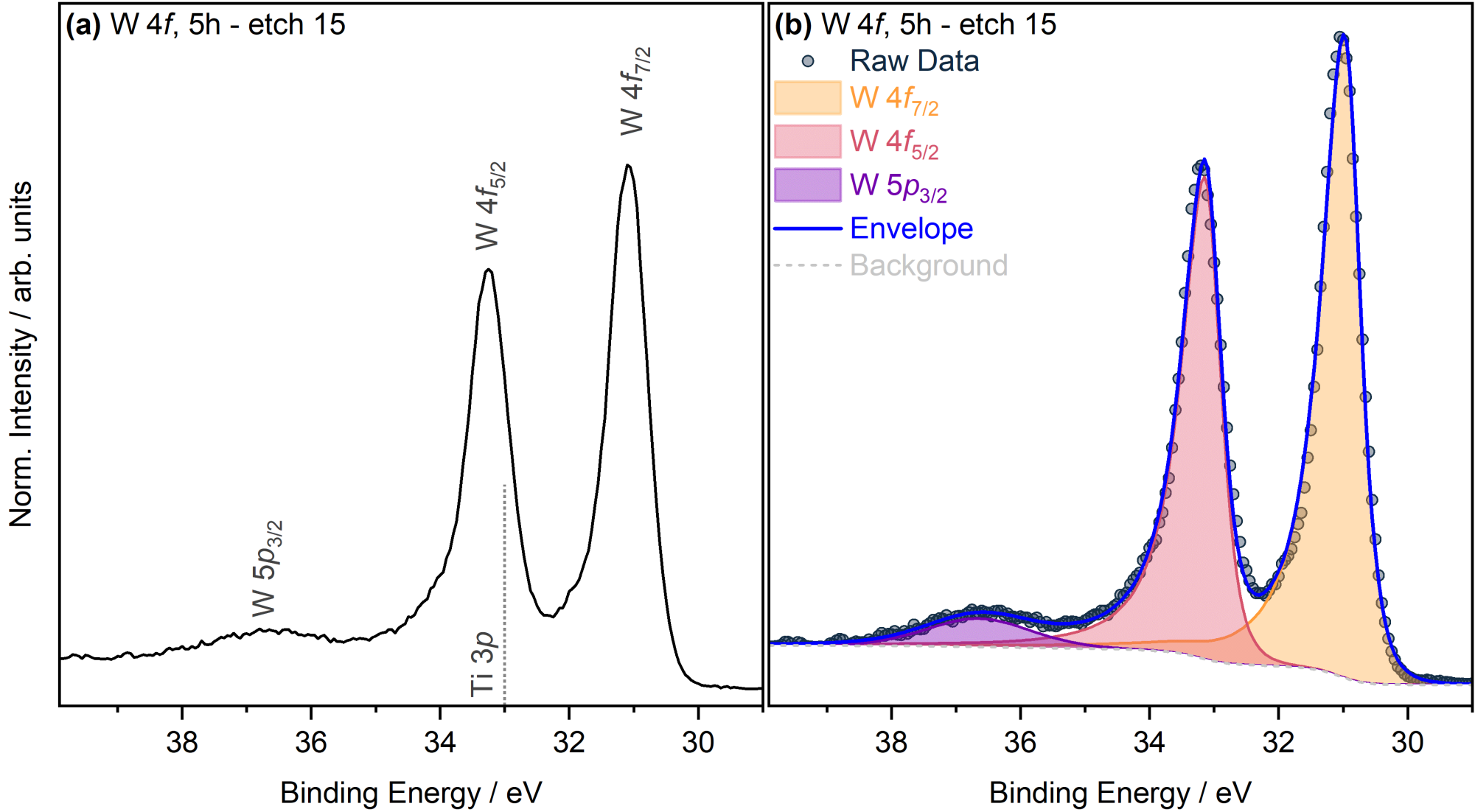}
    \caption{Example of the W~4\textit{f} core level peak fit, including (a) the raw spectrum obtained on the the last etch cycle of the depth profile for the 5~h annealed sample and (b) the resultant peak fit.}
    \label{fig:Quant}
\end{figure}

Peak fit analysis of the depth profile data was conducted in CasaXPS, implementing a Shirley-type background and applying the RASFs taken from the Kratos-modified Scofield photoionisation cross section database (values used are listed in Table~\ref{Kratos}). When fitting the W~4\textit{f} core level region, the asymmetric LA function was used for the 4\textit{f} doublet peaks and a GL(30) function for the 5\textit{p}\textsubscript{3/2} peak. An example of this fit is displayed in Fig.~\ref{fig:Quant}. Constraining the 4\textit{f} doublet to have the same line shape and FWHM resulted in a good fit with the area ratio determined to be 0.75, matching the expected degeneracy ratio. This suggests that the Ti~3\textit{p} intensity is too small and can be omitted from the peak fit analysis. If it was present in any significant capacity one would expect the 4\textit{f} area ratio to deviate from 0.75 and/or the use of two peaks would lead to a poor fit.\par

\begin{table}[h!]
    \caption{\label{Kratos}F~1\textit{s} Kratos library relative atomic sensitivity factors}
    \begin{tabular}{cc}
    \hline
    Core level & RASF  \\
    \hline \hline
    Cu~2\textit{p}\textsubscript{3/2} & 3.547 \\
    O~1\textit{s} & 0.78 \\
    Ti~2\textit{p} & 2.00 \\
    C~1\textit{s} & 0.278 \\
    W~4\textit{d} & 4.42 \\
    W~4\textit{f} & 3.52 \\
    \hline
    \end{tabular}
\end{table}

\begin{figure}[H]
\centering
    \includegraphics[keepaspectratio, width=0.7\textwidth]{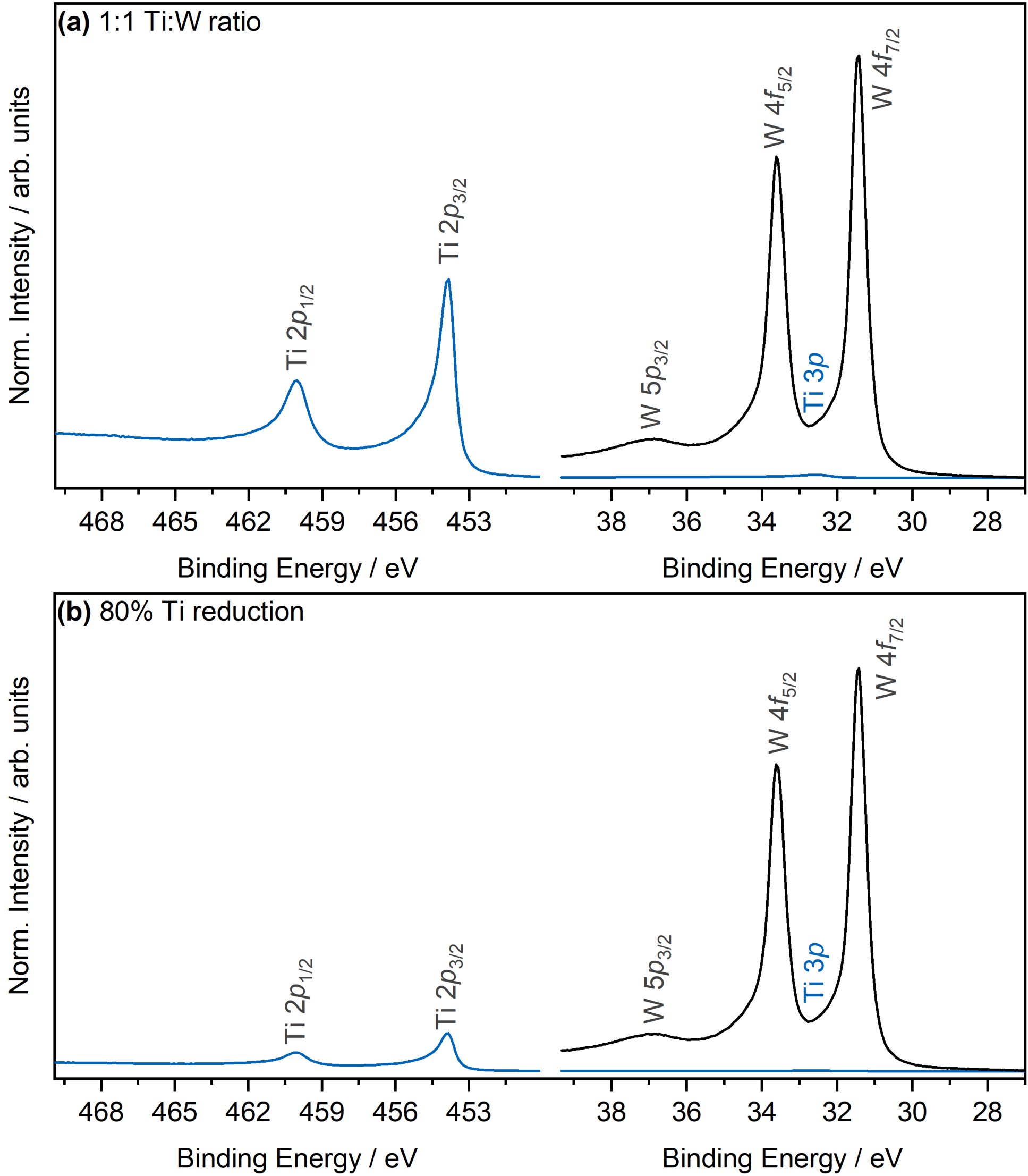}
    \caption{Ti~2\textit{p}, Ti~3\textit{p} and W~4\textit{f} reference spectra collected on Ti and W polycrystalline metal foils, measured after in-situ sputtering using a Thermo K-Alpha laboratory SXPS instrument. The core levels are scaled to resemble the intensities if a (a) 1:1 Ti:W ratio was present and (b) a 0.2:0.8 Ti:W ratio was present.}
    \label{fig:Ti_W}
\end{figure}

To truly rule out the possibility that the Ti~3\textit{p} interferes with the W~4\textit{f} region we measured these spectra on titanium and tungsten metal foils, respectively. Fig.~\ref{fig:Ti_W} displays these core levels along with the Ti~2\textit{p}. Fig.~\ref{fig:Ti_W}(a) displays the expected intensity of the three core levels if a 1:1 Ti:W ratio was studied. To obtain these spectra, the Ti~2\textit{p} and W~4\textit{f} core levels were normalised to their respective areas and then their intensity was corrected by applying appropriate RASFs. The Ti~3\textit{p} core level was first normalised to the total area of the Ti~2\textit{p} region (after the subtraction of a Shirley-type background) and then corrected for intensity with the appropriate RASFs. Based on this, the Ti~3\textit{p} core level intensity relative to the W~4\textit{f} is already very small. Taking into account that the ratio of Ti:W for these systems was expected to be approximately 20:80 gives Fig.~\ref{fig:Ti_W}(b). This shows that the Ti~3\textit{p} is now indistinguishable, and appears to be truly unlikely of interfering with the W~4\textit{f} core level.  

\newpage

\section{\label{DP}Complete Depth Profile}

\begin{figure}[ht!]
\centering
    \includegraphics[keepaspectratio, width=0.4\textwidth]{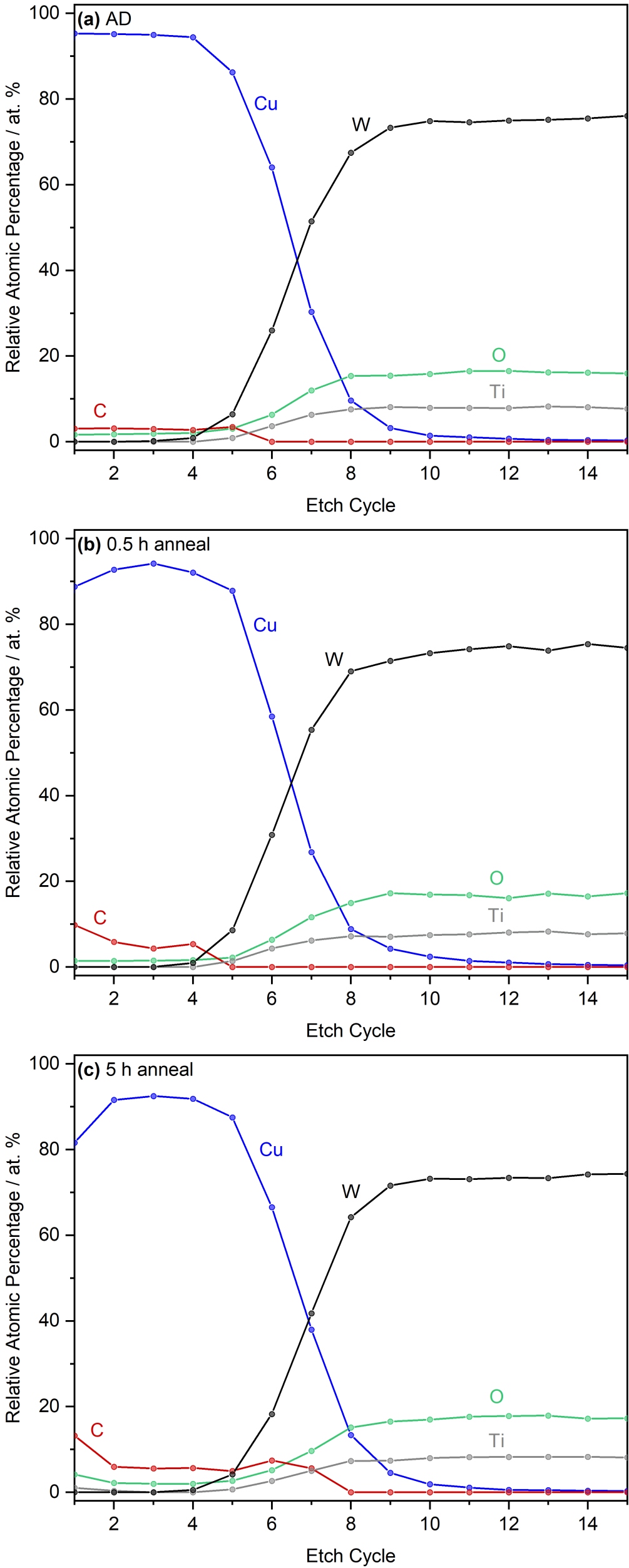}
    \caption{Complete elemental depth profiles collected for the (a) as-deposited, (b) 0.5~h annealed and (c) 5~h annealed samples. Quantification from Etch 0 (i.e. as-recieved) is not included due to the high level of adventitious carbon.}
    \label{fig:DP}
\end{figure}

Fig.~\ref{fig:DP} displays the complete depth profile, including O and C signals, collected using the laboratory-based SXPS instrument. 

\newpage

    \section{\label{sec:Survey_Spectra_Interface}Survey Spectra Collected at TiW/Cu Interface with Synchrotron SXPS and HAXPES Measurements}
    
\begin{figure}[H]
\centering
    \includegraphics[keepaspectratio, width= 0.8\textwidth]{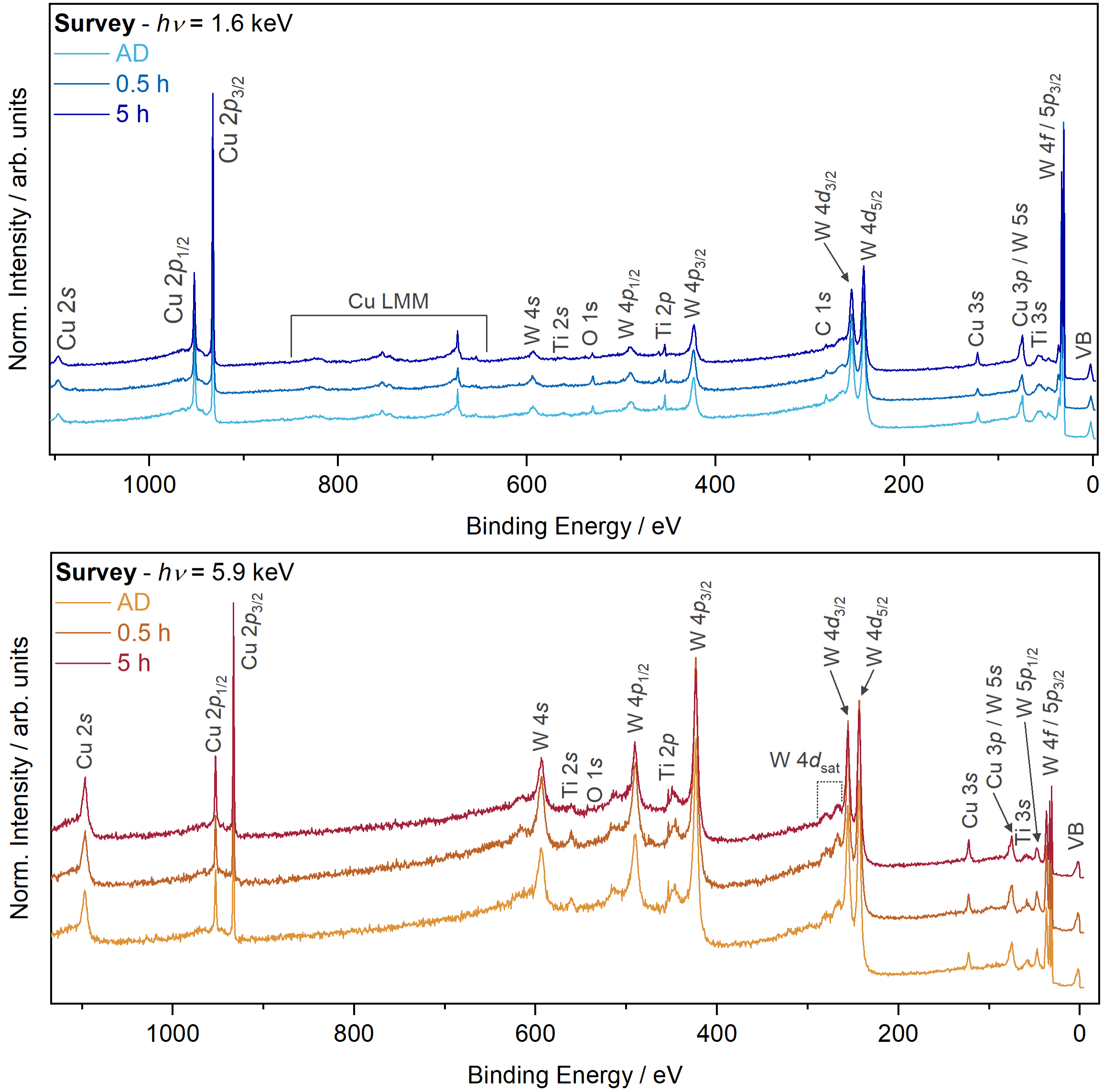}
    \caption{Synchrotron-based SXPS (top) and HAXPES (bottom) survey spectra collected at the TiW/Cu interface after in-situ thinning. Spectra are vertically offset and normalised to the maximum intensity peak to aid with comparison.}
    \label{fig:Interface_Survey}
\end{figure}

Fig.~\ref{fig:Interface_Survey} displays the survey spectra collected using SXPS ($h\nu$ = 1.6~keV) and HAXPES ($h\nu$ = 5.9~keV) at the TiW/Cu interface after the in-situ thinning of the copper metallisation. All observed core and Auger lines are indicated.

\newpage

    \section{\label{sec:Cross Section}Photoionisation Cross Sections of the Key Core Level Spectra}

\begin{figure}[H]
\centering
    \includegraphics[keepaspectratio, width=0.6\textwidth]{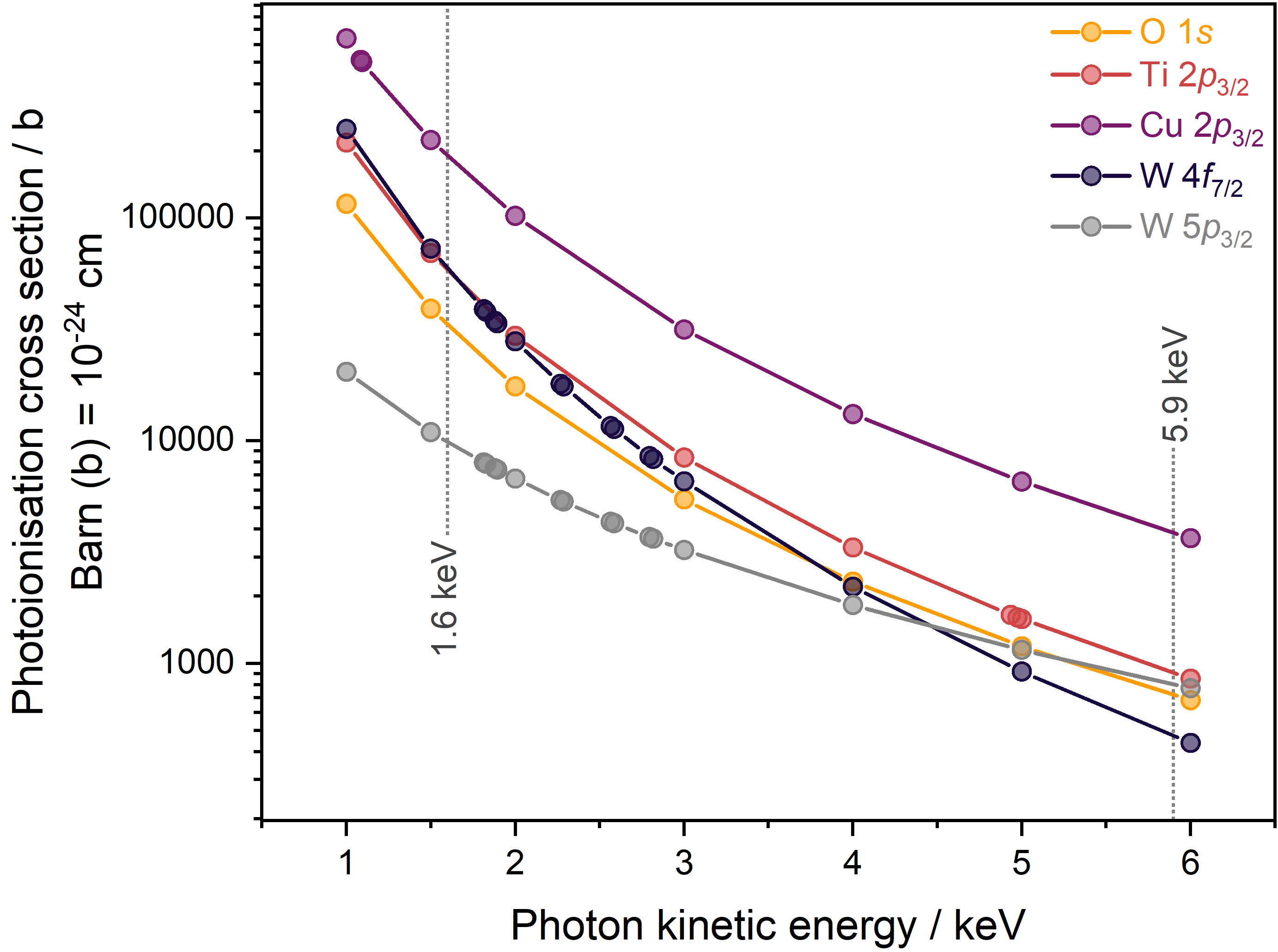}
    \caption{Total electron photoionisation cross sections of O~1\textit{s}, Ti~2\textit{p}\textsubscript{3/2}, Cu~2\textit{p}\textsubscript{3/2}, W~4\textit{f}\textsubscript{7/2} and W~5\textit{p}\textsubscript{3/2} orbitals.~\cite{Scofield1973TheoreticalKeV, Kalha20} Guide lines for the 1.6~keV and 5.9~keV photon energies used in this work are included.}
    \label{fig:Etch}
\end{figure}

\newpage

\section{\label{sec:Differences}Normalisation of Interface SXPS data}

\begin{figure}[H]
\centering
    \includegraphics[keepaspectratio, width=\textwidth]{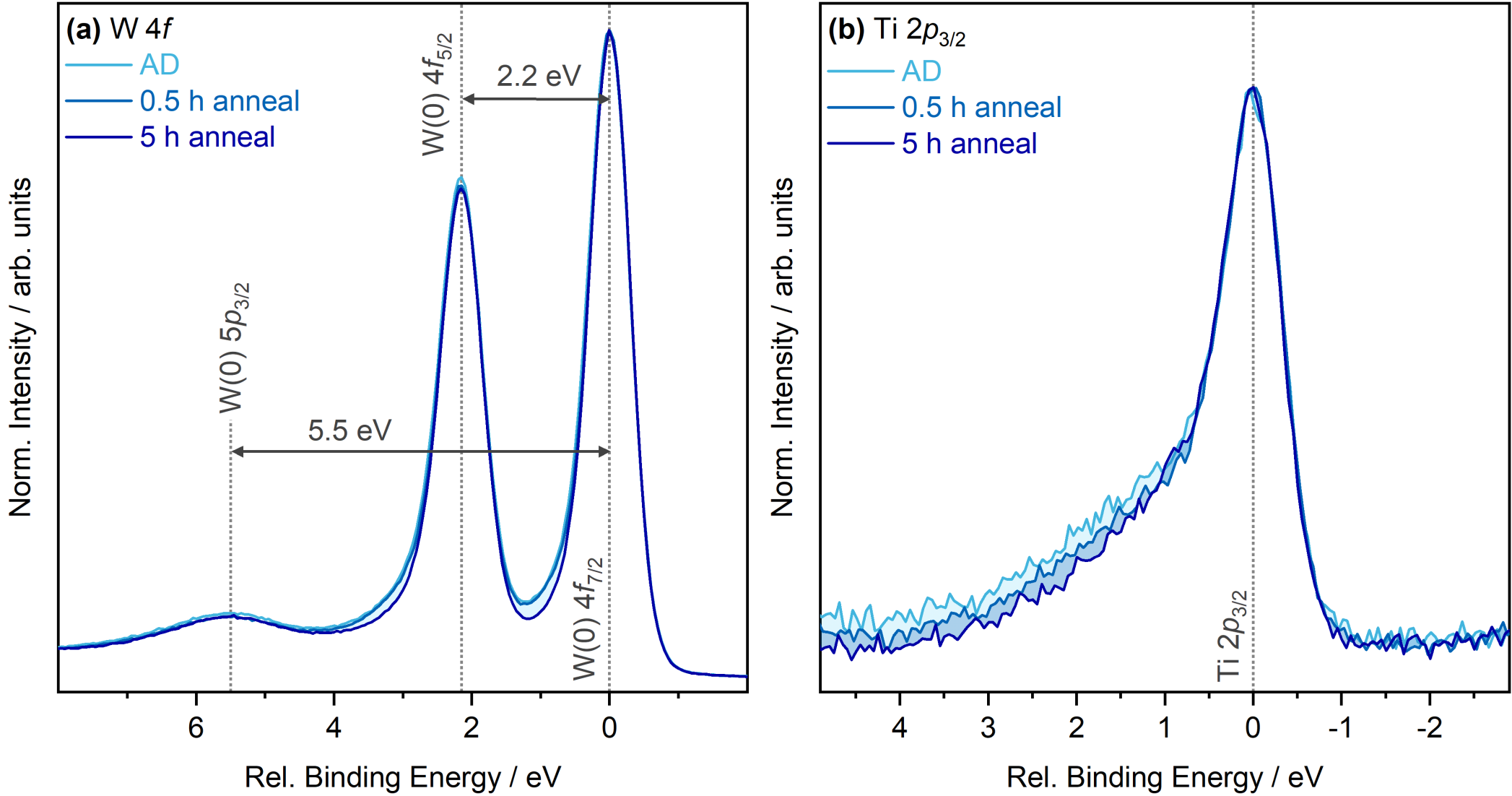}
    \caption{SXPS ($h\nu$ = 1.6~keV) core level spectra collected at the TiW/Cu interface after the in-situ thinning of the copper capping layer, including the (a) W~4\textit{f} and (b) Ti~2\textit{p} core levels. Spectra are plotted on a relative BE scale and are normalised to their maximum intensity.}
    \label{fig:Interface_Norm}
\end{figure}

Fig.~\ref{fig:Interface_Norm} displays the SXPS W~4\textit{f} and Ti~2\textit{p} core level spectra collected at the TiW/Cu interface after in the in-situ thinning of the copper but these spectra are plotted on a relative binding energy (BE) scale and normalised to their maximum intensity, to aid with observing the changes in spectral line shapes between the samples.

\newpage


    \section{\label{sec:Difference}Ti~2\textit{p} difference plots collected at the TiW/Cu Interface}

\begin{figure}[H]
\centering
    \includegraphics[keepaspectratio, width= \textwidth]{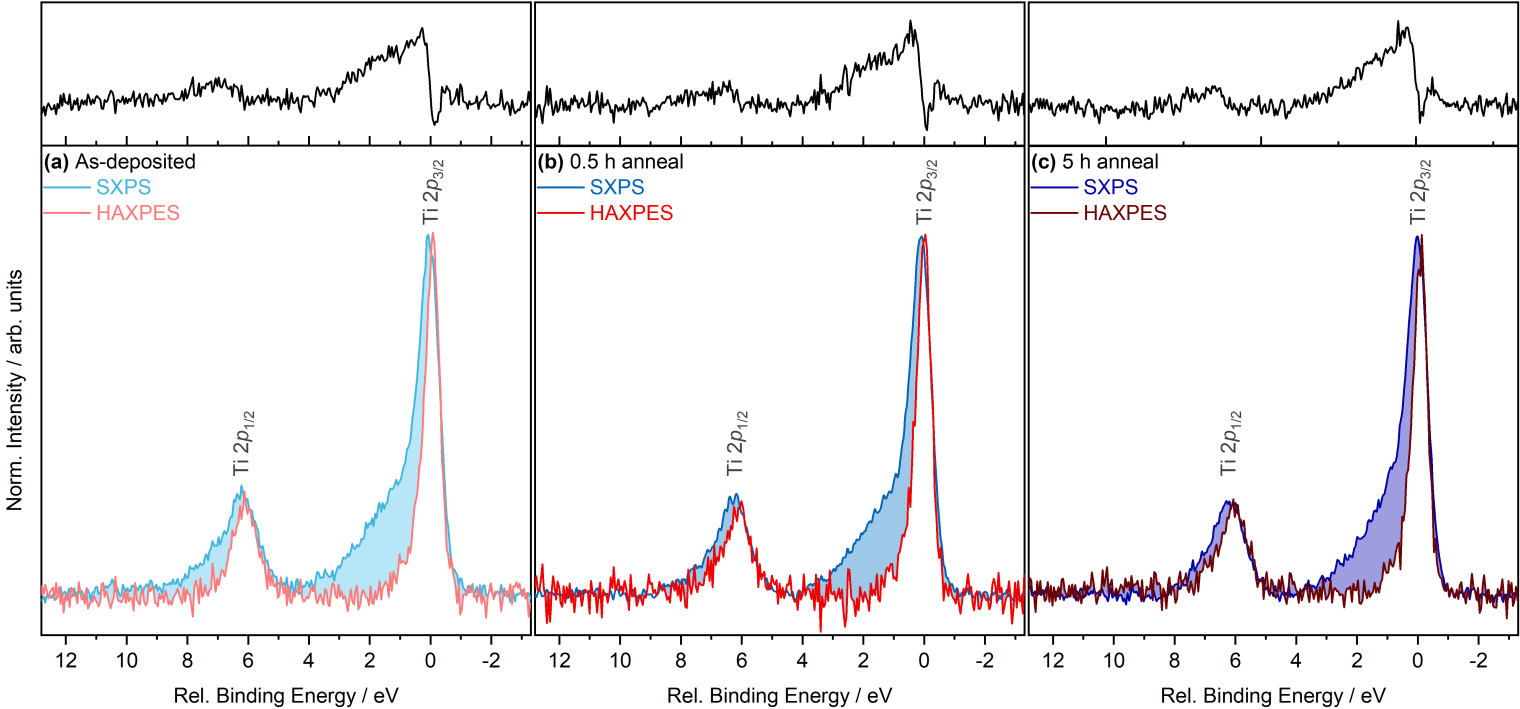}
    \caption{Comparison of the Ti~2\textit{p} spectra collected with SXPS (\textit{h$\nu$} = 1.6~keV) and HAXPES (\textit{h$\nu$} = 5.9~keV). Spectra were normalised to their maximum intensity after the removal of a non-linear curve background. The spectra were additionally aligned so that the lower BE side of the Ti~2\textit{p}\textsubscript{3/2} peak of both spectra were aligned. Difference plots calculated by subtracting the two spectra are displayed above each spectra. Comparison of the collected spectra for the (a) as-deposited, (b) 0.5~h anneal and (c) 5~h samples.}
    \label{fig:Difference}
\end{figure}

Fig.~\ref{fig:Difference} shows the direct comparison of the SXPS and HAXPES recorded Ti~2\textit{p} spectra collected at the TiW/Cu interface, using a spectral subtraction method to aid with unravelling this complex difference. Peak fitting is difficult in this situation due to the asymmetric nature of the main metallic peak and the spin-orbit spitting effects arising from the \textit{p} orbital. The subtraction of the HAXPES core line from the SXPS core line reveals an increase in intensity between the doublet peaks.

\newpage

\section{References}

\bibliography{References_SI}
\bibliographystyle{apsrev4-1.bst}